\documentclass[pra,aps,preprint,amsmath,amssymb]{revtex4}
\usepackage{graphicx}
\usepackage{epsfig}
\include{graphics}
\usepackage{epstopdf} 
\begin{document}

\title{Radiation dynamics in fast soliton collisions in the presence of cubic loss}

\author{Avner Peleg$^{1}$ and Debananda Chakraborty$^{2}$}

\affiliation{$^{1}$ Department of Basic Sciences, Afeka College of Engineering, 
Tel Aviv 69988, Israel}
\affiliation{$^{2}$ Department of Mathematics, New Jersey City University, 
Jersey City, NJ 07305, USA}

\date{\today}

\begin{abstract} 
We study the dynamics of emission of radiation (small-amplitude waves) 
in fast collisions between two solitons of the nonlinear Schr\"odinger (NLS) equation  
in the presence of weak cubic loss. We calculate the radiation dynamics 
by a perturbation technique with two small parameters: 
the cubic loss coefficient $\epsilon_{3}$ and the reciprocal of the group velocity 
difference $1/\beta$. The agreement between the perturbation theory predictions 
and the results of numerical simulations with the full coupled-NLS propagation 
model is very good for large $\beta$ values, and is good for intermediate 
$\beta$ values. Additional numerical simulations with four simplified NLS 
models show that the differences between perturbation theory and simulations 
for intermediate $\beta$ values are due to the effects of Kerr nonlinearity on 
interpulse interaction in the collision. Thus, our study demonstrates that the 
perturbation technique that was originally developed to study radiation dynamics 
in fast soliton collisions in the presence of conservative perturbations can also be 
employed for soliton collisions in the presence of dissipative perturbations.    
\end{abstract}

\maketitle

\section{Introduction}
\label{Introduction}

The cubic nonlinear Schr\"odinger (NLS) equation, which describes wave propagation 
in the presence of cubic (Kerr) nonlinearity and second-order dispersion, 
is one of the most widely used nonlinear wave models in science. 
It was successfully employed to describe water wave dynamics \cite{Zakharov84,Newell85}, 
nonlinear waves in plasmas \cite{Asano69,Horton96}, 
Bose-Einstein condensates \cite{Dalfovo99,BEC2008}, 
and propagation of pulses of light through nonlinear 
optical waveguides \cite{Agrawal2001,Hasegawa95,Iannone98}.  
The fundamental NLS solitons are the most important solutions 
of the cubic NLS equation due to their stability and shape preserving properties. 
Owing to these properties, fundamental NLS solitons are being considered 
for applications in many nonlinear optical waveguide systems,  
including optical waveguide communication links, pulsed waveguide lasers, 
and optical switches \cite{Agrawal2001,Iannone98,Mollenauer2006,Agrawal2001B}.

As mentioned above, a major potential application for fundamental 
NLS solitons is in nonlinear optical waveguide communication links. 
The rates of transmission of information in optical waveguide links can be significantly 
enhanced by sending many pulse sequences through the same 
waveguide \cite{Agrawal2001,Iannone98,Mollenauer2006}. 
In this multisequence transmission method, the pulses in each sequence propagate 
with the same central frequency and group velocity, but the central frequency 
and group velocity are different for pulses from different 
sequences \cite{Agrawal2001,Iannone98,Mollenauer2006}. 
Since the pulses in each sequence have the same central frequency, 
each sequence is called a frequency channel, and multisequence transmission 
is also called multichannel transmission. Furthermore, since pulses from different 
sequences propagate with different group velocities, 
intersequence (interchannel) pulse collisions are very frequent, 
and can therefore lead to severe transmission degradation. 
For this reason, many research efforts have been devoted for studying 
intersequence pulse collisions in general \cite{Agrawal2001,Tkach97,Essiambre2010}, 
and intersequence soliton collisions in particular \cite{Agrawal2001,Hasegawa95,Iannone98,Mollenauer2006}.

Almost all intersequence collisions in multisequence soliton transmission are complete 
collisions, i.e., collisions in which the two solitons are well-separated before and 
after the collision \cite{Mollenauer2006,MM98}. 
In addition, most intersequence collisions in these systems 
are fast, that is, the difference between the central frequencies 
(and group velocities) of the pulses is much larger than the spectral 
width of each pulse \cite{Mollenauer2006,MM98,PNH2017B}. 
For this reason, in the current paper, we focus attention on the 
investigation of complete fast two-soliton collisions. In an ideal optical waveguide, 
in which the only processes affecting the propagation are due to 
second-order dispersion and Kerr nonlinearity, a complete two-soliton collision 
is elastic in the sense that the soliton's amplitude, frequency, and shape do 
not change due to the collision. Furthermore, no radiation is emitted by the 
colliding solitons. The only effects of a complete fast two-soliton collision in 
an ideal waveguide are a phase shift proportional to $1/|\beta|$, 
and a position shift proportional to $1/(|\beta|\beta)$, where $\beta$ 
is the frequency difference between the solitons \cite{MM98,CP2005,PCG2003}.

However, in real nonlinear optical waveguides, the elastic nature of soliton collisions  
breaks down due to the presence of additional physical processes, 
other than second-order dispersion and Kerr nonlinearity. 
In typical soliton-based multisequence optical waveguide systems, 
these additional physical processes can be regarded as small perturbations to the 
cubic NLS equation \cite{Agrawal2001,Iannone98,Mollenauer2006,CP2005,PCG2003}. 
In this case, a complete fast two-soliton collision 
can lead to emission of radiation, changes in the soliton's amplitude, 
frequency, and shape, and additional phase and position shifts. 
The magnitude of the collision-induced changes in the soliton parameters 
and of the amplitude of the emitted radiation is typically proportional 
to $\epsilon/|\beta|^{n}$, where $\epsilon$ is a small parameter characterizing 
the strength of the perturbation, and $n$ is a nonnegative integer 
\cite{CP2005,PCG2003,Malomed89,Malomed86,Malomed91a,
Malomed91b,Kumar98,SP2004}.
Collision-induced emission of radiation is of special importance, since 
it can lead to transmission degradation and to transmission destruction 
in the following ways. (1) In multisequence transmission, the radiation 
emitted by solitons from a given sequence resonantly interacts with other 
sequences \cite{CPN2016,PNT2016}. The resonant interaction leads to 
generation of peaks in the Fourier transforms of the pulse sequences 
(radiative sidebands) and to corruption and destruction 
of the soliton patterns \cite{CPN2016,PNT2016}. 
(2) The emitted radiation leads to long-range interaction between 
solitons from the same sequence \cite{CCDG2003}. The long-range interaction might cause  
large relative position shifts between solitons from the same sequence 
and even undesirable intrasequence soliton collisions \cite{CCDG2003}. 
(3) The collision-induced emitted radiation can undergo unstable growth 
in the presence of Kerr nonlinearity and develop into new (unwanted) 
solitons \cite{Agrawal2001}. For this reason, it is important to investigate 
and analyze the dynamics of radiation emitted in complete fast two-soliton collisions.

In previous works, we developed a perturbation technique that describes 
the effects of weak perturbations on complete fast 
two-soliton collisions \cite{PCG2003,SP2004,PCG2004}. 
The perturbation technique is based on the fact that for these collisions, 
$\epsilon$ and $1/|\beta|$ are two small parameters. 
In the first step of the perturbation procedure, we find a soliton solution 
to the  single-pulse propagation problem, either in an exact form \cite{SP2004}, 
or in an approximate form, by a perturbation expansion 
with respect to $\epsilon$ \cite{PCG2003,PCG2004,PC2003}. 
We then look for a solution of the collision problem in the form 
of a sum of the two soliton solutions of the single-pulse propagation 
problems for solitons 1 and 2, plus a term that describes the collision-induced effects. 
The collisional term is also a sum of two terms $\phi_{1}$ and $\phi_{2}$, 
which oscillate with the frequencies of solitons 1 and 2, and which describe 
the collision effects on these solitons. 
We then substitute the two-pulse solution into the perturbed NLS equation 
and obtain evolution equations for $\phi_{1}$ and $\phi_{2}$, 
which we solve by expanding $\phi_{1}$ and $\phi_{2}$ in perturbation series 
with respect to both $\epsilon$ and $1/|\beta|$ \cite{PCG2003,SP2004,PCG2004}. 
The collision-induced changes in the four soliton parameters are calculated by 
projecting the equations for $\phi_{1}$ and $\phi_{2}$ on the four localized 
eigenfunctions of the linear operator ${\cal L}$, which describes 
the evolution of small perturbations about the NLS soliton, 
and by integrating the equations with respect to propagation distance $z$ \cite{PCG2003,SP2004,PCG2004}. Furthermore, the collision-induced radiation 
dynamics is calculated by projecting the equations for $\phi_{1}$ and $\phi_{2}$ 
on the nonlocalized eigenfunctions of the linear operator ${\cal L}$, 
and by integrating with respect to $z$ \cite{PCG2003,SP2004,PCG2004}.

The perturbation technique of Refs. \cite{PCG2003,PCG2004,SP2004} was 
originally developed to treat fast two-soliton collisions in the presence of 
conservative perturbations. Indeed, only in this case, the first step of the 
perturbation procedure (finding soliton solutions to the single-pulse 
propagation problem) can be carried out. The perturbation technique 
was first employed for studying fast two-soliton collisions in the presence 
of conservative perturbations due to third-order dispersion 
\cite{PCG2003,PCG2004} and quintic nonlinearity \cite{SP2004}.      
Later on, it was shown that the perturbation approach can also be 
employed for calculating changes in the four soliton parameters 
in fast two-soliton collisions in the presence of weak dissipative perturbations, 
such as nonlinear loss \cite{PNC2010,PC2012b} 
and delayed Raman response \cite{CP2005,P2004,NP2010b}. 
The main idea behind this important extension was that in fast collisions 
in the presence of dissipative perturbations, the changes in the 
solitons amplitude and frequency can be approximately described as jumps in the values 
of these parameters. It follows that the collision-induced changes 
in soliton parameters can be calculated from the numerical simulations results 
by subtracting the parameter changes due to single-pulse propagation 
from the total numerical parameter shifts \cite{PNC2010,PC2012b,NP2010b}. 
This enabled accurate comparison between numerical simulations results 
and perturbation theory predictions for the changes in soliton parameters 
up to third order of the perturbation expansion \cite{PNC2010,PC2012b,NP2010b}. 
However, the situation is very different for radiation dynamics 
in fast soliton collisions in the presence of dissipative perturbations. 
Indeed, since the radiation profile changes both 
with time and with propagation distance, it is very difficult to accurately 
subtract radiation-induced changes in the solitons shapes due to single-pulse propagation 
from the solitons shapes obtained in numerical simulations. 
The dependence of radiation dynamics on soliton amplitudes, 
which change during the collision, adds to the difficulty. 
As a result, it is very difficult to compare the predictions 
of the perturbation theory and the numerical simulations results for 
radiation dynamics in fast soliton collisions in the presence of dissipative perturbations. 
For this reason, despite of its importance, analysis of radiation dynamics in 
these collisions has not been addressed before.

In the current paper, we address this important problem and provide 
the first analysis of radiation emission in fast collisions between NLS solitons 
in the presence of weak dissipative perturbations, considering the cubic loss perturbation 
as a central example. We assume an optical waveguide setup, in which the 
effects of cubic loss on single-pulse propagation can be neglected \cite{cubic_loss_1}. 
Therefore, the dynamics of the fast two-soliton collision is described by a perturbed 
coupled-NLS model, where the perturbation terms are due to the effects of cubic loss 
on two-pulse interaction. Since the effects of cubic loss on single-pulse propagation 
are negligible, the soliton solutions of the single-pulse propagation problems are 
simply the fundamental soliton solutions of the unperturbed NLS equation. 
Thus, the problem of calculating the soliton solutions of the single-pulse 
propagation problems in the presence of dissipation is circumvented, 
and the perturbation technique of Refs. \cite{PCG2003,PCG2004,SP2004} 
can be used.

We employ this perturbation technique to calculate the collision-induced 
radiation dynamics and the pulse profile. We then compare the perturbation 
theory predictions with results of numerical simulations with the full coupled-NLS 
propagation model for two values of the frequency difference $\beta$, 
$\beta=20$ and $\beta=10$, representing large and intermediate frequency 
difference values, respectively. For large $\beta$ values,  
we obtain very good agreement between the numerical simulations results 
and the perturbation theory predictions.  
For intermediate $\beta$ values, we obtain good agreement between 
the simulations results and the perturbation theory predictions, 
but the agreement is not as good as the one obtained for large $\beta$ values. 
To gain further insight into the reasons for the differences between the 
perturbation theory and the simulations with the full perturbed coupled-NLS model 
for intermediate $\beta$ values, we carry out numerical simulations 
with four simplified NLS models. These additional simulations show that the main 
reason for the observed differences is due to additional emission of radiation, 
which is induced by the effects of Kerr nonlinearity on interpulse interaction in the collision. 
Furthermore, the results of the additional simulations with the simplified NLS models 
provide the first strong evidence that stochastic perturbed NLS models, 
which were used in Refs. \cite{CP2005,CP2008,PC2012a} to describe 
multisequence soliton transmission in broadband nonlinear optical 
waveguide systems, correctly capture the collision-induced radiation dynamics 
in these systems.

We choose to study soliton collisions in the presence of cubic loss, 
since cubic loss is important in many nonlinear optical waveguide systems, 
and is therefore a central example for nonlinear dissipative perturbations. 
The optical waveguide's cubic loss can arise due to two-photon absorption (2PA) 
or due to gain/loss saturation \cite{Boyd2008,Agrawal2007a,Dekker2007,Borghi2017}. 
Propagation of optical pulses in the presence of 2PA or cubic loss 
has been studied in many previous works \cite{Malomed89,Stegeman89,Silberberg90,Aceves92,Kivshar95,Tsoy2001,
Silberberg2008,PCDN2009,PNC2010,Gaeta2012}.
The subject gained renewed interest in recent years due to the importance of 2PA 
in silicon nanowaveguides, which are expected to play a major role in many applications 
in optoelectronic devices \cite{Agrawal2007a,Dekker2007,Borghi2017,Gaeta2008,Soref2006}.             
In the current paper, we assume that the effects of cubic loss on single-pulse 
propagation are much weaker compared with the effects of cubic loss on interpulse interaction. 
This situation can be realized, for example, in certain nonlinear semiconductor waveguides, 
in which 2PA associated with the simultaneous absorption of two photons with 
the same wavelength (degenerate 2PA) is much weaker than 2PA associated 
with the simultaneous absorption of two photons with different wavelengths 
(nondegenerate 2PA) \cite{Hagan2002,Hagan2011,Rauscher97}.

The remainder of the paper is organized in the following manner. 
In Sec. \ref{theory}, we present the calculation of the collision-induced 
radiation dynamics by the perturbation theory. We also show that it is 
possible to obtain the radiation dynamics by analyzing an equivalent 
single-pulse propagation problem. In Sec. \ref{simu}, we present the 
results of numerical simulations with the full coupled--NLS model and 
with four simpler NLS models, and compare these results with the 
perturbation theory predictions. Our conclusions are summarized 
in Sec. \ref{conclusions}. In Appendix \ref{appendA}, we present 
a summary of the adiabatic perturbation theory for the fundamental NLS soliton.  
Appendix \ref{appendB} is devoted to the derivation of the four simpler 
NLS propagation models used in Sec.  \ref{simu}.

\section{Perturbative calculation of collision-induced radiation dynamics}
\label{theory}

\subsection{Propagation equation and perturbation approach}
\label{theory_1}
We consider the dynamics of a fast collision between two solitons 
in a nonlinear optical waveguide with weak cubic loss. 
We assume that the effects of cubic loss on single-pulse 
propagation are negligible compared with the effects of cubic loss on interpulse interaction,  
a situation that can be realized, for example, in certain 
nonlinear semiconductor waveguides  \cite{Hagan2002,Hagan2011,Rauscher97}.  
In addition, we assume that linear loss is compensated for by linear gain, generated by 
distributed Raman amplification \cite{Islam2004,Agrawal2005,Jalali2003,Lipson2004,Cohen2005b}.  
Under these assumptions, the dynamics of the collision can be described by the following 
perturbed coupled-NLS model \cite{PNC2010,Agrawal2007a}: 
\begin{eqnarray} &&
i\partial_z\psi_{1}+\partial_{t}^2\psi_{1}+2|\psi_{1}|^2\psi_{1}
+4|\psi_{2}|^2\psi_{1}=-2i\epsilon_{3}|\psi_{2}|^2\psi_{1},
\nonumber \\&&
i\partial_z\psi_{2}+\partial_{t}^2\psi_{2}
+2|\psi_{2}|^2\psi_{2}+4|\psi_{1}|^2\psi_{2}=
-2i\epsilon_{3}|\psi_{1}|^2\psi_{2},  
\label{rad1}
\end{eqnarray}           
where $\psi_{1}$ and $\psi_{2}$ are the envelopes of the electric fields of the pulses, 
$z$ is propagation distance, $t$ is time, and $\epsilon_{3}$ is the cubic loss 
coefficient, which satisfies $0<\epsilon_{3} \ll 1$ \cite{dimensions1}. 
The terms $\partial_{t}^{2}\psi_{j}$ with $j=1,2$ on the left hand side 
of Eq. (\ref{rad1}) are due to second-order dispersion, while the terms 
$2|\psi_{j}|^2\psi_{j}$ describe the effects of Kerr nonlinearity 
on single-pulse propagation. In addition, the terms $4|\psi_{2}|^2\psi_{1}$ 
and $4|\psi_{1}|^2\psi_{2}$ describe the effects of Kerr nonlinearity 
on interpulse interaction. The terms $-2i\epsilon_{3}|\psi_{2}|^2\psi_{1}$ 
and $-2i\epsilon_{3}|\psi_{1}|^2\psi_{2}$ describe the effects of cubic loss 
on interpulse interaction. The terms $-i\epsilon_{3}|\psi_{j}|^2\psi_{j}$ with $j=1,2$ 
are neglected, since the effects of cubic loss on single-pulse propagation 
are assumed to be very weak. As a result, the single-pulse propagation problems 
for the two pulses are described by the unperturbed NLS equations  $i\partial_z\psi_{j}+\partial_{t}^2\psi_{j}+2|\psi_{j}|^2\psi_{j}=0$ for $j=1,2$.  
The fundamental soliton solutions of these equations are given by: 
\begin{eqnarray} 
\psi_{sj}(t,z)=\Psi_{j}(x_{j})\exp(i\chi_{j})=
\eta_{j}\exp(i\chi_{j})\mbox{sech}(x_{j}),
\label{rad2}
\end{eqnarray}
where $x_{j}=\eta_{j}\left(t-y_{j}+2\beta_{j} z\right)$, 
$\chi_{j}=\alpha_{j}-\beta_{j}(t-y_{j})+
\left(\eta_{j}^2-\beta_{j}^{2}\right)z$, 
and $\eta_{j}$, $\beta_{j}$, $y_{j}$, and $\alpha_{j}$ are the amplitude, 
frequency, position, and phase of the $j$th soliton.


We consider a complete fast collision between two fundamental solitons of 
the unperturbed NLS equation in the presence of weak cubic loss. 
For simplicity and without loss of generality, 
we take the initial soliton frequencies as $\beta_{1}(0)=0$ 
and $\beta_{2}(0)=\beta$. The complete collision assumption means 
that the two solitons are well separated at $z=0$ and at the 
final propagation distance $z=z_{f}$.  
The fast collision assumption means that the frequency difference 
between the solitons is much larger than the soliton spectral width, 
i.e., $|\beta_{2}(0)-\beta_{1}(0)|=|\beta|\gg 1$. These conditions are 
satisfied for almost all collisions in long-distance soliton-based 
multichannel transmission experiments in optical fibers \cite{MM98,Nakazawa2000}.

We study the collision dynamics by employing the perturbation approach
that was developed in Refs. \cite{PCG2003,PCG2004,SP2004} for analyzing radiation dynamics 
in fast two-soliton collisions in the presence of conservative perturbations. 
Following this perturbation approach, we look for a solution of Eq. (\ref{rad1}) 
in the form 
\begin{eqnarray}&&
\!\!\!\!\!\!\!
\psi_{j}(t,z)=\psi_{j0}(t,z)+\phi_{j}(t,z), 
\label{rad3}
\end{eqnarray}       
where $j=1,2$, $\psi_{j0}$ are the solutions of Eq. (\ref{rad1}) without 
the interpulse interaction terms, and $\phi_{j}$ represent small 
corrections to $\psi_{j0}$ due to the collision. 
In the absence of the interpulse interaction terms, Eq. (\ref{rad1}) reduces to 
a pair of uncoupled unperturbed NLS equations. Therefore, in the current problem, 
the $\psi_{j0}$ are simply the fundamental soliton solutions of the unperturbed NLS equation, 
i.e., $\psi_{j0}(t,z)=\psi_{sj}(t,z)$ for $j=1,2$.       
Note that finding the solutions of the single-pulse propagation problems 
in the presence of cubic loss becomes unnecessary because of the omission 
of the terms $-i\epsilon_{3}|\psi_{j}|^2\psi_{j}$ from the propagation equation 
(due to the assumed weakness of cubic loss effects on single-pulse propagation). 
In this manner, the application of the perturbation approach 
of Refs. \cite{PCG2003,PCG2004,SP2004} to soliton collisions in the presence 
of nonlinear dissipation is enabled, despite of the fact that the approach was originally 
developed for treating fast soliton collisions in the presence of conservative perturbations.  
We substitute the ansatz (\ref{rad3}) together with the relations 
$\psi_{j0}(t,z)=\psi_{sj}(t,z)=\Psi_{j}(x_{j})\exp(i\chi_{j})$ and
$\phi_{j}(t,z)=\Phi_{j}(t,z)\exp(i\chi_{j})$ into Eq. (\ref{rad1}) 
and obtain equations for the $\Phi_{j}$ \cite{PCG2003,SP2004,PNC2010}. 
We focus attention on the calculation of $\Phi_{1}$, 
since the calculation of $\Phi_{2}$ is similar. 
The equation for $\Phi_{1}$ is 
\begin{eqnarray} &&
i\partial_z\Phi_{1}+
\left[(\partial_t^2-\eta_{1}^{2})\Phi_{1}
+4|\Psi_{1}|^2\Phi_{1}+2\Psi_{1}^{2}\Phi_{1}^{\ast}\right]
\nonumber\\ &&
=-4\left[|\Psi_{2}|^2\Psi_{1}+|\Psi_{2}|^2\Phi_{1}+
\Psi_{1}\left(\Psi_{2}\Phi_{2}^{\ast}+
\Psi_{2}^{\ast}\Phi_{2}\right)\right]
\nonumber\\ &&
-2i\epsilon_{3}\left[|\Psi_{2}|^2\Psi_{1}+
|\Psi_{2}|^2\Phi_{1}+
\Psi_{1}\left(\Psi_{2}\Phi_{2}^{\ast}+
\Psi_{2}^{\ast}\Phi_{2}\right)
\right].
\label{rad4}
\end{eqnarray}

We solve Eq. (\ref{rad4}) and the corresponding equation for $\Phi_{2}$ 
by expanding the $\Phi_{j}$ in perturbation series with respect to 
$\epsilon_{3}$ and $1/\beta$. More specifically, the expansions of the $\Phi_{j}$ 
are 
\begin{eqnarray} &&
\Phi_{j}(t,z)=\Phi_{j1}^{(0)}(t,z)+\Phi_{j1}^{(1)}(t,z)+
\Phi_{j2}^{(0)}(t,z)
\nonumber \\ &&
+\Phi_{j2}^{(1)}(t,z)+\dots,
\label{rad5}
\end{eqnarray}    
where the first subscript in $\Phi_{jk}^{(m)}$ stands for the pulse index, 
the second subscript indicates the combined order with respect to 
both $\epsilon_{3}$ and $1/\beta$, and the superscript represents 
the order in $\epsilon_{3}$. Substitution of the expansions (\ref{rad5})     
into Eq. (\ref{rad4}) yields linear equations for the $\Phi_{1k}^{(m)}$, 
which can be integrated with respect to $z$. The collision-induced changes 
in the four soliton parameters are calculated by projecting both sides of the 
resulting equations on the four localized eigenfunctions of the linear operator ${\cal L}$, 
which describes the evolution of small perturbations about the fundamental 
soliton of the unperturbed NLS equation (see Appendix \ref{appendA} and 
Refs. \cite{PCG2003,SP2004,CCDG2003,Kaup90,Kaup91} 
for a description of the operator ${\cal L}$ and its eigenfunctions). 
The dynamics of the collision-induced radiation emitted by soliton 1   
is calculated by projecting the equations for the $\Phi_{1k}^{(m)}$ 
on the nonlocalized eigenfunctions of ${\cal L}$. 
This calculation is described in sections \ref{theory_2} and \ref{theory_3}.

The only effect of the collision on soliton 1 in order $1/\beta$ is a phase shift, 
which is given by \cite{MM98,CP2005,PCG2003}: 
\begin{equation} 
\Delta\alpha_{11}^{(0)}=4\eta_{2}/|\beta|. 
\label{rad5_add1}
\end{equation}       
In addition, the only collision-induced effect on soliton 1 in order $1/\beta^{2}$ 
is a position shift, which is given by \cite{MM98,CP2005,PCG2003}: 
\begin{equation} 
\Delta y_{12}^{(0)}=4\eta_{2}/(\beta|\beta|). 
\label{rad5_add2}
\end{equation}  
Both effects are caused by the term $4|\psi_{2}|^2\psi_{1}$ 
in Eq. (\ref{rad1}) \cite{MM98,PCG2003}. That is, the collision-induced effects 
in orders $1/\beta$ and $1/\beta^{2}$ are due to Kerr-induced interpulse 
interaction and not due to cubic loss. In addition, there are no terms of order 
$\epsilon_{3}\beta$ on the right hand side of Eq. (\ref{rad4}) 
and therefore, there are no collision-induced effects in order $\epsilon_{3}$.         
We therefore start the detailed description of the perturbative calculations by
considering the effects of the collision in order $\epsilon_{3}/\beta$.

\subsection{Basic calculation of radiation dynamics in order $\epsilon_{3}/\beta$}
\label{theory_2}
In order $\epsilon_{3}/\beta$, Eq. (\ref{rad4}) reduces to 
\begin{eqnarray} &&
\partial_{z}\Phi_{12}^{(1)}=
-2\epsilon_{3}|\Psi_{2}|^{2}\Psi_{1}=
-\frac{2\epsilon_{3}\eta_{1}\eta_{2}^{2}}
{\cosh(x_{1})\cosh^{2}(x_{2})}.
\label{rad6}
\end{eqnarray}    
We denote by $\Delta z_{c}$ and $z_{c}$ the collision interval and the 
collision distance. The collision interval $\Delta z_{c}$ is the interval along 
which the envelopes of the colliding solitons overlap. It can be estimated by 
$\Delta z_{c}=1/(2|\beta|)$. The collision distance $z_{c}$ is the distance 
at which the maxima of $|\psi_{j}(t,z)|$ coincide. This distance is given by 
$z_{c}=[y_{2}(0)-y_{1}(0)]/(2\beta)$, where $y_{1}(0)$ and $y_{2}(0)$ are the 
initial positions of the solitons. In a fast collision, the collision takes place in a small 
interval $[z_{c}-\Delta z_{c}/2,z_{c}+\Delta z_{c}/2]$ around $z_{c}$. 
Thus, to calculate the collision-induced effects in order $\epsilon_{3}/\beta$, 
we integrate both sides of Eq. (\ref{rad6}) over the collision region. 
This integration yields: 
\begin{eqnarray} &&
\Phi_{12}^{(1)}(t,z_{c}+\Delta z_{c}/2)=
-\frac{2\epsilon_{3}\eta_{1}\eta_{2}^{2}}{\cosh(x_{1})}
\int_{z_{c}-\Delta z_{c}/2}^{z_{c}+\Delta z_{c}/2} 
\!\! \frac{dz'}{\cosh^{2}(x_{2})},
\label{rad7}
\end{eqnarray}          
where we used $\Phi_{12}^{(1)}(t,z_{c}-\Delta z_{c}/2) \simeq 0$. 
Since the integrand on the right hand side of Eq. (\ref{rad7}) 
is sharply peaked at a small interval around $z_{c}$,  
we can extend the integral's limits to $-\infty$ and $\infty$. 
The integration yields the following expression for 
$\Phi_{12}^{(1)}(t,z_{c}+\Delta z_{c}/2)$:  
\begin{eqnarray} &&
\Phi_{12}^{(1)}(t,z_{c}+\Delta z_{c}/2)=
-\frac{2\epsilon_{3}\eta_{1}\eta_{2}}
{|\beta|\cosh(x_{1})}.
\label{rad8}
\end{eqnarray}

We now write $\Phi_{12}^{(1)}(t,z_{c}+\Delta z_{c}/2)$ and its complex 
conjugate in a vector form and expand this vector in terms of the eigenfunctions 
of the linear operator ${\cal L}$: 
\begin{eqnarray} &&
{\Phi_{12}^{(1)}(t,z_{c}+\Delta z_{c}/2)
\choose \Phi_{12}^{(1)\ast}(t,z_{c}+\Delta z_{c}/2)}=
-\frac{2\epsilon_{3}\eta_{1}\eta_{2}}
{|\beta|\cosh(x_{1})}{1 \choose 1}
\nonumber \\ &&
=\sum_{j=0}^{3} \tilde a_{j}f_{j}(x_{1})
+{v_{12b}(t,z_{c}+\Delta z_{c}/2)
\choose v_{12b}^{\ast }(t,z_{c}+\Delta z_{c}/2)}. 
\label{rad9}
\end{eqnarray}        
In Eq. (\ref{rad9}), $\tilde a_{j}$ with $j=0, \dots, 3$ are constants, 
$f_{j}(x_{1})$ with  $j=0, \dots, 3$ are the four localized eigenfunctions 
of ${\cal L}$, defined by Eq. (\ref{perturbation5}), and $v_{12b}(t,z)$ is the radiation 
emitted by soliton 1 in this basic version of the perturbative calculation. 
The radiation part $v_{12b}(t,z)$ is expressed by the following expansion 
in the nonlocalized eigenfunctions $f_{s}(x_{1})$ and $\bar f_{s}(x_{1})$
of the operator ${\cal L}$: 
\begin{eqnarray} &&
{v_{12b}(t,z_{c}+\Delta z_{c}/2)
\choose v_{12b}^{\ast }(t,z_{c}+\Delta z_{c}/2)}=
\nonumber \\ &&
\int_{-\infty}^{+\infty} \frac{ds}{2\pi }
\left[a_{s}(z_{c}+\Delta z_{c}/2) f_{s}(x_{1})
+a_{s}^{\ast}(z_{c}+\Delta z_{c}/2) \bar f_{s}(x_{1})\right],
\label{rad9_add1}
\end{eqnarray}           
where $-\infty < s < \infty$, and the eigenfunctions are defined 
in Eqs. (\ref{perturbation7}) and (\ref{perturbation8}).   
The collision-induced changes in the four soliton parameters are calculated 
by projecting both sides of Eq. (\ref{rad9}) on the four localized 
eigenfunctions of  ${\cal L}$. This calculation shows that the only effect 
of the collision on the soliton parameters in order $\epsilon_{3}/\beta$ 
is an amplitude shift, $\Delta\eta_{12}^{(1)}$, 
which is given by \cite{PNC2010}:  
\begin{eqnarray} &&
\Delta\eta_{12}^{(1)}=
-4\epsilon_{3}\eta_{1}\eta_{2}/|\beta|.
\label{rad10}
\end{eqnarray}                     
Furthermore, the radiation part of $\Phi_{12}^{(1)}$ at $z=z_{c}+\Delta z_{c}/2$,  
$v_{12b}(t,z_{c}+\Delta z_{c}/2)$, is obtained by 
projecting both sides of Eq. (\ref{rad9}) on the nonlocalized 
eigenfunctions of  ${\cal L}$. This calculation yields the following 
expression for $a_{s}(z_{c}+\Delta z_{c}/2)$:  
\begin{eqnarray} &&
a_{s}(z_{c}+\Delta z_{c}/2)=
\frac{2\pi\epsilon_{3}\eta_{1}\eta_{2}(s+i)^{2}}
{|\beta|(s^{2}+1)\cosh(\pi s/2)}.
\label{rad11}
\end{eqnarray}

Outside of the collision interval, that is, for distances $z>z_{c}+\Delta z_{c}/2$, 
the two solitons are no longer overlapping. As a result, in this post-collision interval, 
the term $-2i\epsilon_{3}|\Psi_{2}|^2\Psi_{1}$ and all other interpulse interaction 
terms on the right hand side of Eq. (\ref{rad4}) are exponentially small and can 
be neglected. Thus, for $z>z_{c}+\Delta z_{c}/2$, the equation describing the 
evolution of $\Phi_{12}^{(1)}$ is: 
\begin{eqnarray} &&
\partial_z \Phi_{12}^{(1)}
-i\left[(\partial_{t}^2-\eta_{1}^{2})\Phi_{12}^{(1)}
+4|\Psi_{1}|^2\Phi_{12}^{(1)}
+2\Psi_{1}^{2}\Phi_{12}^{(1)\ast}\right]=0.
\label{rad12}
\end{eqnarray}    
This equation and its complex conjugate can be written as 
\begin{eqnarray}
\partial _{z} {\Phi_{12}^{(1)} \choose  \Phi_{12}^{(1)}}-
i\eta_{1}^{2}{\cal {L}}{\Phi_{12}^{(1)} \choose  \Phi_{12}^{(1)}}=0.
\label{rad12_add1} 
\end{eqnarray}
Since $\Phi_{12}^{(1)}(t,z)=\sum_{j=0}^{3} \tilde a_{j}f_{j1}(x_{1})
+v_{12b}(t,z)$, the equation describing radiation dynamics in the post-collision 
interval is 
\begin{eqnarray}
\partial _{z} {v_{12b} \choose v_{12b}^{\ast}}-
i\eta_{1}^{2}{\cal {L}}{v_{12b} \choose v_{12b}^{\ast}}=0.
\label{rad13} 
\end{eqnarray} 
We solve Eq. (\ref{rad13}) with an initial condition 
at $z=z_{c}+\Delta z_{c}/2$, which is given by Eqs. (\ref{rad9_add1}) and 
(\ref{rad11}). For this purpose, we first expand $v_{12b}(t,z)$ in 
the nonlocalized eigenfunctions of ${\cal L}$
\begin{eqnarray} &&
{v_{12b}(t,z)\choose v_{12b}^{\ast }(t,z)}=
\int_{-\infty}^{+\infty} \frac{ds}{2\pi }
\left[a_{s}(z) f_{s}(x_{1})+a_{s}^{\ast}(z) \bar f_{s}(x_{1})\right].
\label{rad14}
\end{eqnarray}                    
Projecting both sides of Eq. (\ref{rad13}) on the nonlocalized 
eigenfunctions of ${\cal L}$, while using the expansion (\ref{rad14}),   
we obtain the following equation for the expansion coefficients
\begin{equation}
\frac{d a_{s}(z)}{dz} - i\eta_{1}^{2}(s^{2}+1)a_{s}(z)=0.
\label{rad15}
\end{equation}   
Integrating Eq. (\ref{rad15}) with respect to $z$ and using the initial 
condition (\ref{rad11}), we arrive at 
\begin{eqnarray}&&
a_{s}(z \geq z_{c}+\Delta z_{c}/2) =
\frac{2\pi\epsilon_{3}\eta_{1}\eta_{2}(s+i)^{2}}
{|\beta|(s^{2}+1)\cosh(\pi s/2)}
\exp\left[i\eta_{1}^{2}(s^{2}+1)(z-z_{c})\right].
\label{rad16}
\end{eqnarray}  
Thus, the dynamics of the collision-induced radiation emitted by soliton 1  
in the post-collision interval is fully described by Eqs. (\ref{rad14}) 
and (\ref{rad16}).

The envelope of the electric field of soliton 1 in the post-collision 
interval is given by: $\psi_{1b}(t,z)=\psi_{10}(t,z)+\phi_{1b}(t,z)$, 
where $\phi_{1b}(t,z)$ describes the effects of the collision on soliton 1 in the 
basic version of the perturbation theory. $\psi_{1b}(t,z)$ can be written as: 
\begin{eqnarray}&&
\psi_{1b}(t,z)=\left \{\frac{\eta_{1}}{\cosh(x_{1})} 
- \Delta\eta_{12}^{(1)}\frac{\left[x_{1} \tanh(x_{1}) -1\right]}{\cosh(x_{1})}
+ v_{12b}(t,z) \right \}
\nonumber \\ &&
\times \exp \left \{i \left[ \chi_{1}(t,z) + \Delta\alpha_{11}^{(0)} \right] \right \},
\label{rad17}
\end{eqnarray}   
where $x_{1}=\eta_{1}[t-y_{1}(z)]$, 
$\chi_{1}(t,z)=\alpha_{1}(0)+ \eta_{1}^2 z$, 
$\Delta\alpha_{11}^{(0)}$ is given by Eq. (\ref{rad5_add1}), 
$\Delta\eta_{12}^{(1)}$ is given by Eq. (\ref{rad10}), 
and $\alpha_{1}(0)$ is the initial phase of soliton 1. 
In addition, $y_{1}(z) = y_{1}(0)+\Delta y_{12}^{(0)} + y_{1}^{(C)}(z)$, 
where $\Delta y_{12}^{(0)}$ is given by Eq. (\ref{rad5_add2}), and 
$y_{1}^{(C)}(z)=[40\epsilon_{3}\eta_{1}^{2}\eta_{2}(z-z_{c})]/(3|\beta|\beta)$  
is the position shift arising from the collision-induced frequency shift 
[see Eq. (\ref{simple_NLS_1A}) in Appendix \ref{appendB}].

\subsection{Taking into account propagation of radiation in the collision interval}
\label{theory_3}
The perturbative calculation described in Sec. \ref{theory_2} is based 
on neglecting the effects of propagation of radiation in the collision interval. 
More accurately, in writing Eq. (\ref{rad6}), we neglected the terms 
$-i(\partial_{t}^2-\eta_{1}^{2})\Phi_{12}^{(1)}$, 
$-4i|\Psi_{1}|^2\Phi_{12}^{(1)}$, and $-2i\Psi_{1}^{2}\Phi_{12}^{(1)\ast}$, 
which describe these effects, since these terms are of 
order $\epsilon_{3}/\beta^{2}$. However, for $|t|>z$,  
the radiation profile can receive significant contributions 
from fast moving waves with group velocities larger than $|\beta|$.  
Therefore, it might be important to include the contributions of 
these fast moving waves to the radiation profile by taking into account 
propagation of radiation in the collision interval 
(see also Ref. \cite{SP2004}, where a similar calculation was carried out 
for fast collisions between NLS solitons in the presence of quintic nonlinearity). 
We therefore turn to describe an improved perturbation approach 
that achieves this goal.

Following the calculation in Ref. \cite{SP2004}, we denote by $\Phi_{12c}^{(1)}$ 
the part of $\Phi_{1}$ that describes the collision-induced effects in order $\epsilon_{3}/\beta$
{\it and} the effects of propagation of radiation in the collision interval. 
The equation for the evolution of $\Phi_{12c}^{(1)}$ is: 
\begin{eqnarray} &&
\partial_z \Phi_{12c}^{(1)}
-i\left[(\partial_{t}^2-\eta_{1}^{2})\Phi_{12c}^{(1)}
+4|\Psi_{1}|^2\Phi_{12c}^{(1)}
+2\Psi_{1}^{2}\Phi_{12c}^{(1)\ast}\right]
=-2\epsilon_{3}|\Psi_{2}|^{2}\Psi_{1}.
\label{rad21}
\end{eqnarray}    
This equation and its complex conjugate can be written as 
\begin{eqnarray}
\partial _{z} {\Phi_{12c}^{(1)}\choose\Phi_{12c}^{(1)\ast}}-
i\eta_{1}^{2}{\cal L}{\Phi_{12c}^{(1)}\choose\Phi_{12c}^{(1)\ast}}
=-\frac{2\epsilon_{3}\eta_{1}\eta_{2}^{2}}
{\cosh(x_{1})\cosh^{2}(x_{2})}{1 \choose 1}.
\label{rad22} 
\end{eqnarray}        
We express $\Phi_{12c}^{(1)}$ in the form 
\begin{eqnarray} &&
{\Phi_{12c}^{(1)}(t,z) \choose \Phi_{12c}^{(1)\ast}(t,z)}
=\sum_{j=0}^{3} \tilde a^{(c)}_{j}f_{j}(x_{1})
+{v_{12c}(t,z) \choose v_{12c}^{\ast }(t,z)},  
\label{rad23}
\end{eqnarray}        
where the radiation part $v_{12c}(t,z)$ is expanded in the nonlocalized eigenfunctions 
of ${\cal L}$
\begin{eqnarray} &&
{v_{12c}(t,z)\choose v_{12c}^{\ast }(t,z)}=
\int_{-\infty}^{+\infty} \frac{ds}{2\pi }
\left[a^{(c)}_{s}(z) f_{s}(x_{1})+a^{(c)\ast}_{s}(z) \bar f_{s}(x_{1})\right].
\label{rad24}
\end{eqnarray}                     
The expansion of the right hand side of Eq. (\ref{rad22}) in the 
eigenfunctions of ${\cal L}$ is 
\begin{eqnarray} &&
-\frac{2\epsilon_{3}\eta_{1}\eta_{2}^{2}}
{\cosh(x_{1})\cosh^{2}(x_{2})}{1 \choose 1}=
\sum_{j=0}^{3}{\tilde b_{j}}f_{j}(x_{1})
\nonumber \\ &&
+\int_{-\infty}^{+\infty} \frac{{\rm d}s}{2\pi }
\left[b_{s}(z)f_{s}(x_{1})
+b_{s}^{\ast}(z)\bar f_{s}(x_{1})\right],
\label{rad25}
\end{eqnarray}      
where 
\begin{equation}
b_{s}(z)=2\epsilon_{3}\eta_{1}\eta_{2}^{2}
\int_{-\infty}^{\infty} 
\frac{dx_{1}\,f_{s}^{+}(x_{1})\pmb{\sigma_{3}}}
{\cosh(x_{1})\cosh^{2}(x_{2})}{1 \choose 1}.
\label{rad26}
\end{equation}

We substitute the expansions (\ref{rad23})-(\ref{rad26}) into 
Eq. (\ref{rad22}) and project both sides of the resultant equation 
on the nonlocalized eigenfunctions of ${\cal L}$. 
This calculation yields the following equation for the $a^{(c)}_{s}$  
coefficients: 
\begin{equation}
\frac{d a^{(c)}_{s}(z)}{dz} - i\eta_{1}^{2}(s^{2}+1)a^{(c)}_{s}(z)=b_{s}(z).
\label{rad27}
\end{equation}       
The solution of Eq. (\ref{rad27}) on the interval $[z_{c}-\Delta z_{c}/2, z]$ is 
\begin{equation}
a_{s}^{(c)}(z)=\int_{z_{c}-\Delta z_{c}/2}^{z} dz' \,b_{s}(z')
\exp\left[i\eta_{1}^{2}(s^{2}+1)(z-z')\right], 
\label{rad28}
\end{equation}
where $a_{s}^{(c)}(z_{c}-\Delta z_{c}/2) \simeq 0$ is used.  
The integrand on the right hand side of Eq. (\ref{rad28}) 
is sharply peaked at a small interval around $z_{c}$.
Therefore, we can extend the integral's limits to $-\infty$ and $\infty$ 
and obtain
\begin{equation}
a_{s}^{(c)}(z)=\int_{-\infty}^{\infty} dz' \,b_{s}(z')
\exp\left[i\eta_{1}^{2}(s^{2}+1)(z-z')\right]. 
\label{rad29}
\end{equation} 
Substituting Eq. (\ref{rad26}) into Eq. (\ref{rad29}) and carrying out 
the integrations we arrive at: 
\begin{eqnarray} &&
a^{(c)}_{s}(z)=
\frac{-2\pi^{2}\epsilon_{3}\eta_{1}\eta_{2}}{4|\beta|AB}
\frac{(s+i)^{2}(s-B)}
{\sinh\left[\frac{\pi(s^{2}+1)}{4A}\right]
\cosh\left[\frac{\pi(s^{2}+1-2Bs)}{4B}\right]}
\nonumber \\ &&
\times \exp\left[i\eta_{1}^{2}(s^{2}+1)(z-z_{c})\right],
\label{rad30}
\end{eqnarray}
where $A=\eta_{2}\beta/\eta_{1}^{2}$ and $B=\beta/\eta_{1}$. 
It is straightforward to show that in the limit $|\beta| \gg 1$, 
Eq. (\ref{rad30}) reduces to Eq. (\ref{rad16}).   
Equations (\ref{rad24}) and (\ref{rad30}) describe the evolution of 
the radiation profile $|v_{12c}(t,z)|$ in the improved perturbation approach.
The envelope of the electric field of soliton 1 in the improved perturbation approach 
is $\psi_{1c}(t,z)=\psi_{10}(t,z)+\phi_{1c}(t,z)$,  where $\phi_{1c}(t,z)$ describes 
the effects of the collision in this version of the perturbation approach. 
In the post-collision interval, $\psi_{1c}(t,z)$ can be written as: 
\begin{eqnarray}&&
\psi_{1c}(t,z)=\left \{\frac{\eta_{1}}{\cosh(x_{1})} 
- \Delta\eta_{12}^{(1)}\frac{\left[x_{1} \tanh(x_{1}) -1\right]}{\cosh(x_{1})}
+ v_{12c}(t,z) \right \}
\nonumber \\ &&
\times \exp \left \{i \left[ \chi_{1}(t,z) + \Delta\alpha_{11}^{(0)} \right] \right \},
\label{rad30_add1}
\end{eqnarray}   
where $x_{1}$, $\chi_{1}(t,z)$, and $\Delta\alpha_{11}^{(0)}$ were 
defined in the last paragraph of Sec. \ref{theory_2}.

\begin{figure}[ptb]
\begin{center}
\begin{tabular}{cc}
\epsfxsize=9.0cm  \epsffile{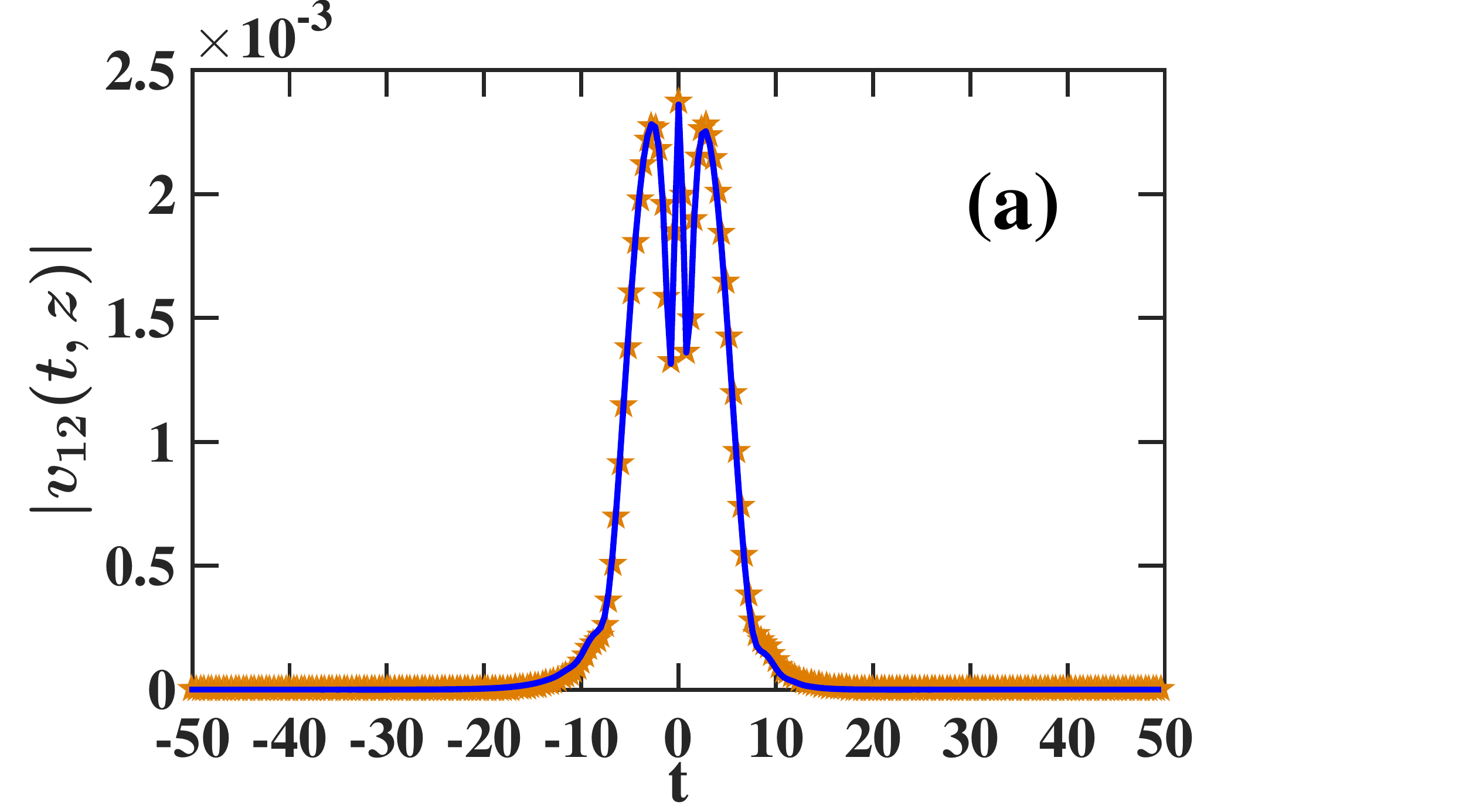} \\
\epsfxsize=9.0cm  \epsffile{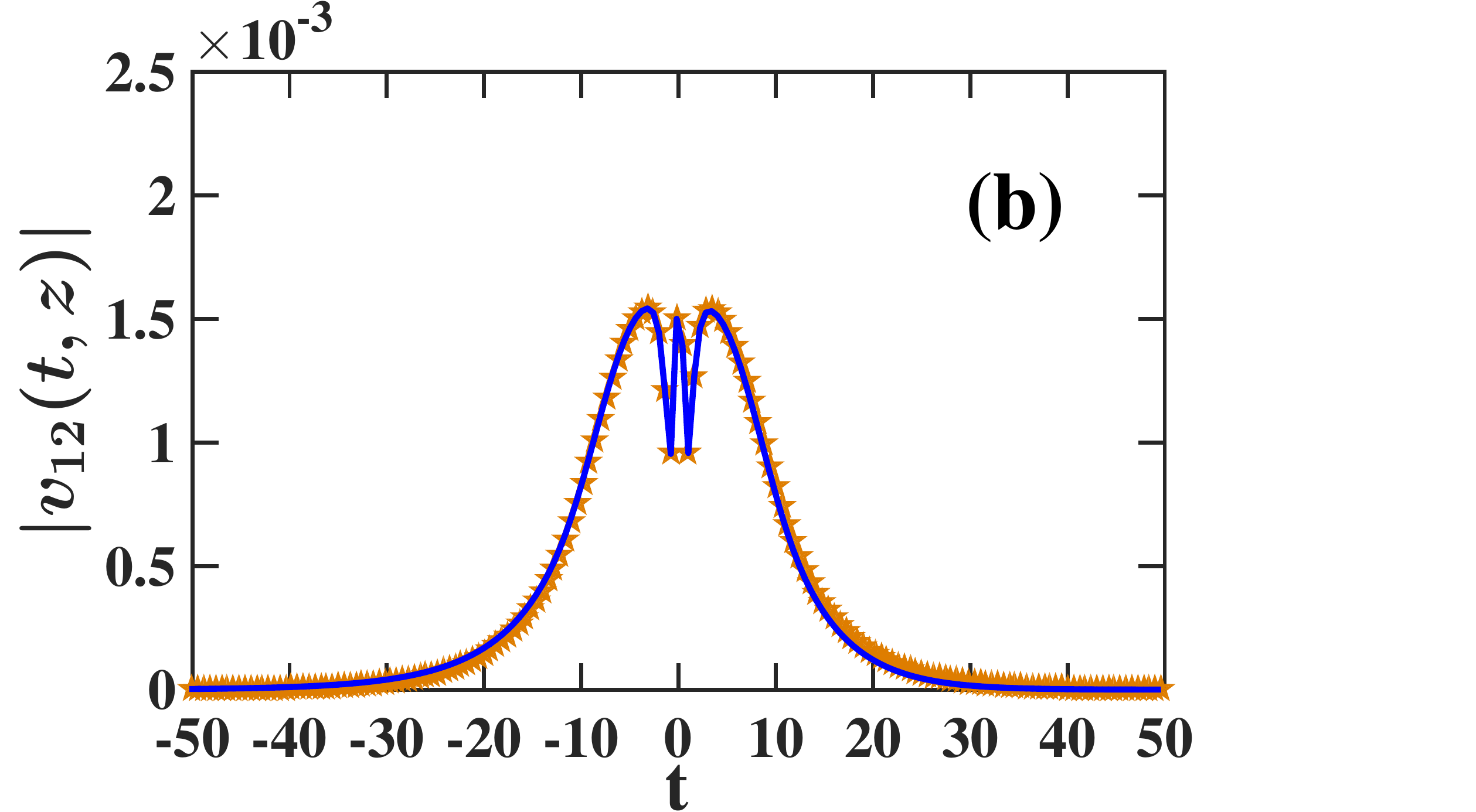} \\
\epsfxsize=9.0cm  \epsffile{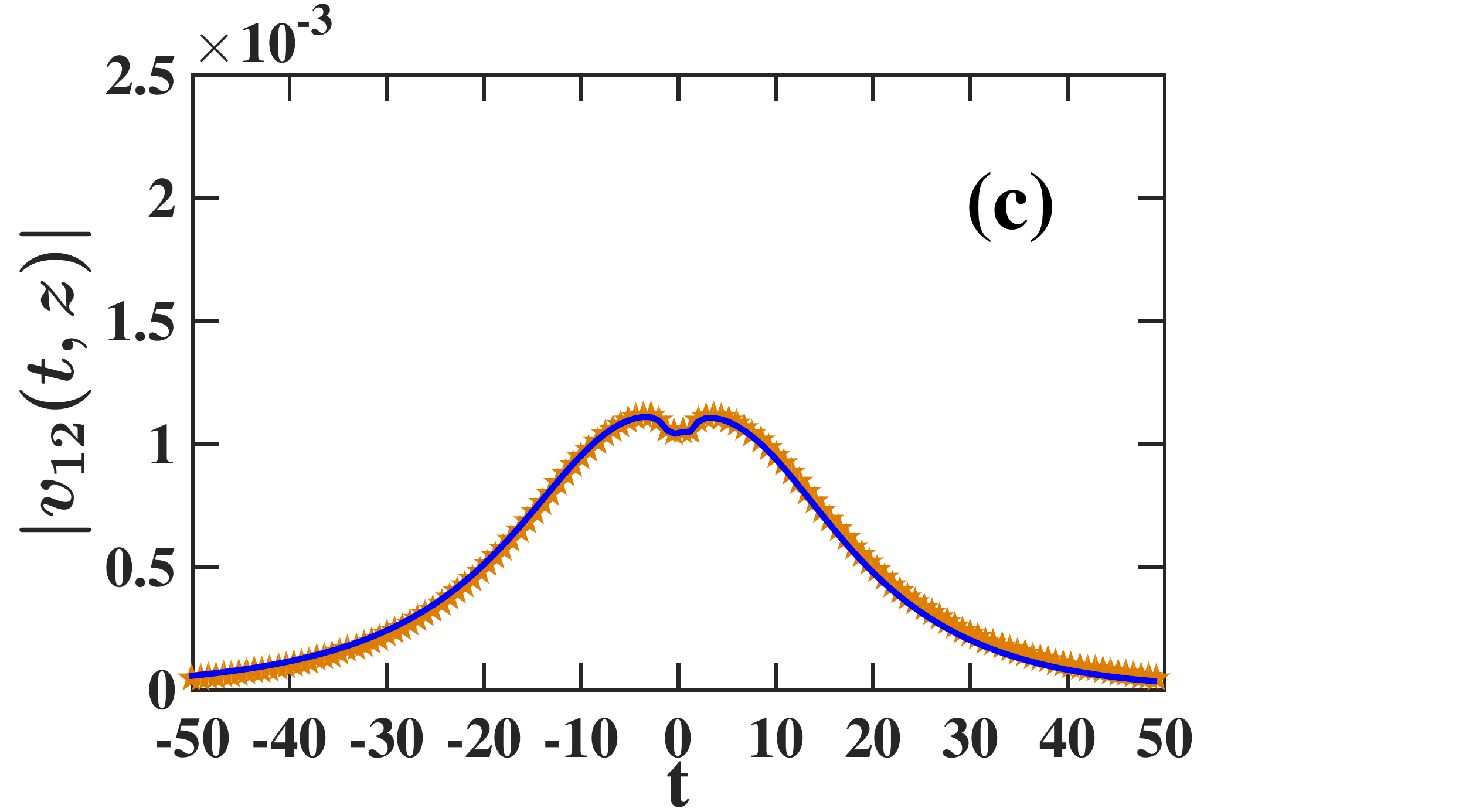} 
\end{tabular}
\end{center}
\caption{The $t$ dependence of the collision-induced radiation 
profiles obtained with the basic and improved versions of the 
perturbation approach $|v_{12b}(t,z)|$ and $|v_{12c}(t,z)|$ at 
$z=z_{c}+2$ (a), $z=z_{c}+5$ (b), and $z=z_{c}+10$ (c). 
The parameter values are $\beta=10$, $\epsilon_{3}=0.02$, 
and $\eta_{1}=\eta_{2}=1$. The orange stars represent 
$|v_{12b}(t,z)|$, as obtained with Eqs. (\ref{rad14}) and (\ref{rad16}).       
The solid blue line represents $|v_{12c}(t,z)|$, as obtained 
with Eqs. (\ref{rad24}) and (\ref{rad30}).}                
 \label{fig1}
\end{figure}

\begin{figure}[ptb]
\begin{center}
\begin{tabular}{cc}
\epsfxsize=9.0cm  \epsffile{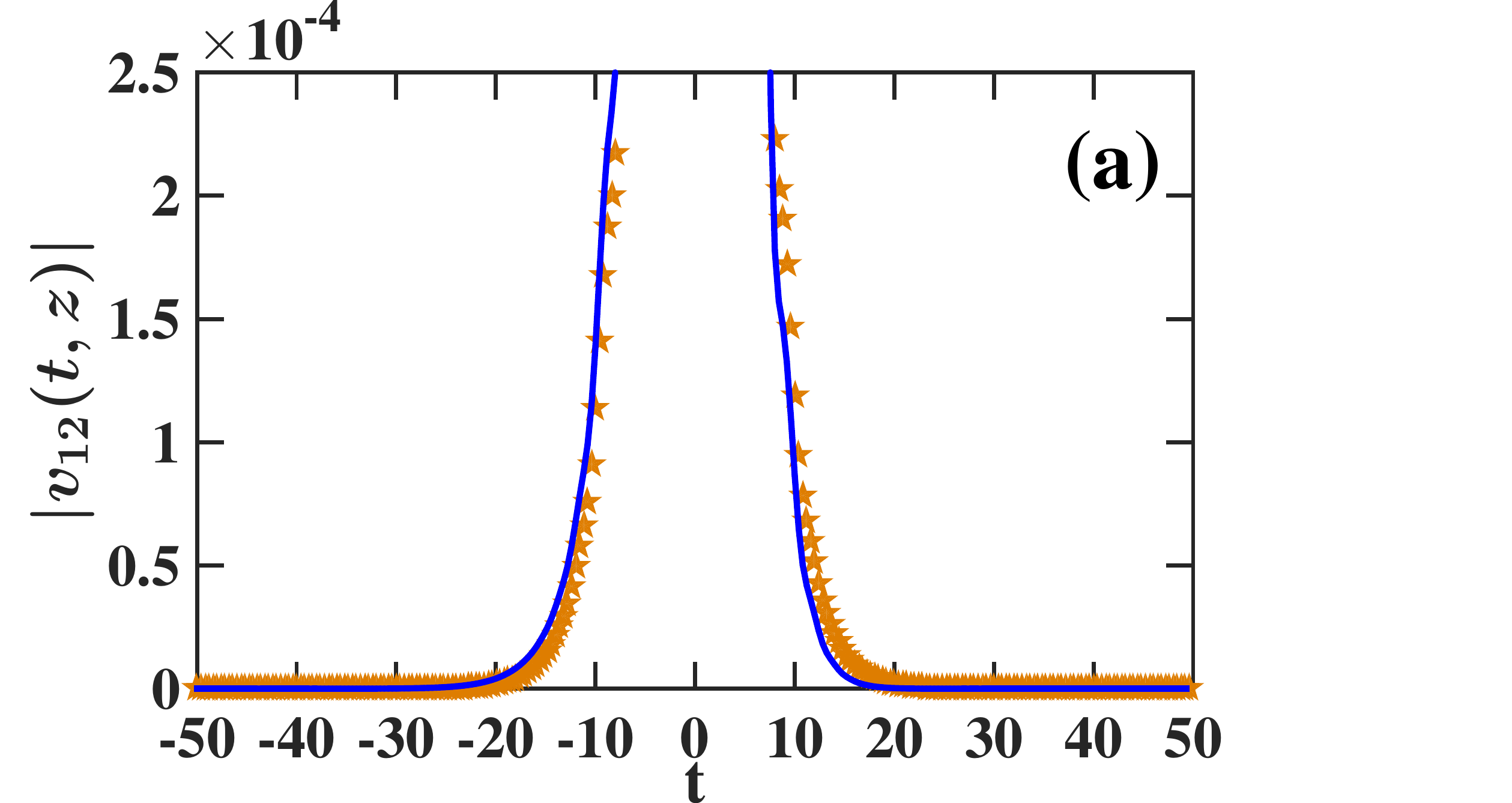} \\
\epsfxsize=9.0cm  \epsffile{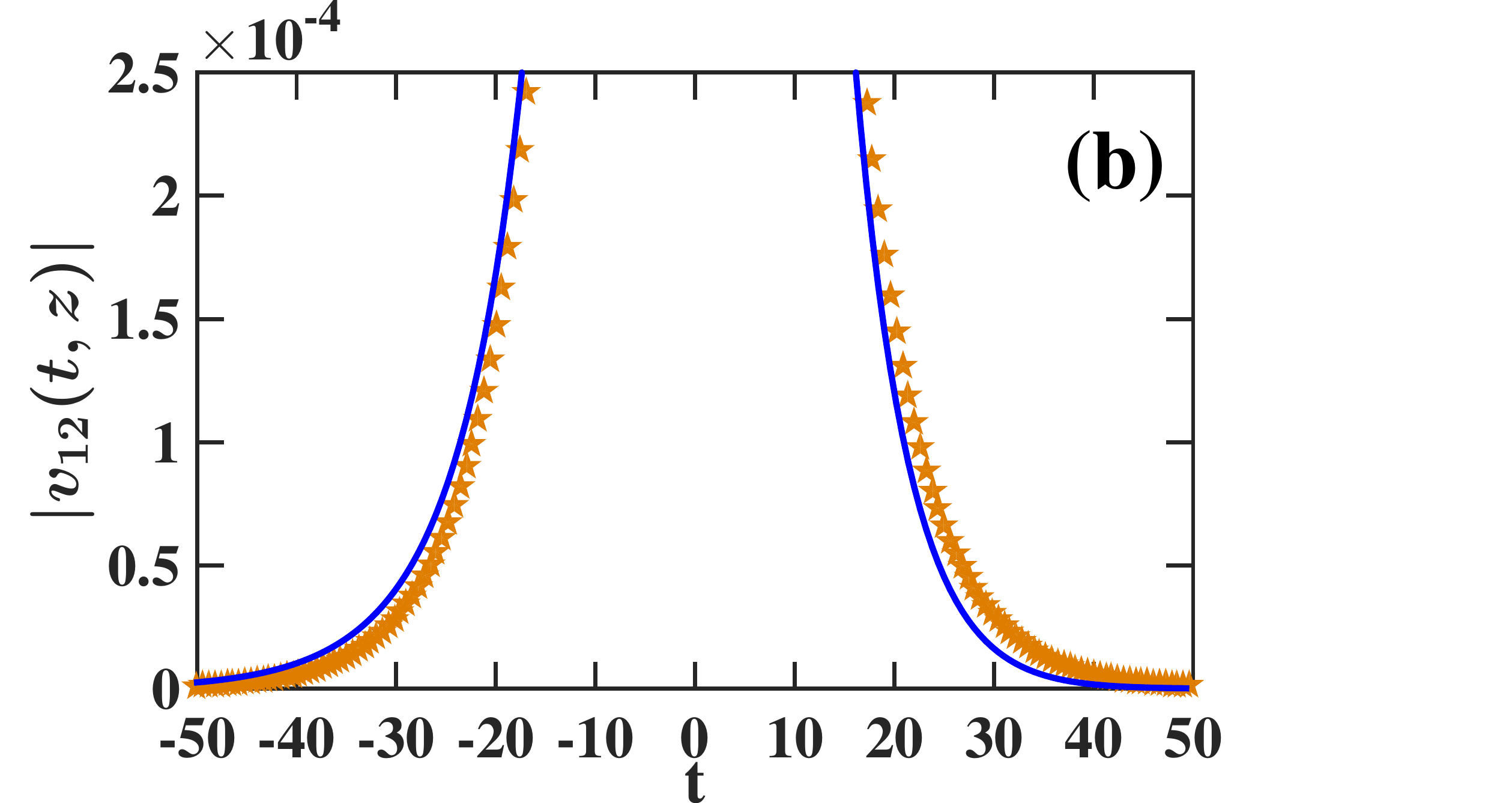} \\
\epsfxsize=9.0cm  \epsffile{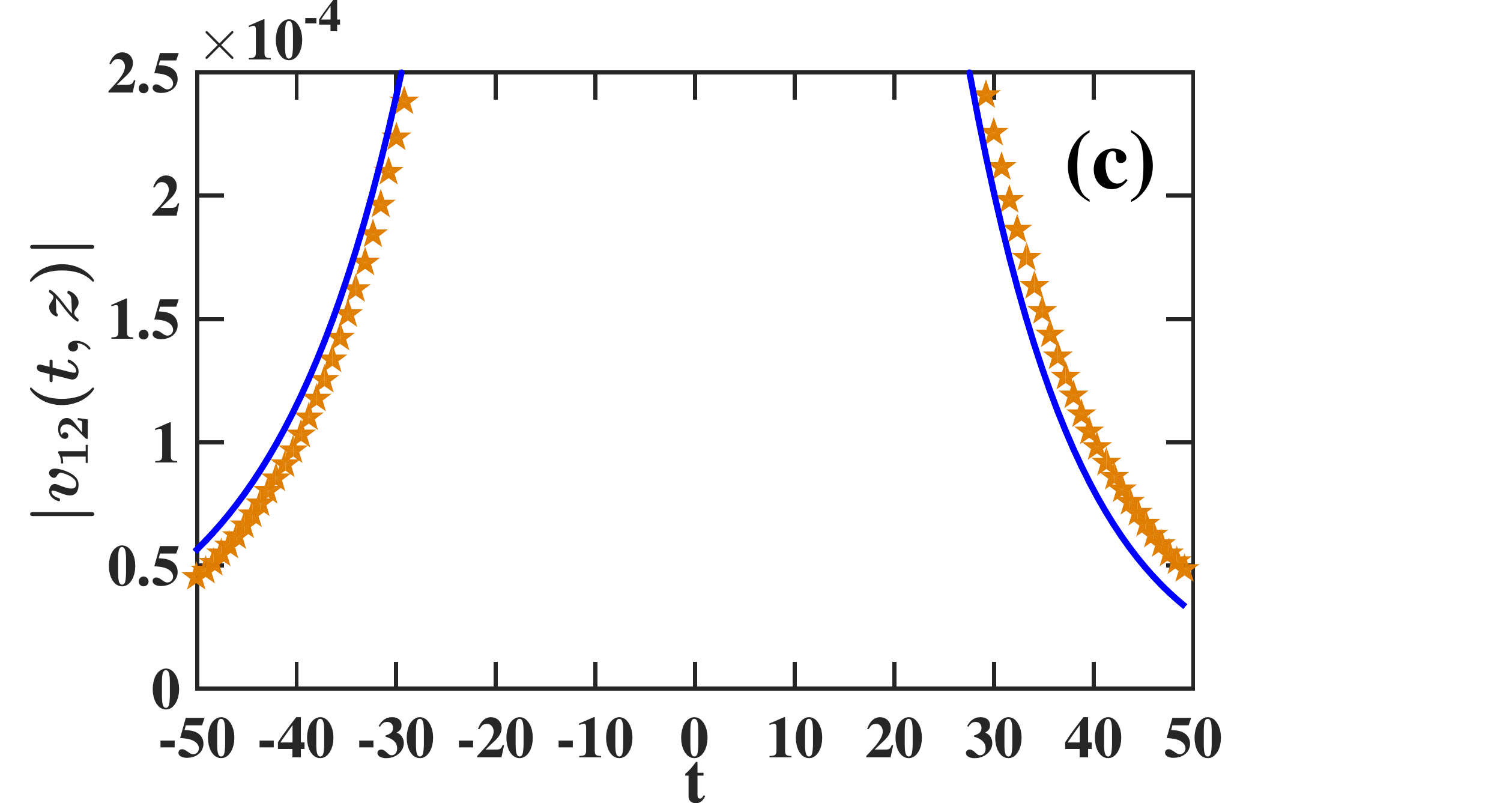} 
\end{tabular}
\end{center}
\caption{Magnified versions of the graphs in Fig. \ref{fig1} for small values
of $|v_{12b}(t,z)|$ and $|v_{12c}(t,z)|$. The symbols, distances, 
and physical parameter values are the same as in Fig. \ref{fig1}.}         
 \label{fig2}
\end{figure}

\begin{figure}[ptb]
\begin{center}
\begin{tabular}{cc}
\epsfxsize=9.0cm  \epsffile{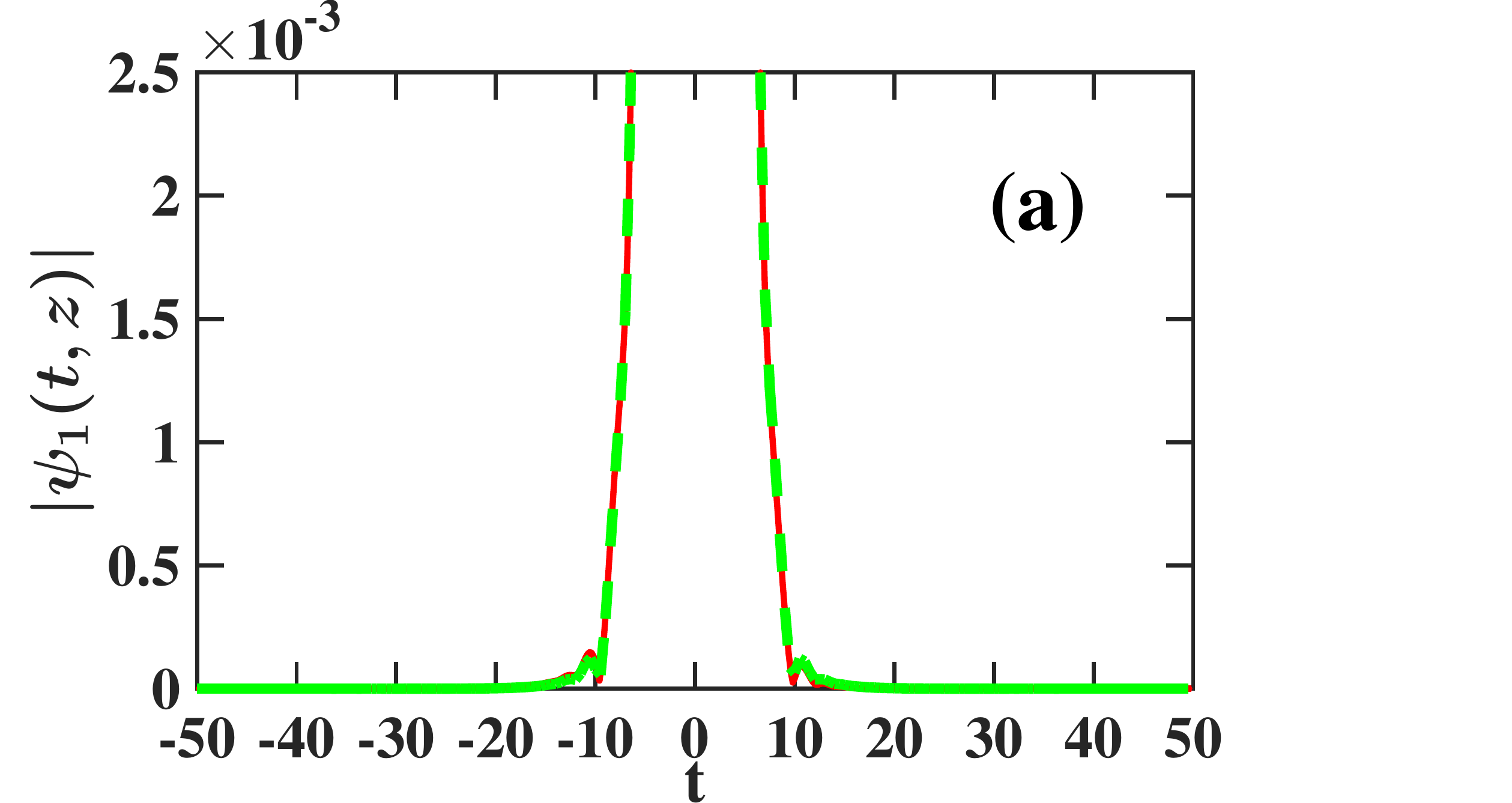} \\
\epsfxsize=9.0cm  \epsffile{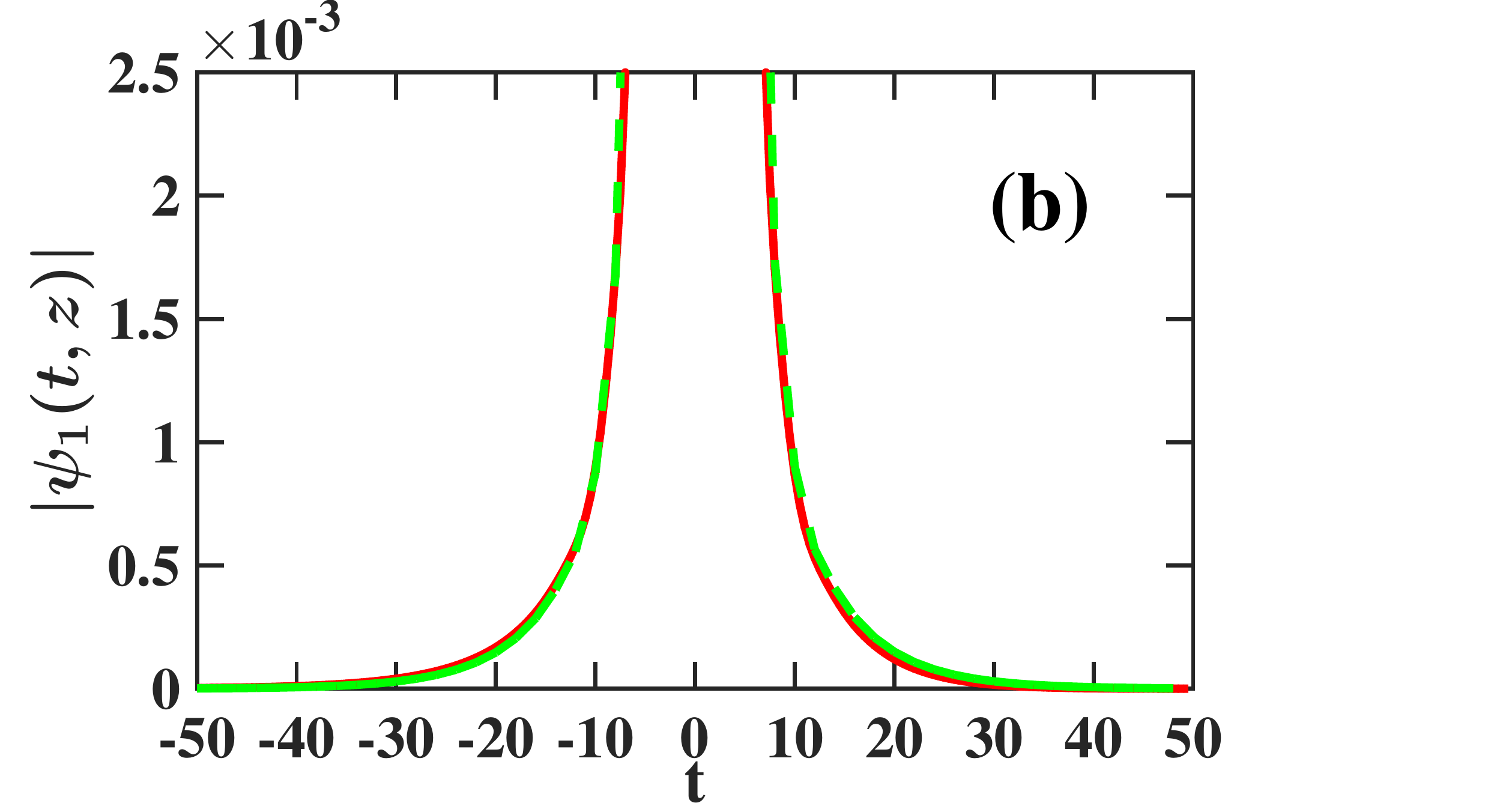} \\
\epsfxsize=9.0cm  \epsffile{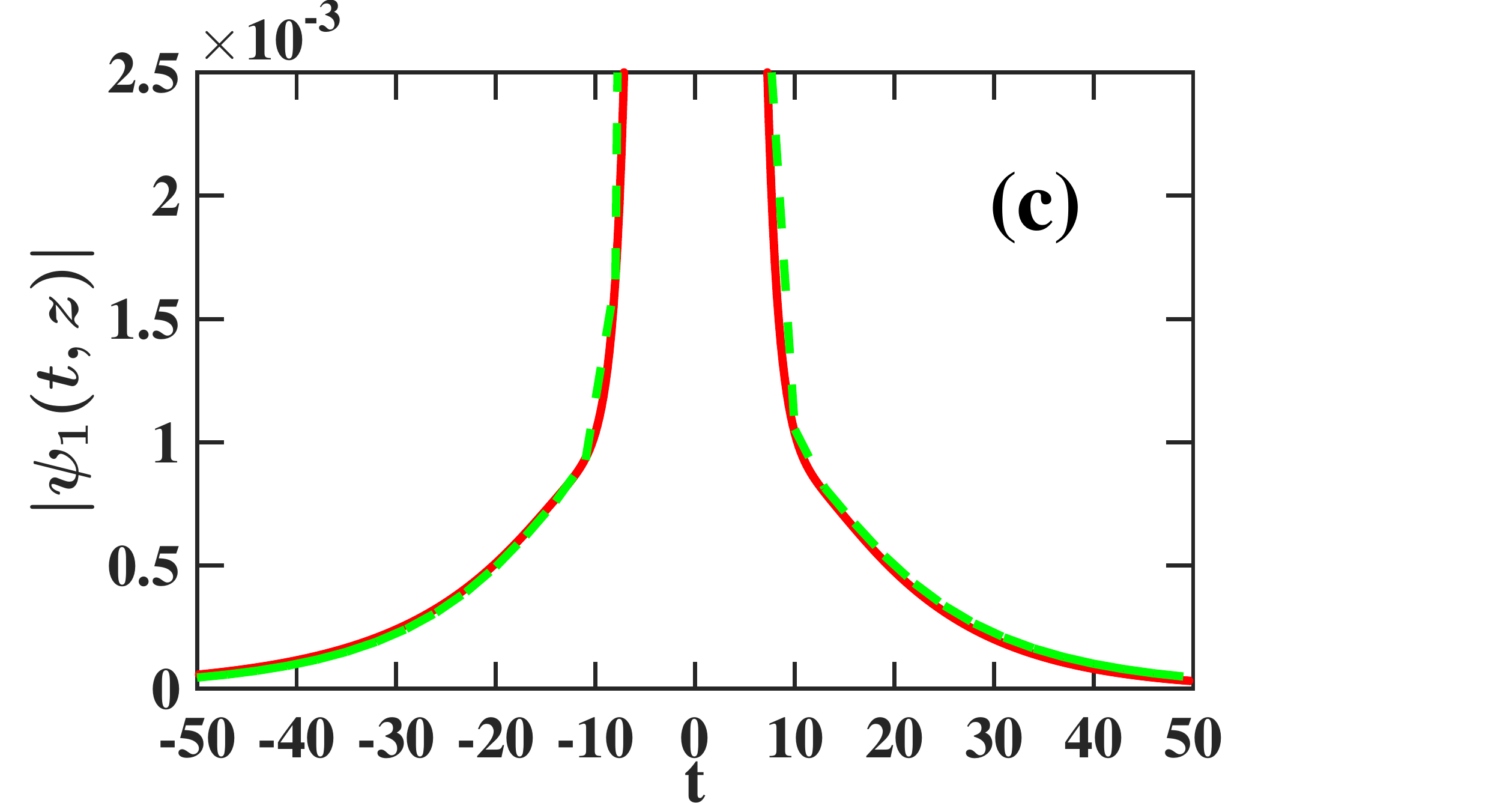} 
\end{tabular}
\end{center}
\caption{The $t$ dependence of the pulse profiles 
obtained with the basic and improved versions of the 
perturbation approach $|\psi_{1b}(t,z)|$ and $|\psi_{1c}(t,z)|$ at 
$z=z_{c}+2$ (a), $z=z_{c}+5$ (b), and $z=z_{c}+10$ (c). 
The parameter values are the same as in Fig. \ref{fig1}.  
The dashed green and solid red lines represent 
$|\psi_{1b}(t,z)|$ and $|\psi_{1c}(t,z)|$, respectively.}                
 \label{fig3}
\end{figure}

\begin{figure}[ptb]
\begin{center}
\begin{tabular}{cc}
\epsfxsize=9.0cm  \epsffile{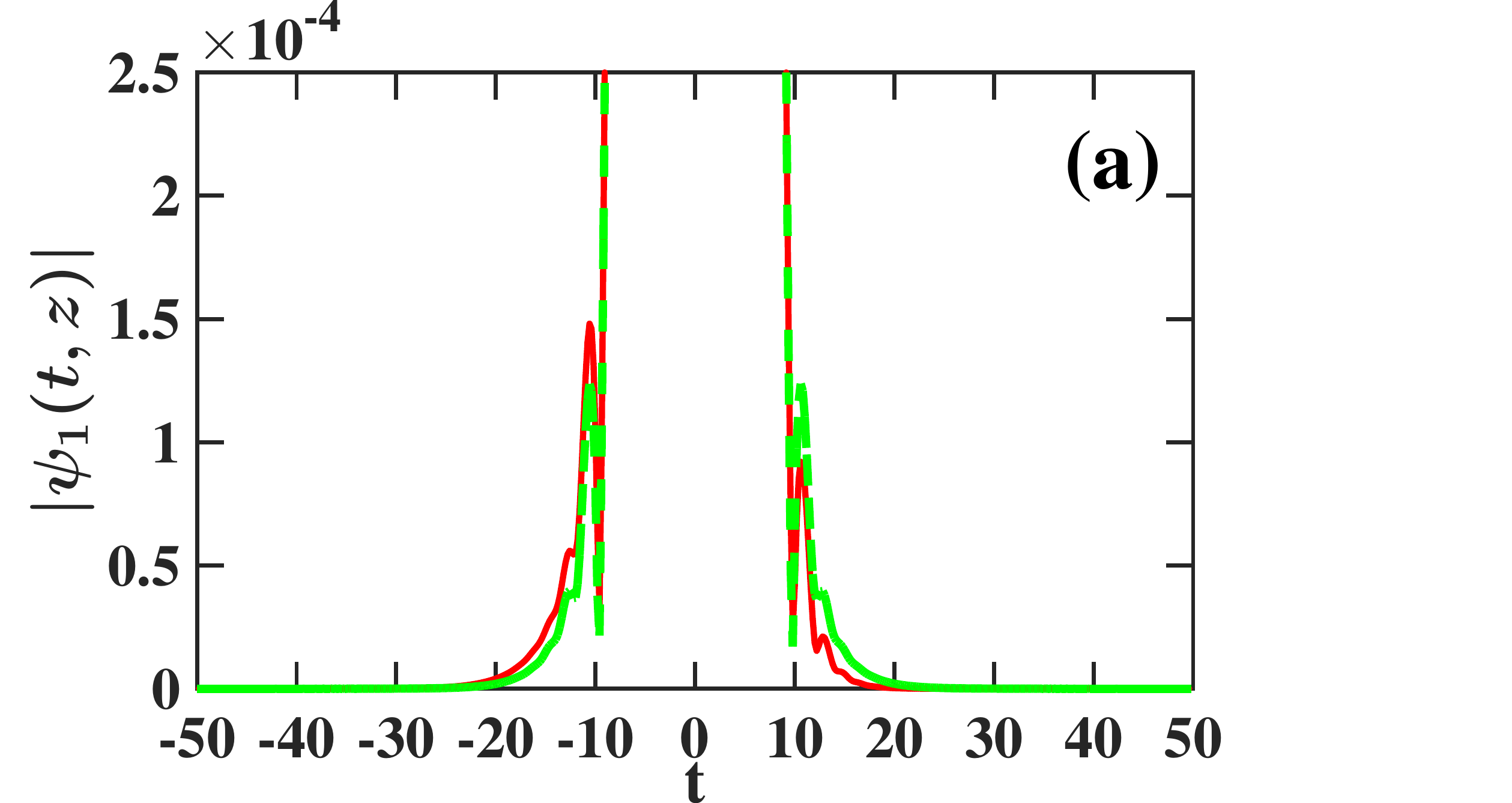} \\
\epsfxsize=9.0cm  \epsffile{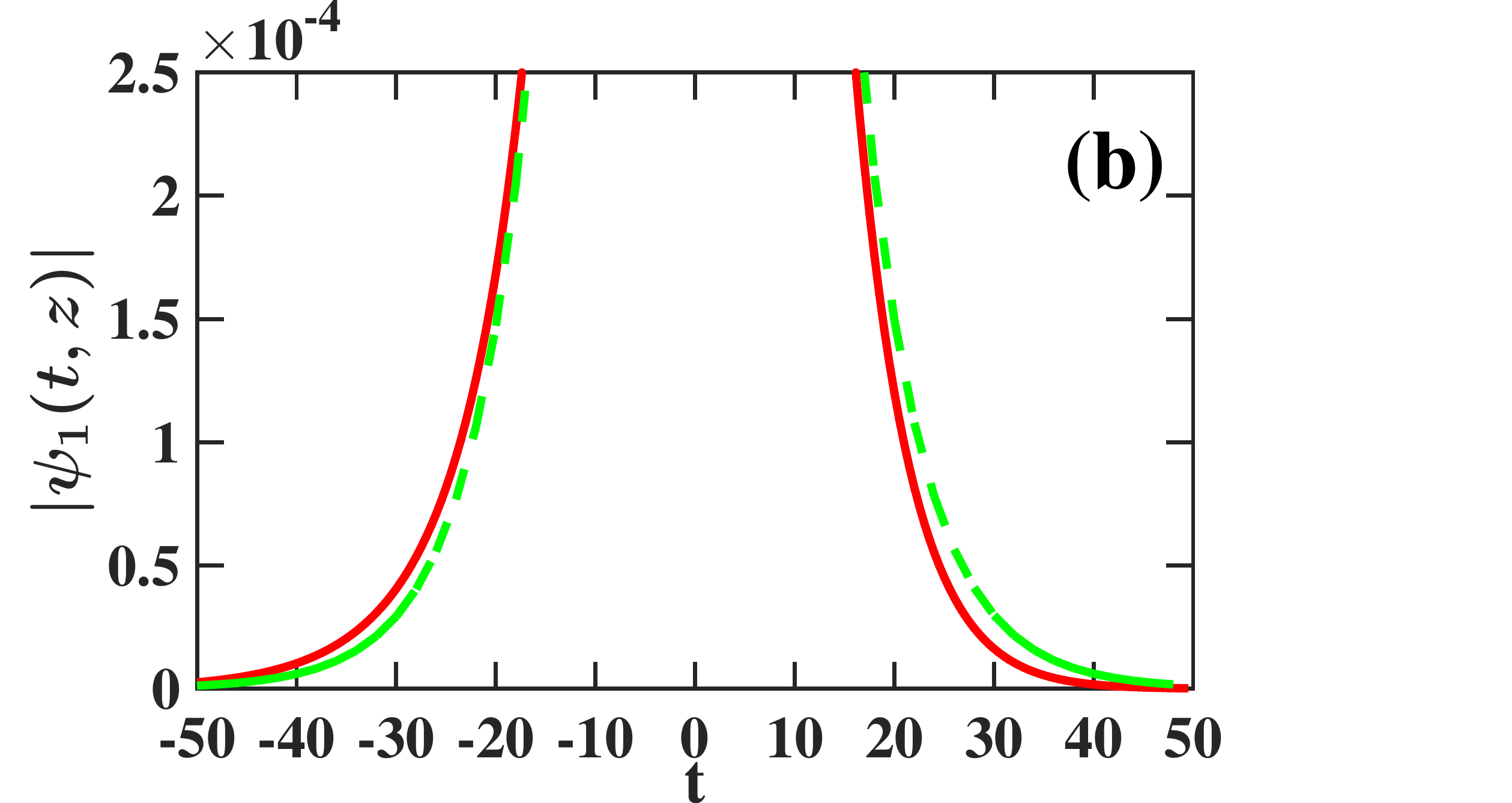} \\
\epsfxsize=9.0cm  \epsffile{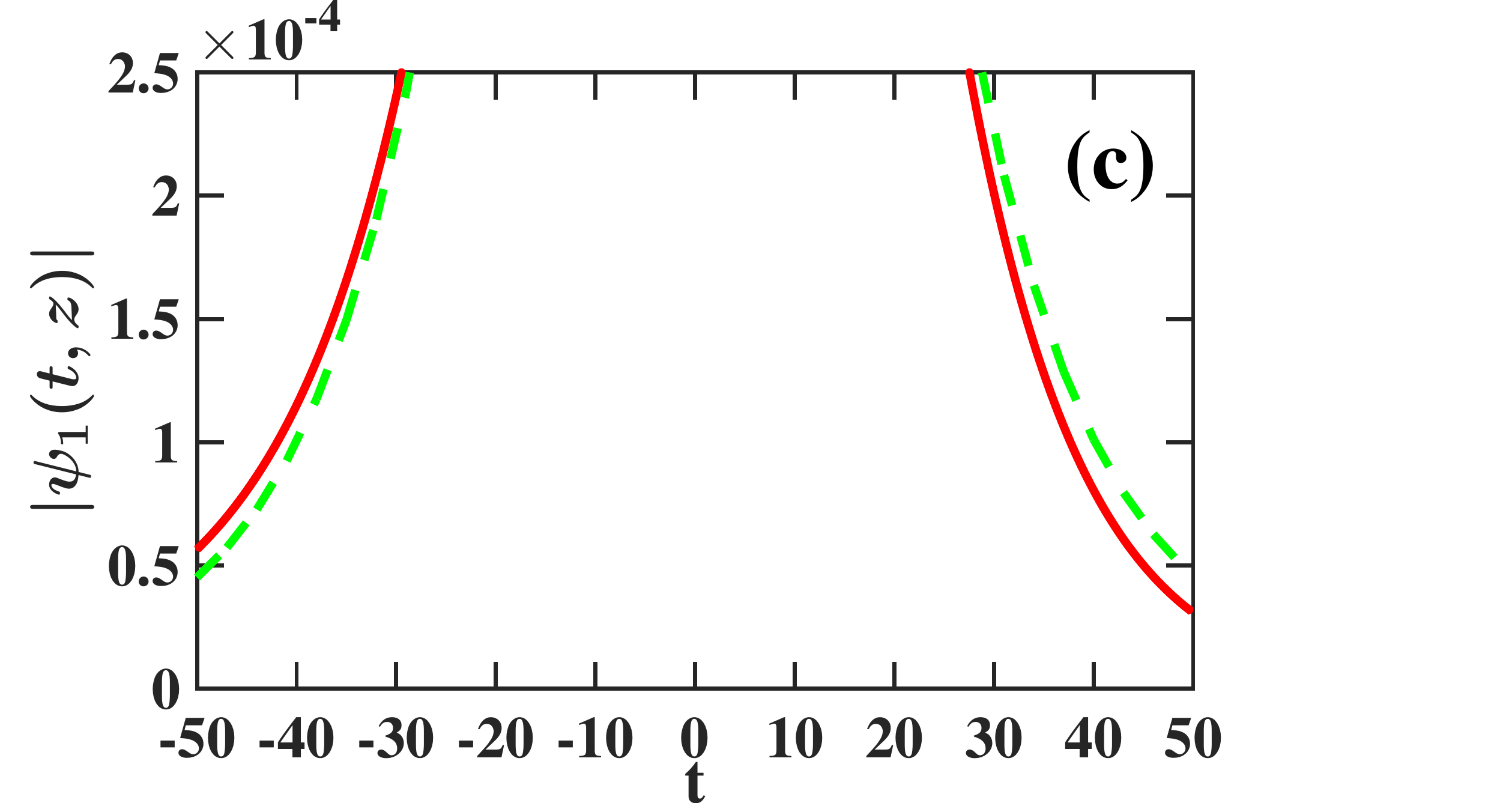} 
\end{tabular}
\end{center}
\caption{Magnified versions of the graphs in Fig. \ref{fig3} for small values
of $|\psi_{1b}(t,z)|$ and $|\psi_{1c}(t,z)|$. The symbols, distances, 
and physical parameter values are the same as in Fig. \ref{fig3}.}                
 \label{fig4}
\end{figure}

The time dependence of $|v_{12b}(t,z)|$ and $|v_{12c}(t,z)|$ at 
$z=z_{c}+2$, $z=z_{c}+5$, and $z=z_{c}+10$ is shown in 
Figs. \ref{fig1} and \ref{fig2} in the case where $\beta=10$, 
$\epsilon_{3}=0.02$, and $\eta_{1}=\eta_{2}=1$.                   
In addition, Figs. \ref{fig3} and \ref{fig4} show the profiles of the 
total envelope of the electric field $|\psi_{1b}(t,z)|$ and 
$|\psi_{1c}(t,z)|$ obtained with the 
two versions of the perturbation theory at the same distances 
and for the same physical parameter values. 
As can be seen from Figs. \ref{fig1} and \ref{fig3}, the radiation profiles 
$|v_{12b}(t,z)|$ and $|v_{12c}(t,z)|$ are very close 
at the three distances, and the pulse profiles $|\psi_{1b}(t,z)|$ 
and $|\psi_{1c}(t,z)|$ are also very close. In addition, it can be seen 
from Figs. \ref{fig2} and \ref{fig4} that the differences between the radiation 
and pulse profiles obtained with the two versions of the perturbation 
approach in the time interval $[-50,50]$ are of order $10^{-5}$. 
The tails of the radiation and pulse profiles obtained with the basic 
perturbation approach are larger than the ones obtained with the 
improved approach for $t \gg 1$ and are smaller for $t \ll -1$.  
The magnitude of the differences between the two theoretical predictions  
can be explained by noting that the improved perturbative 
calculation incorporates effects of order $\epsilon_{3}/\beta^{2}$, 
which are not taken into account in the basic perturbative calculation. 
For the parameter values used in the example,  
$\epsilon_{3}/\beta^{2}=2 \times 10^{-5}$, which is 
of the same order of magnitude as the differences between 
the two predictions for the radiation and pulse profiles that 
are observed in Figs. \ref{fig2} and \ref{fig4}.

\subsection{Radiation dynamics in the equivalent single-pulse propagation problem}
\label{theory_4}
In this subsection we provide a simpler treatment of the fast two-soliton 
collision in the presence of weak cubic loss that involves a single perturbed NLS equation. 
This simpler approach gives an equivalent description of radiation dynamics 
in order $\epsilon_{3}/\beta$, which is the leading order of the perturbation 
theory for radiation emission effects. 
Our reasons for providing the simpler equivalent treatment of the collision problem 
are the following. (a) The simpler description helps in verifying the identification 
of the physical effects that are involved in radiation emission. 
In particular, by comparing the results of numerical simulations of the 
full coupled-NLS model (\ref{rad1}) with results of the perturbed NLS equation 
for the equivalent single-pulse propagation problem we can determine 
if collision-induced changes in the shape of soliton 2 (distortion) are important 
for the dynamics of the radiation emitted by soliton 1. In addition, we can determine 
if physical processes other than cubic loss, such as interpulse interaction due 
to Kerr nonlinearity, have an important effect on radiation dynamics. 
(b) The single-pulse propagation problem can be formulated in a general 
manner, such that it describes propagation of soliton 1 in the presence of 
a generic fast and localized variation in the effective linear gain-loss 
coefficient. This generalization enables the calculation of radiation dynamics 
due to a wide class of fast and localized dissipative processes. 
The fast two-soliton collision discussed in Secs. \ref{theory_1} - \ref{theory_3}   
is just one example of these processes. 
(c) The perturbed NLS equation for the equivalent single-soliton 
propagation problem can be regarded as an intermediate model 
between the coupled-NLS equation (1) for the full single two-soliton collision problem 
and stochastic perturbed NLS models, which describe soliton propagation  
in multisequence nonlinear optical waveguide systems 
(see Refs. \cite{CP2005,CP2008,PC2012a} for studies of the latter models). 
Therefore, by showing that the perturbed NLS equation for 
the equivalent single-soliton propagation problem accurately describes 
the dynamics of radiation in the full two-soliton collision problem, we provide 
strong evidence that the collision-induced radiation dynamics is accurately captured 
by stochastic perturbed NLS models for soliton propagation in 
multisequence nonlinear optical waveguide systems.

Consider the general version of the single-pulse propagation problem, 
in which the soliton propagates in the presence of second-order dispersion, 
Kerr nonlinearity, and a fast and localized variation in the effective linear 
gain-loss coefficient. The change in the linear gain-loss coefficient can be due 
an actual change in the optical waveguide's properties or due to an external 
process, such as a fast collision between the soliton and another optical pulse. 
Thus, the propagation is described by the following perturbed NLS equation: 
\begin{eqnarray} &&
i\partial_z\psi_{1}+\partial_{t}^2\psi_{1}+2|\psi_{1}|^2\psi_{1}=
i\epsilon_{l}g_{l}(t,z)\psi_{1},
\label{rad31}
\end{eqnarray}           
where $0 < |\epsilon_{l}| \ll 1$, and $g_{l}(t,z)$ is a real-valued function, 
which describes the fast and localized variation in the effective linear 
gain-loss coefficient. More specifically, $g_{l}(t,z)$ is of the form 
\begin{equation}
g_{l}(t,z) = g(x_{l}) \;\;\; \mbox{with} \;\;\; x_{l} = t - t_{l} + \bar b z,
\label{rad32}
\end{equation}              
where $g(x_{l})$ is sharply-peaked at $x_{l}=0$, 
$t_{l}$ and $\bar b$ are constants, and $|\bar b - 2\beta_{1}(0)| \gg 1$.

Let us obtain the expression for $g_{l}(t,z)$ for a fast two-soliton 
collision in the presence of weak cubic loss. For this purpose, we first note 
that the main effects of the collision on soliton 1 that are due to cubic loss, 
i.e., the collision-induced effects in order $\epsilon_{3}/\beta$, can 
be described by adding a term of the form 
$i \exp(i \chi_{1})\partial_{z}\Phi_{12}^{(1)}$ to the unperturbed NLS 
equation for soliton 1. From Eq. (\ref{rad6}) it follows that 
this term is of the form          
\begin{eqnarray} &&
i \exp(i \chi_{1})\partial_{z}\Phi_{12}^{(1)}=
-\frac{2i\epsilon_{3}\eta_{2}^{2}}
{\cosh^{2}(x_{2})}\psi_{1}.
\label{rad33}
\end{eqnarray}                                        
Therefore, the perturbed NLS equation for the equivalent 
single-soliton propagation problem is given by 
\begin{eqnarray} &&
i\partial_z\psi_{1}+\partial_{t}^2\psi_{1}+2|\psi_{1}|^2\psi_{1}=
-\frac{2i\epsilon_{3}\eta_{2}^{2}}
{\cosh^{2}(x_{2})}\psi_{1}.
\label{rad34}
\end{eqnarray}                    
Equating the right hand sides of Eqs. (\ref{rad34}) and (\ref{rad31}) 
we find that in a fast two-soliton collision in the presence of weak cubic loss  
$\epsilon_{l}=-\epsilon_{3}$ and $g_{l}(t,z)=2\eta_{2}^{2}/\cosh^{2}(x_{2})$.

We now show that the equations for radiation dynamics obtained with 
the simpler perturbed NLS model (\ref{rad34}) are identical to the equations 
obtained in Secs.  \ref{theory_2} and \ref{theory_3} for the full 
two-soliton collision problem. For simplicity and without loss of generality, 
we take the initial frequencies as $\beta_{1}(0)=0$ and $\beta_{2}(0)=\beta$. 
We assume that $0 < \epsilon_{3} \ll 1$ 
and $|\beta|\gg 1$ and look for a solution of  Eq. (\ref{rad34}) 
in the form 
\begin{eqnarray}&&
\!\!\!\!\!\!\!
\psi_{1}(t,z)=\psi_{s1}(t,z)+\tilde\phi_{1}(t,z), 
\label{rad35}
\end{eqnarray}            
where $\psi_{s1}(t,z)$ is the single-soliton solution of the unperturbed NLS 
equation and $\tilde\phi_{1}(t,z)$ describes collision-induced changes to 
$\psi_{s1}(t,z)$. We substitute the ansatz (\ref{rad35}) into Eq. (\ref{rad34}) 
and keep terms up to order $\epsilon_{3}/\beta$ in the equation. 
We obtain: 
\begin{eqnarray} &&
i\partial_z \tilde\phi_{1}+\partial_{t}^2\tilde\phi_{1}+
4|\psi_{s1}|^2\tilde\phi_{1}+2\psi_{s1}^2\tilde\phi_{1}^{\ast}=
-\frac{2i\epsilon_{3}\eta_{2}^{2}}
{\cosh^{2}(x_{2})}\psi_{s1}.
\label{rad36}  
\end{eqnarray} 
Next, we substitute $\psi_{s1}(t,z)=\Psi_{1}(x_{1})\exp(i\chi_{1})$ and
$\tilde\phi_{1}(t,z)=\tilde\Phi_{1}(t,z)\exp(i\chi_{1})$ into Eq. (\ref{rad36}). 
This substitution yields the following equation for $\tilde\Phi_{1}$:          
\begin{eqnarray} &&
i\partial_z \tilde\Phi_{1}+(\partial_{t}^2 - \eta_{1}^{2})\tilde\Phi_{1}+
4\Psi_{1}^{2}\tilde\Phi_{1}+2\Psi_{1}^{2}\tilde\Phi_{1}^{\ast}=
-\frac{2i\epsilon_{3}\eta_{1}\eta_{2}^{2}}
{\cosh(x_{1})\cosh^{2}(x_{2})}.
\label{rad37}  
\end{eqnarray}                
Equation (\ref{rad37}) is solved by expanding $\tilde\Phi_{1}$
in a perturbation series with respect to $\epsilon_{3}$ and $1/\beta$:   
\begin{eqnarray} &&
\tilde\Phi_{1}(t,z)=\tilde\Phi_{11}^{(0)}(t,z)+\tilde\Phi_{11}^{(1)}(t,z)+
\tilde\Phi_{12}^{(0)}(t,z)
\nonumber \\ &&
+\tilde\Phi_{12}^{(1)}(t,z)+\dots,
\label{rad38}
\end{eqnarray}     
where the subscripts and superscript notations are similar to the 
ones in Eq. (\ref{rad5}).

The equation describing the collision-induced effects in order $\epsilon_{3}/\beta$, 
which is the leading order, is: 
\begin{eqnarray} &&
\partial_{z} \tilde\Phi_{12}^{(1)}=
-\frac{2\epsilon_{3}\eta_{1}\eta_{2}^{2}}
{\cosh(x_{1})\cosh^{2}(x_{2})}.
\label{rad39}
\end{eqnarray}     
This equation is identical to Eq. (\ref{rad6}), which was obtained 
with the basic version of the perturbative calculation for the full two-soliton 
collision problem. As a result, the equations for radiation dynamics obtained in the 
equivalent single-soliton propagation problem in order $\epsilon_{3}/\beta$ 
are identical to the equations obtained in Sec. \ref{theory_2} with the basic 
version of the perturbation approach for the full two-soliton collision problem.

We can also obtain an improved approximation for radiation dynamics 
by taking into account the effects of propagation of radiation in the collision 
interval, which are described by the $O(\epsilon_{3}/\beta^{2})$ terms  
$(\partial_{t}^2-\eta_{1}^{2})\tilde\Phi_{1}$, 
$4\Psi_{1}^{2}\tilde\Phi_{1}$, and $2\Psi_{1}^{2}\tilde\Phi_{1}$ in Eq. (\ref{rad37}). 
For this purpose, we denote by $\tilde\Phi_{12c}^{(1)}$ the part of $\tilde\Phi_{1}$ 
that describes the collision effects in order $\epsilon_{3}/\beta$ 
{\it and} the effects of propagation of radiation in the collision interval. 
$\tilde\Phi_{12c}^{(1)}$ satisfies the following equation: 
\begin{eqnarray} &&
\partial_z \tilde\Phi_{12c}^{(1)}
-i\left[(\partial_{t}^2-\eta_{1}^{2})\tilde\Phi_{12c}^{(1)}
+4\Psi_{1}^{2}\tilde\Phi_{12c}^{(1)}
+2\Psi_{1}^{2}\tilde\Phi_{12c}^{(1)\ast}\right]
=-\frac{2\epsilon_{3}\eta_{1}\eta_{2}^{2}}
{\cosh(x_{1})\cosh^{2}(x_{2})}.
\label{rad40}
\end{eqnarray}    
Equation (\ref{rad40}) is identical to Eq. (\ref{rad21}) that was obtained 
with the improved perturbative calculation for the full two-soliton 
collision problem. Therefore, the equations for radiation dynamics 
obtained with the improved approximation in the equivalent single-soliton 
propagation problem are identical to the equations obtained in 
Sec. \ref{theory_3} with the improved perturbation approach 
for the full two-soliton collision problem.

We point out that the same perturbation approaches, 
which were described in Secs. \ref{theory_2} and \ref{theory_3},   
can be used for analyzing radiation dynamics in the general problem 
of single-soliton propagation in the presence of a fast and localized variation 
in the linear gain-loss coefficient. 
Indeed, substituting the ansatz (\ref{rad35}) into the general 
perturbed NLS model (\ref{rad31}), we arrive at the following equations 
for $\tilde\Phi_{12}^{(1)}$ and $\tilde\Phi_{12c}^{(1)}$ in the general 
propagation problem:  
\begin{eqnarray} &&
\partial_{z} \tilde\Phi_{12}^{(1)}=
\frac{\epsilon_{l}\eta_{1}g(x_{l})}
{\cosh(x_{1})}
\label{rad41}
\end{eqnarray}       
and 
\begin{eqnarray} &&
\partial_z \tilde\Phi_{12c}^{(1)}
-i\left[(\partial_{t}^2-\eta_{1}^{2})\tilde\Phi_{12c}^{(1)}
+4\Psi_{1}^{2}\tilde\Phi_{12c}^{(1)}
+2\Psi_{1}^{2}\tilde\Phi_{12c}^{(1)\ast}\right]
=\frac{\epsilon_{l}\eta_{1}g(x_{l})}{\cosh(x_{1})}.
\label{rad42}
\end{eqnarray}    
Equations (\ref{rad41}) and  (\ref{rad42}) have a form similar 
to Eqs. (\ref{rad6}) and  (\ref{rad21}) that were analyzed in 
Secs. \ref{theory_2} and \ref{theory_3}, respectively. 
Therefore, the equations for radiation dynamics in the general 
single-soliton propagation problem, described by Eq. (\ref{rad31}), 
can be obtained by employing the perturbation approaches 
of Secs. \ref{theory_2} and \ref{theory_3}.

\section{Numerical simulations}
\label{simu}

\subsection{Introduction}
\label{simu_1}

The analytic predictions for radiation dynamics obtained in 
Secs. \ref{theory_2} - \ref{theory_4} are based on several 
simplifying approximations. In particular, the perturbative calculations 
in these sections only take into account the $O(\epsilon_{3}/\beta)$ 
effects of cubic loss and the $O(\epsilon_{3}/\beta^{2})$ effects 
associated with propagation of radiation in the collision interval. 
This means that the perturbative calculations neglect the 
distortion of soliton 2 and the effect of interpulse interaction due 
to Kerr nonlinearity on radiation dynamics.       
Since the validity and accuracy of these approximations depend  
on the values of the physical parameters, it is important to check the analytic 
predictions obtained in Secs. \ref{theory_2} - \ref{theory_4} 
by numerical simulations with the full coupled-NLS model (\ref{rad1}).

To gain further insight into the physical processes that determine radiation 
dynamics and into the reasons for differences between the analytic predictions  
and results of simulations with Eq. (\ref{rad1}), we carry out numerical 
simulations with four additional simpler propagation models.     
The first additional model is a modified version of the perturbed NLS equation 
(\ref{rad34}), which takes into account collision-induced radiation emission 
due to both cubic loss and Kerr nonlinearity, and position shifts due to both 
Kerr nonlinearity and the collision-induced frequency shift. 
As we show in Appendix \ref{appendB}, this model has the form: 
\begin{eqnarray} &&
i\partial_z\psi_{1}+\partial_{t}^2\psi_{1}+2|\psi_{1}|^2\psi_{1}=
-\frac{2i\epsilon_{3}\eta_{2}^{2}}
{\cosh^{2}(x_{2})}\psi_{1}
-\frac{4\eta_{2}^{2}}{\cosh^{2}(x_{2})}\psi_{1}
-iC_{1}(z)\partial_{t}\psi_{1},
\label{rad44}
\end{eqnarray}          
where $C_{1}(z)$ is given by   
\begin{eqnarray} &&
C_{1}(z)=
\left\{ \begin{array}{l l}
0 &  \mbox{for} \;\;   z<z_{c} \, ,\\
\frac{40\epsilon_{3}\eta_{1}^{2}\eta_{2}}{3|\beta|\beta}
& \mbox{for} \;\;   z \ge z_{c} \,.\\
\end{array} \right. 
\label{rad45}
\end{eqnarray}     
The second term on the right hand side of Eq. (\ref{rad44}) describes 
radiation emission and the collision-induced position shift due to Kerr nonlinearity.    
The third term on the right hand side of Eq. (\ref{rad44}) describes the position 
shift caused by the collision-induced frequency shift. The most important difference 
between Eq. (\ref{rad44}) and the full coupled-NLS model (\ref{rad1}) is that 
Eq. (\ref{rad44}) neglects the distortion of soliton 2 and its effects on the collision.

The second simplified NLS model is a variation of Eq. (\ref{rad44}), 
in which the second term on the right hand side is replaced by a term that describes 
only the Kerr-induced position shift and neglects the Kerr-induced radiation 
emission in the collision. In Appendix \ref{appendB}, we show that this model 
takes the form: 
\begin{eqnarray} &&
i\partial_z\psi_{1}+\partial_{t}^2\psi_{1}+2|\psi_{1}|^2\psi_{1}=
-\frac{2i\epsilon_{3}\eta_{2}^{2}}
{\cosh^{2}(x_{2})}\psi_{1}
-iC_{2}(z)\partial_{t}\psi_{1}
-iC_{1}(z)\partial_{t}\psi_{1},
\label{rad46}
\end{eqnarray}          
where $C_{2}(z)$ is given by   
\begin{eqnarray} &&
C_{2}(z)=
\left\{ \begin{array}{l l}
0 &  \mbox{for} \;\;   z<z_{c} - \frac{1}{2|\beta|} \, ,\\
\frac{4\eta_{2}}{|\beta|}
& \mbox{for} \;\; z_{c} - \frac{1}{2|\beta|} \le  z \le z_{c} + \frac{1}{2|\beta|} \, ,\\ 
0 & \mbox{for} \;\;   z > z_{c}+ \frac{1}{2|\beta|} \,. \\
\end{array} \right. 
\label{rad47}
\end{eqnarray}        
Equation (\ref{rad47}) neglects distortion of soliton 2 and Kerr-induced radiation emission 
in the collision. The latter two effects are taken into account in the full coupled-NLS model.

The third simplified model is the following coupled-NLS model, 
which takes into account distortion of soliton 2 in describing the effects of 
cubic loss on the collision, but neglects distortion of soliton 2 in describing 
the effects of Kerr nonlinearity on the collision. This model also takes into 
account the position shift due to the collision-induced frequency shift. 
As shown in Appendix \ref{appendB}, this third simplified propagation 
model is given by: 
\begin{eqnarray} &&
i\partial_z\psi_{1}+\partial_{t}^2\psi_{1}+2|\psi_{1}|^2\psi_{1}
=-2i\epsilon_{3}|\psi_{2}|^2\psi_{1}
-\frac{4\eta_{2}^{2}}{\cosh^{2}(x_{2})}\psi_{1}
-iC_{1}(z)\partial_{t}\psi_{1},
\nonumber \\&&
i\partial_z\psi_{2}+\partial_{t}^2\psi_{2}
+2|\psi_{2}|^2\psi_{2}=
-2i\epsilon_{3}|\psi_{1}|^2\psi_{2}
-\frac{4\eta_{1}^{2}}{\cosh^{2}(x_{1})}\psi_{2}
-iC_{3}(z)\partial_{t}\psi_{2}, 
\label{rad51}
\end{eqnarray}                 
where $C_{3}(z)$ is    
\begin{eqnarray} &&
C_{3}(z)=
\left\{ \begin{array}{l l}
0 &  \mbox{for} \;\;   z<z_{c} \, ,\\
\frac{-40\epsilon_{3}\eta_{1}\eta_{2}^{2}}{3|\beta|\beta}
& \mbox{for} \;\;   z \ge z_{c} \,.\\
\end{array} \right. 
\label{rad52}
\end{eqnarray}      
The main difference 
between Eq. (\ref{rad51}) and the full coupled-NLS model (\ref{rad1}) is that 
Eq. (\ref{rad51}) neglects the distortion of soliton 2 in describing the effects of 
Kerr nonlinearity on the collision.

The fourth simplified model is a variation of the coupled-NLS model (\ref{rad51}), 
in which the second terms on the right hand sides are replaced by terms that 
describe only the Kerr-induced position shift and neglect the Kerr-induced radiation 
emission. In Appendix \ref{appendB}, we show that this model has the form  
\begin{eqnarray} &&
i\partial_z\psi_{1}+\partial_{t}^2\psi_{1}+2|\psi_{1}|^2\psi_{1}
=-2i\epsilon_{3}|\psi_{2}|^2\psi_{1}
-iC_{2}(z)\partial_{t}\psi_{1}
-iC_{1}(z)\partial_{t}\psi_{1},
\nonumber \\&&
i\partial_z\psi_{2}+\partial_{t}^2\psi_{2}
+2|\psi_{2}|^2\psi_{2}=
-2i\epsilon_{3}|\psi_{1}|^2\psi_{2}
-iC_{4}(z)\partial_{t}\psi_{2}
-iC_{3}(z)\partial_{t}\psi_{2}, 
\label{rad53}
\end{eqnarray}                 
where
\begin{eqnarray} &&
C_{4}(z)=
\left\{ \begin{array}{l l}
0 &  \mbox{for} \;\;   z<z_{c} - \frac{1}{2|\beta|} \, ,\\
-\frac{4\eta_{1}}{|\beta|}
& \mbox{for} \;\; z_{c} - \frac{1}{2|\beta|} \le  z \le z_{c} + \frac{1}{2|\beta|} \, ,\\ 
0 & \mbox{for} \;\;   z > z_{c}+ \frac{1}{2|\beta|} \,. \\
\end{array} \right. 
\label{rad54}
\end{eqnarray}

\subsection{Description and discussion of simulations results}
\label{simu_2}

Equations (\ref{rad1}), (\ref{rad44}), (\ref{rad46}), (\ref{rad51}), 
and (\ref{rad53}) are numerically integrated on a time domain 
$[t_{\mbox{min}},t_{\mbox{max}}]=[-3200,3200]$ 
using the split-step method with periodic boundary conditions 
\cite{Agrawal2001,Yang2010}. The large temporal domain is chosen 
such that the values of $|\psi_{1}(t,z)|$ and $|\psi_{2}(t,z)|$          
at and near the boundaries are negligible throughout the simulation. 
As a result, potential artificial effects due to the radiation leaving the 
computational domain at one boundary and reentering it at another 
boundary are also negligible. The $t$-step and $z$-step of the 
numerical scheme are taken as $\Delta t=0.065$ and $\Delta z =0.0002$.
These values ensure stability of the numerical scheme and provide 
sufficient accuracy for capturing the radiation dynamics.

For concreteness, we present the results of the simulations for two 
sets of values of $\epsilon_{3}$ and $\beta$: (1) $\epsilon_{3}=0.02$ 
and $\beta=20$; (2) $\epsilon_{3}=0.02$ and $\beta=10$. 
This choice enables investigation of the dependence of the radiation 
dynamics on the frequency difference parameter $\beta$. 
More specifically, the results obtained for $\epsilon_{3}=0.02$ and $\beta=20$ 
are representative for large $\beta$ values, while the results obtained for 
$\epsilon_{3}=0.02$ and $\beta=10$ are representative for intermediate $\beta$ values.     
The initial condition for the simulations with Eqs. (\ref{rad1}), (\ref{rad51}), and (\ref{rad53}) 
consists of two NLS solitons of the form (\ref{rad2}) with 
frequencies $\beta_{1}(0)=0$ and $\beta_{2}(0)=\beta=20$ or 10. 
The initial amplitudes and phases of the solitons are 
$\eta_{1}(0)=\eta_{2}(0)=1$ and $\alpha_{1}(0)=\alpha_{2}(0)=0$. 
The initial positions and the final propagation distance are 
$y_{1}(0)=0$, $y_{2}(0)=20$, and $z_{f}=10$. 
For these values, the two solitons are well-separated before and after 
the collision. In addition, the collision distance is $z_{c}=0.5$ for 
$\beta=20$ and $z_{c}=1.0$ for $\beta=10$. 
The initial condition for the simulations with Eqs. (\ref{rad44}) and (\ref{rad46})  
is an NLS soliton of the form (\ref{rad2}) with $\beta_{1}(0)=0$, 
$\eta_{1}(0)=1$, $y_{1}(0)=0$, and $\alpha_{1}(0)=0$. 
The parameters $\beta_{2}$, $\eta_{2}$, and $y_{2}$ in the 
equation are taken as $\beta_{2}=20$ or 10, $\eta_{2}=1$, and $y_{2}=20$.


\begin{figure}[ptb]
\begin{center}
\begin{tabular}{cc}
\epsfxsize=9.5cm  \epsffile{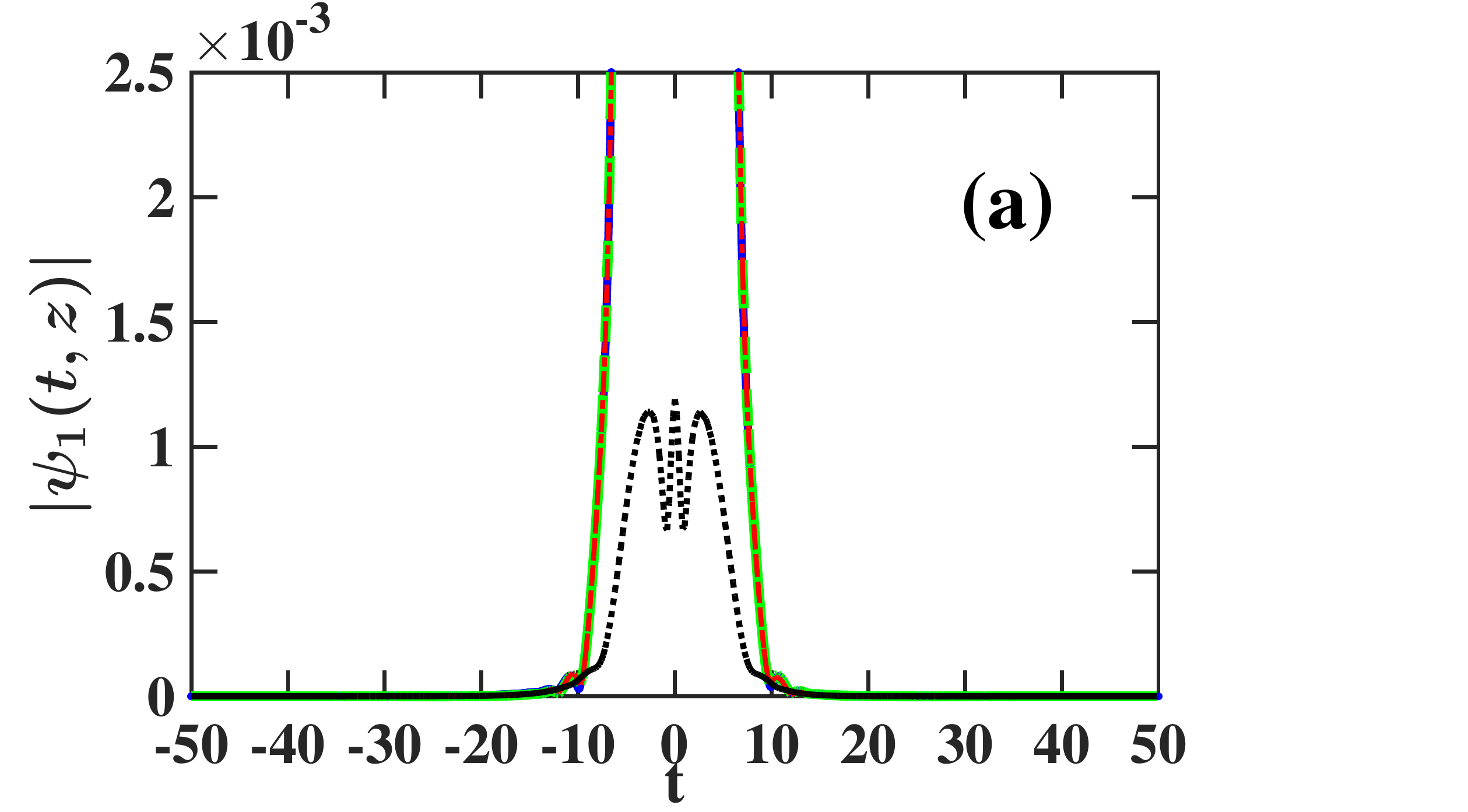} \\
\epsfxsize=9.5cm  \epsffile{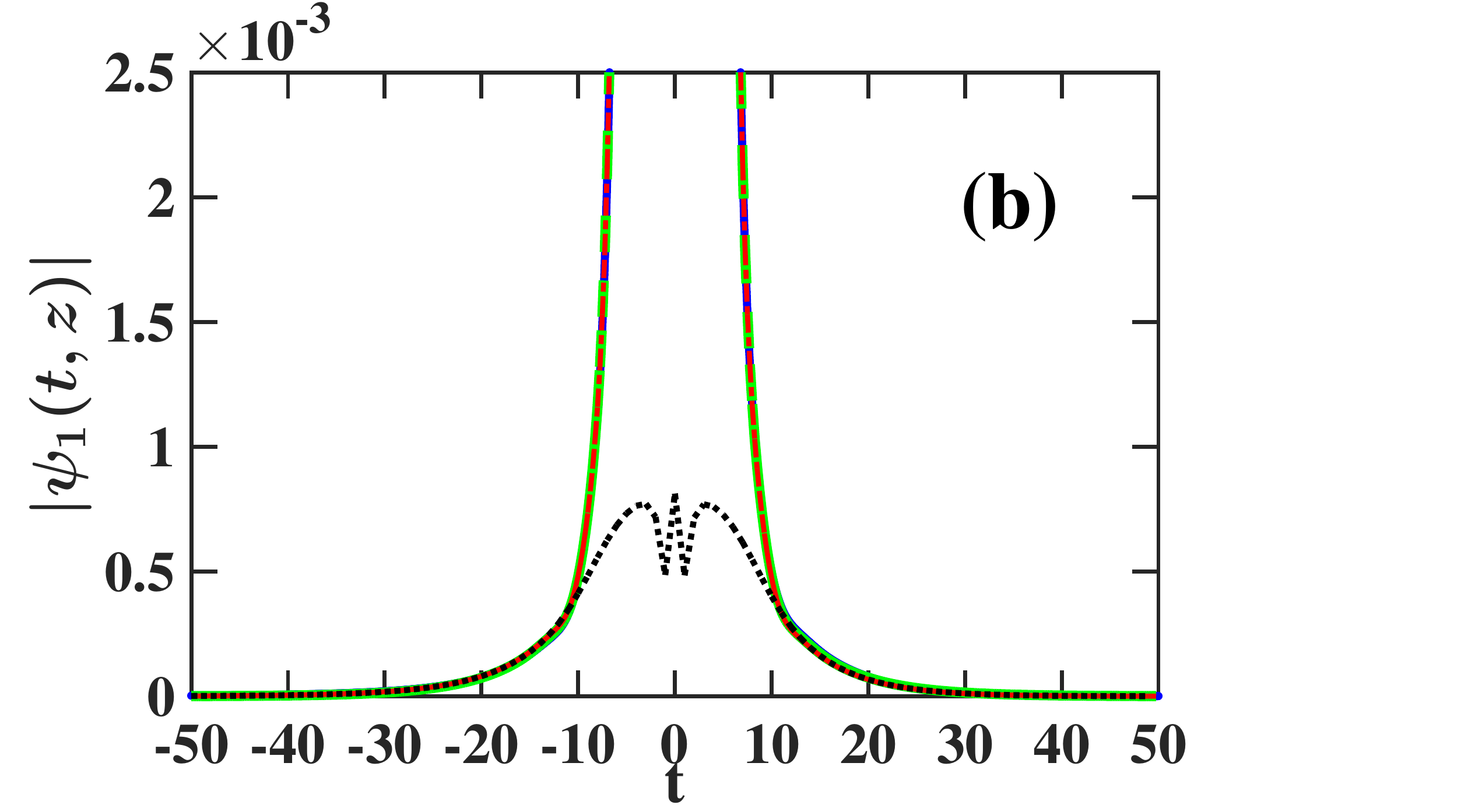} \\
\epsfxsize=9.5cm  \epsffile{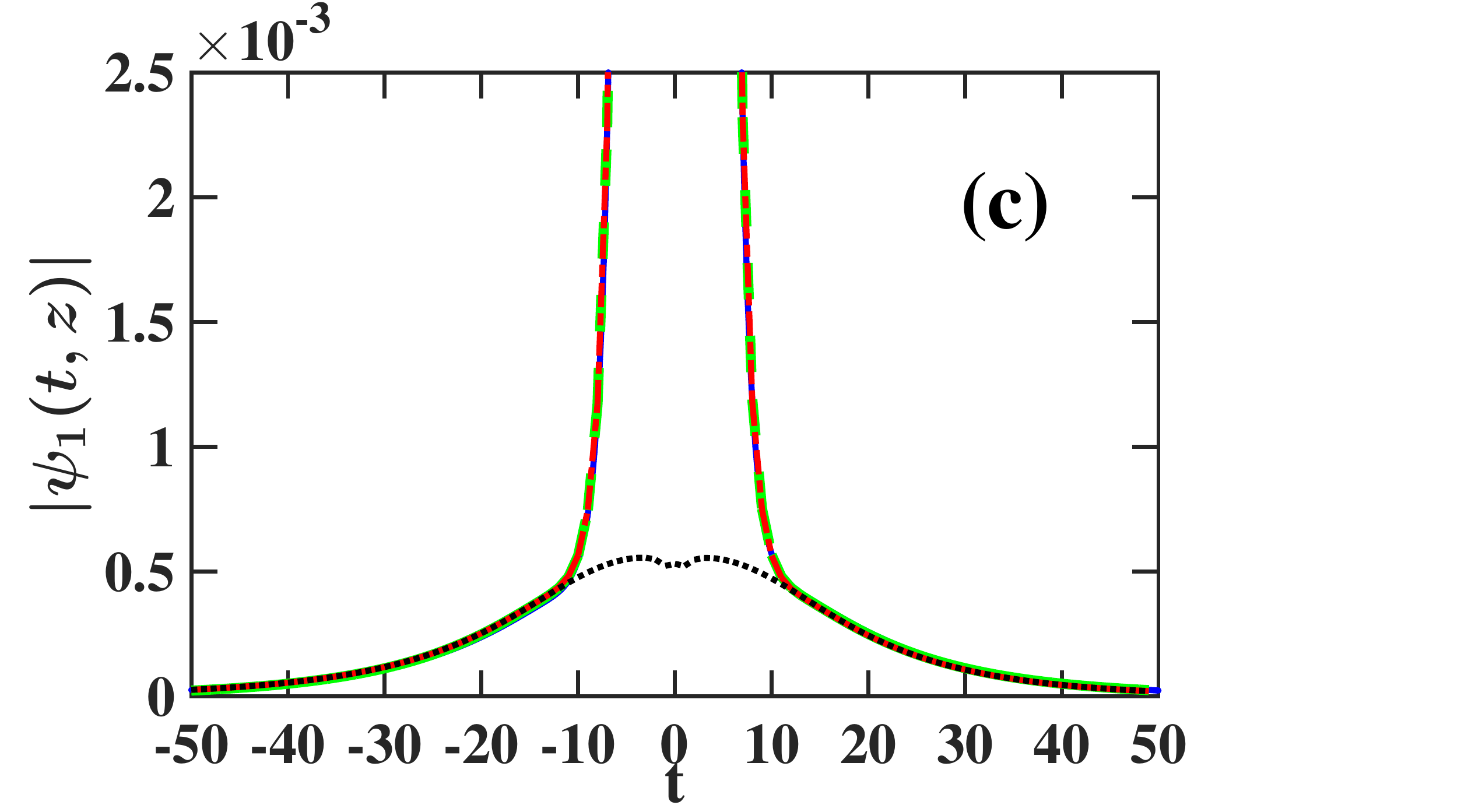} 
\end{tabular}
\end{center}
\caption{A comparison between the perturbation theory's predictions and 
the result of numerical simulation with Eq. (\ref{rad1}) for   
the $t$ dependence of the pulse profile $|\psi_{1}(t,z)|$.
The physical parameter values are $\epsilon_{3}=0.02$ and $\beta=20$ 
and the distances are $z=z_{c}+2$ in (a), $z=z_{c}+5$ in (b), 
and $z=z_{c}+10$ in (c). 
The solid blue curve represents the result obtained by numerical solution 
of Eq. (\ref{rad1}). The dashed green and dashed-dotted red 
curves correspond to the predictions of the basic and improved versions of the 
perturbation approach $|\psi_{1b}(t,z)|$ and $|\psi_{1c}(t,z)|$, respectively. 
The dotted black curve corresponds to the radiation profile $|v_{12c}(t,z)|$ 
obtained with the improved perturbation procedure.}            
 \label{fig5}
\end{figure}

\begin{figure}[ptb]
\begin{center}
\begin{tabular}{cc}
\epsfxsize=9.5cm  \epsffile{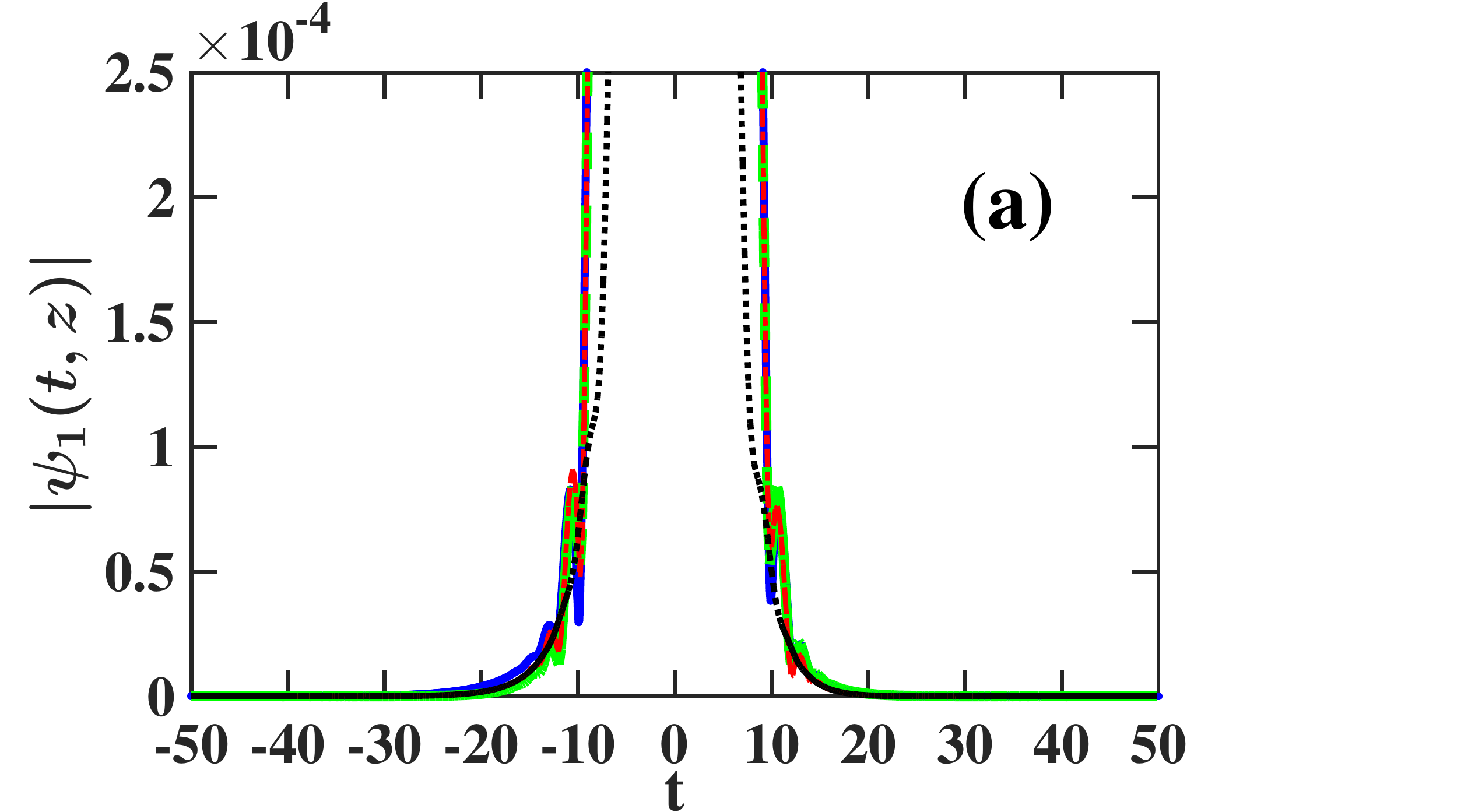} \\
\epsfxsize=9.5cm  \epsffile{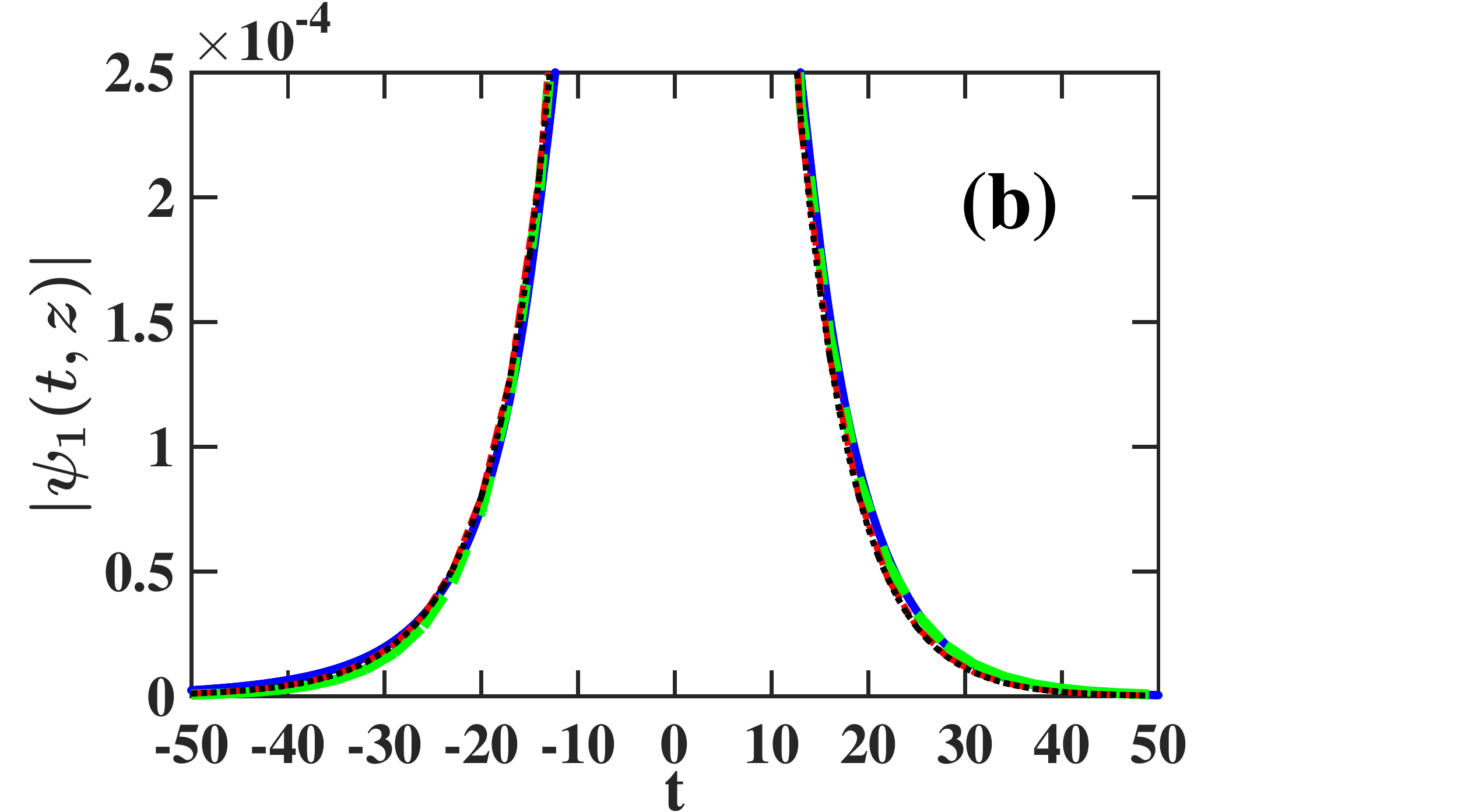} \\
\epsfxsize=9.5cm  \epsffile{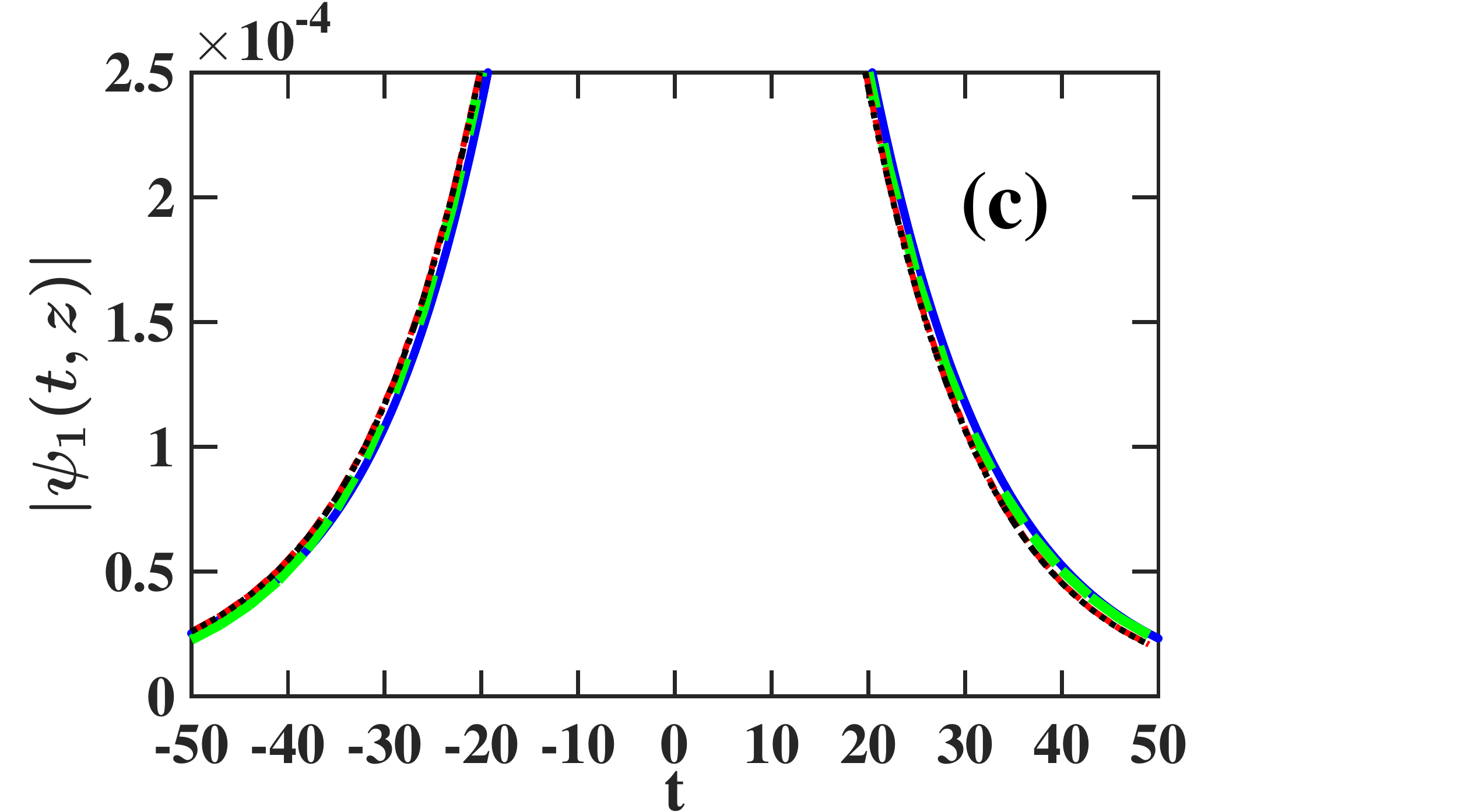} 
\end{tabular}
\end{center}
\caption{Magnified versions of the graphs in Fig. \ref{fig5} for small 
$|\psi_{1}(t,z)|$ values. The symbols are the same as in Fig. \ref{fig5}.}        
 \label{fig6}
\end{figure}

Figure \ref{fig5} shows the $t$ dependence of the pulse profile $|\psi_{1}(t,z)|$ 
obtained in the numerical simulation with Eq. (\ref{rad1}) for $\epsilon_{3}=0.02$ 
and $\beta=20$ at $z=z_{c}+2$, $z=z_{c}+5$, and $z=z_{c}+10$. 
Also shown are the predictions of the simple and improved versions of the 
perturbation approach $|\psi_{1b}(t,z)|$ and $|\psi_{1c}(t,z)|$, and the 
radiation profile $|v_{12c}(t,z)|$ obtained with the improved perturbation approach. 
Figure \ref{fig6} shows magnified versions of the graphs in Fig. \ref{fig5} for small 
$|\psi_{1}(t,z)|$ values. We observe very good agreement between the predictions 
of the two perturbation approaches and the numerical simulation's result at all three 
distances. To further quantify the accuracy of the two predictions of the perturbation theory, 
we calculate the deviations $||\psi_{1}^{(num)}(t,z)| - |\psi_{1b}(t,z)||$ and 
$||\psi_{1}^{(num)}(t,z)| - |\psi_{1c}(t,z)||$, where $\psi_{1}^{(num)}(t,z)$ 
is the envelope of the electric field obtained in the simulation with Eq. (\ref{rad1}). 
We find that the accuracies of the two predictions are comparable. In addition, the prediction of the 
improved perturbation approach for the radiative tails is more accurate 
at short distances, while the prediction of the simple perturbation approach 
is more accurate at long distances. 
For example, at $z=z_{c}+2$, we find that the improved theory's prediction is 
more accurate than the simple theory's prediction in the intervals $-50 \le t < -10.77$, 
and $21.44 < t \le 50.0$ \cite{numerics1}. 
In contrast, at $z=z_{c}+10$, the simple theory's prediction is more accurate 
for $-45.84 < t < -6.28$ and for $6.54 < t \le 50$, while the improved theory's 
prediction is more accurate for $-50 \le t < -45.84$ \cite{numerics2}.         
Moreover, we find that the accuracies of both predictions of the perturbation 
theory for the radiative tails increase with increasing distance. More specifically, 
for $|t| > 6$, the deviations $||\psi_{1}^{(num)}(t,z)| - |\psi_{1b}(t,z)||$ and 
$||\psi_{1}^{(num)}(t,z)| - |\psi_{1c}(t,z)||$ are smaller than $4.77 \times 10^{-5}$ 
and $5.21 \times 10^{-5}$ at $z=z_{c}+2$, and are smaller than $1.59 \times 10^{-5}$ 
and $1.86 \times 10^{-5}$ at $z=z_{c}+10$. 
Thus, our numerical simulation with $\epsilon_{3}=0.02$ and $\beta=20$ 
validate the predictions of both perturbation approaches with high accuracy.


\begin{figure}[ptb]
\begin{center}
\begin{tabular}{cc}
\epsfxsize=9.5cm  \epsffile{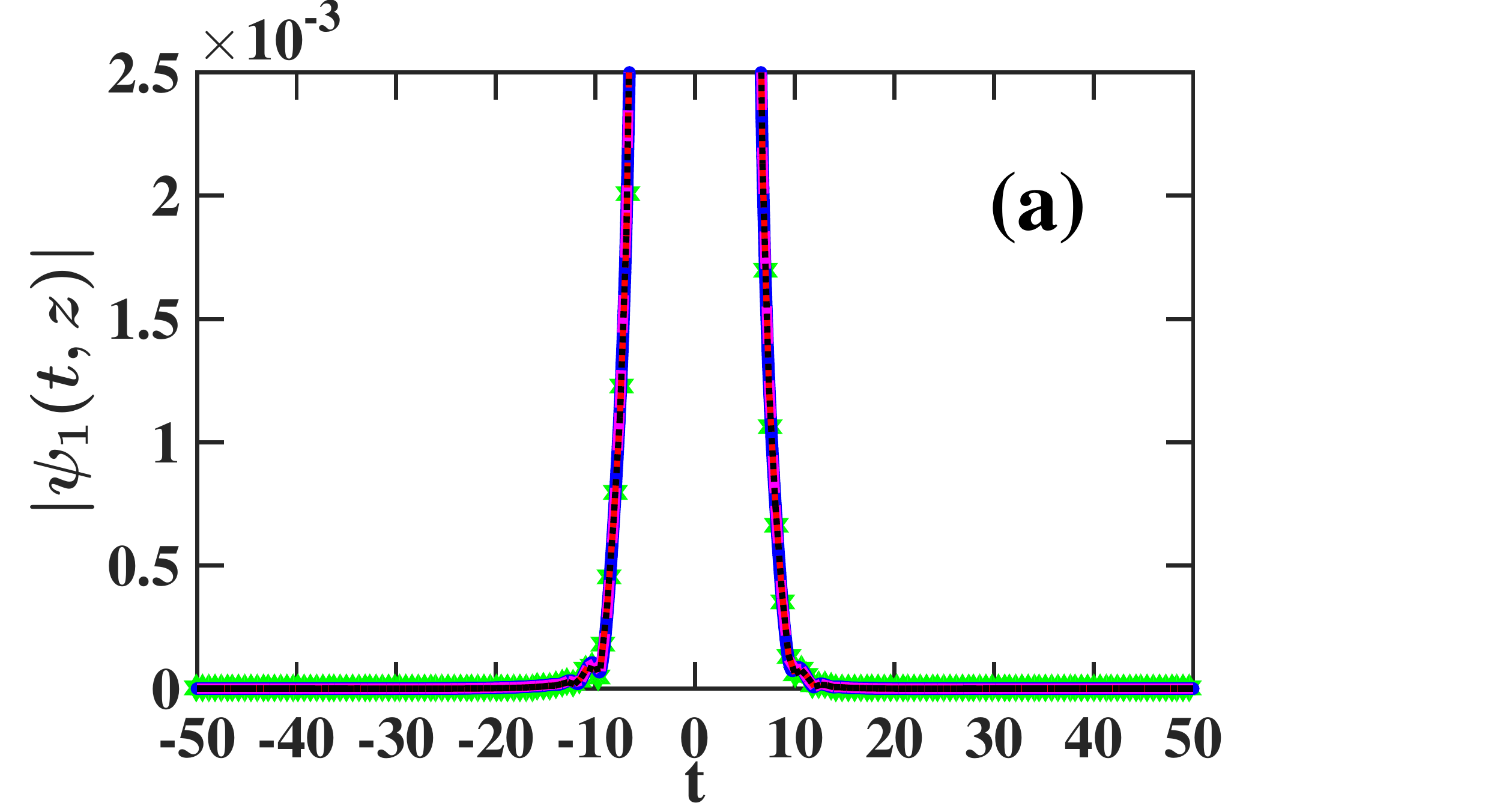} \\
\epsfxsize=9.5cm  \epsffile{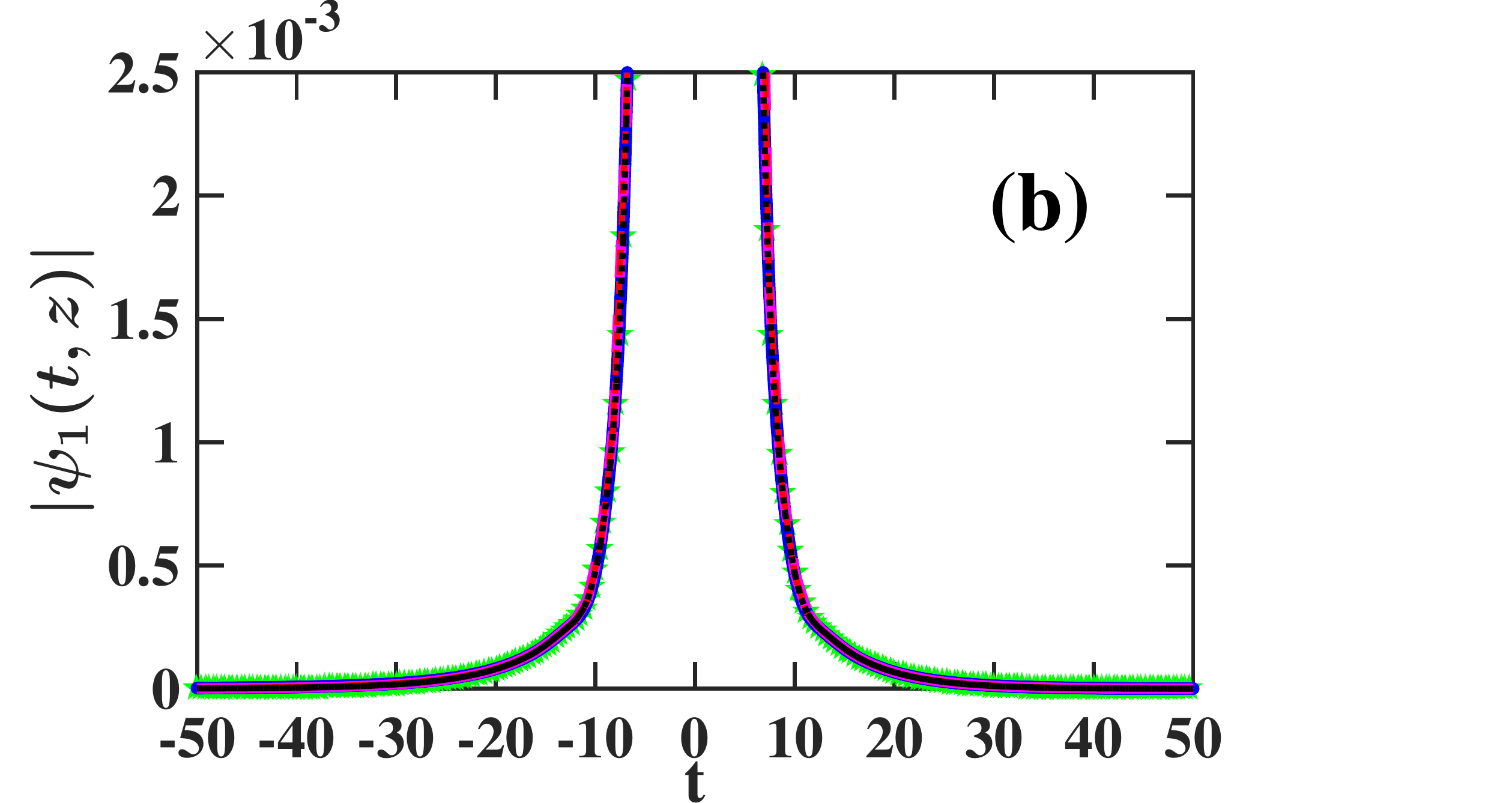} \\
\epsfxsize=9.5cm  \epsffile{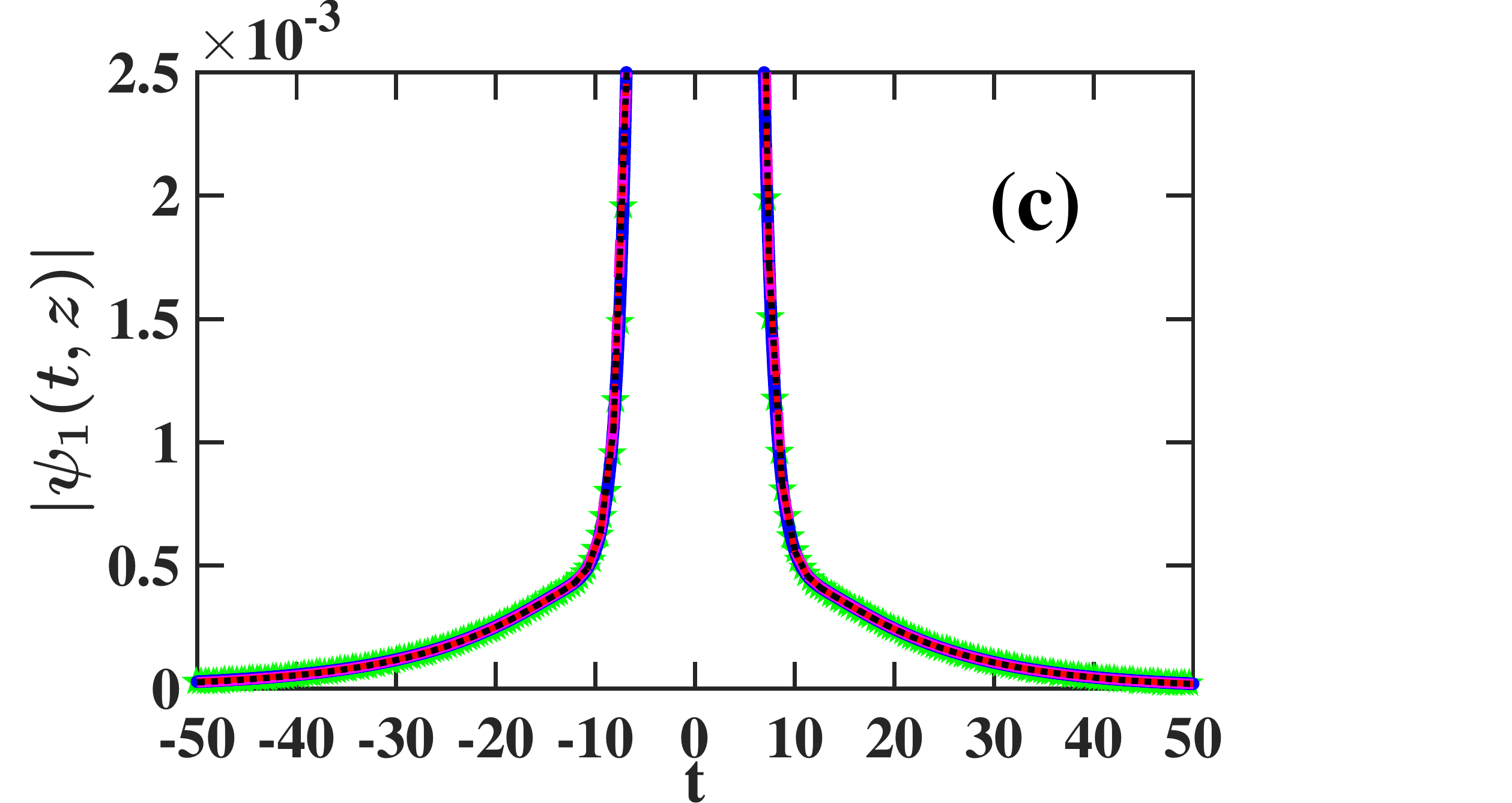} 
\end{tabular}
\end{center}
\caption{A comparison between the $t$ dependences of the pulse profile 
$|\psi_{1}(t,z)|$ obtained by the improved perturbation approach and the 
pulse profiles obtained in numerical simulations with the four simplified NLS 
models (\ref{rad44}), (\ref{rad46}), (\ref{rad51}), and (\ref{rad53}). 
The parameter values are $\epsilon_{3}=0.02$ and $\beta=20$ 
and the distances are $z=z_{c}+2$ in (a), $z=z_{c}+5$ in (b), 
and $z=z_{c}+10$ in (c). The green stars represent the prediction of 
the improved perturbation approach. The solid-blue, dashed magenta, 
dashed-dotted red, and dotted black curves correspond to the pulse 
profiles obtained by numerical solution of Eqs. (\ref{rad44}), 
(\ref{rad46}), (\ref{rad51}), and (\ref{rad53}).} 
 \label{fig_add1}
\end{figure}

\begin{figure}[ptb]
\begin{center}
\begin{tabular}{cc}
\epsfxsize=9.5cm  \epsffile{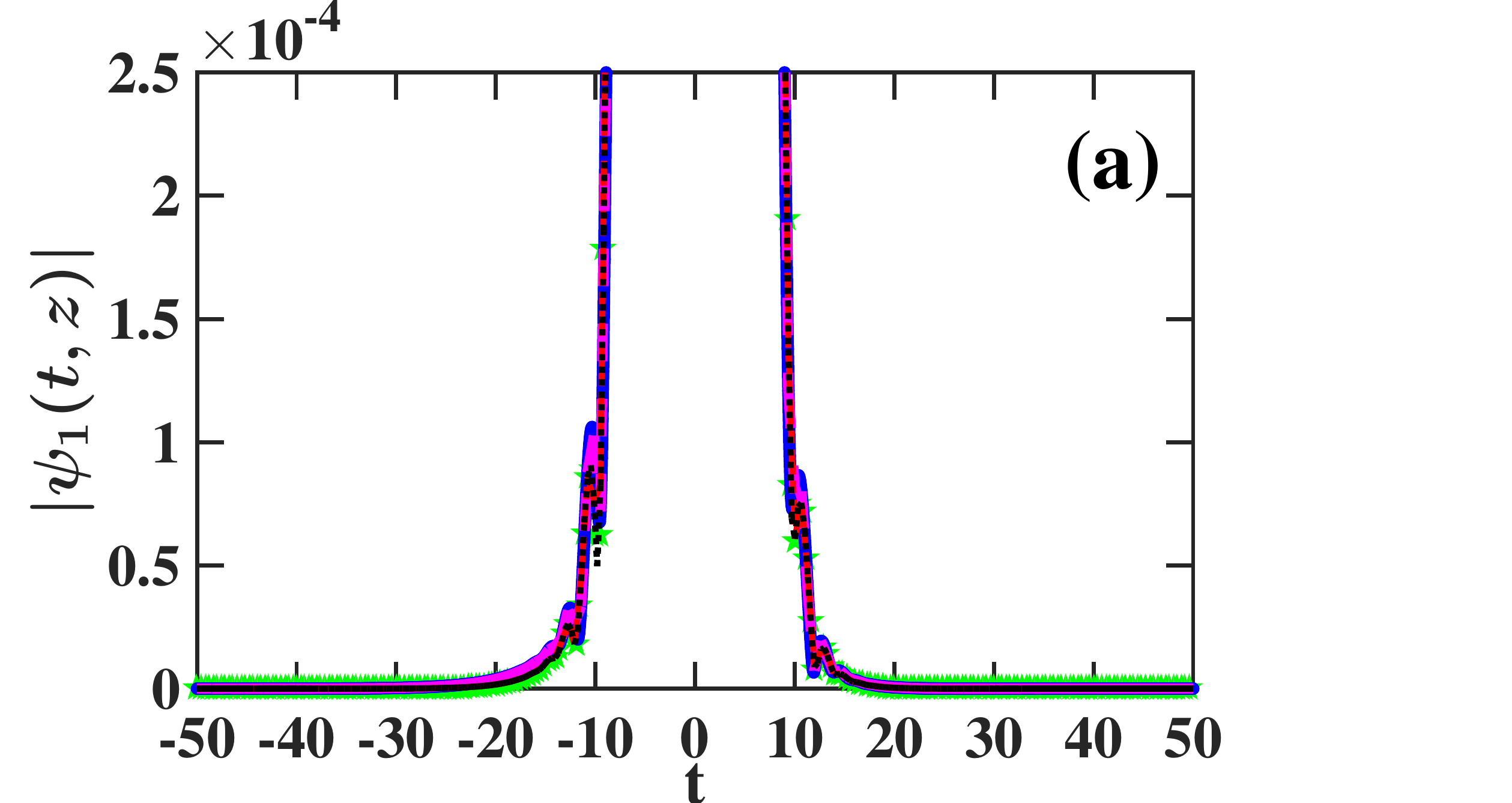} \\
\epsfxsize=9.5cm  \epsffile{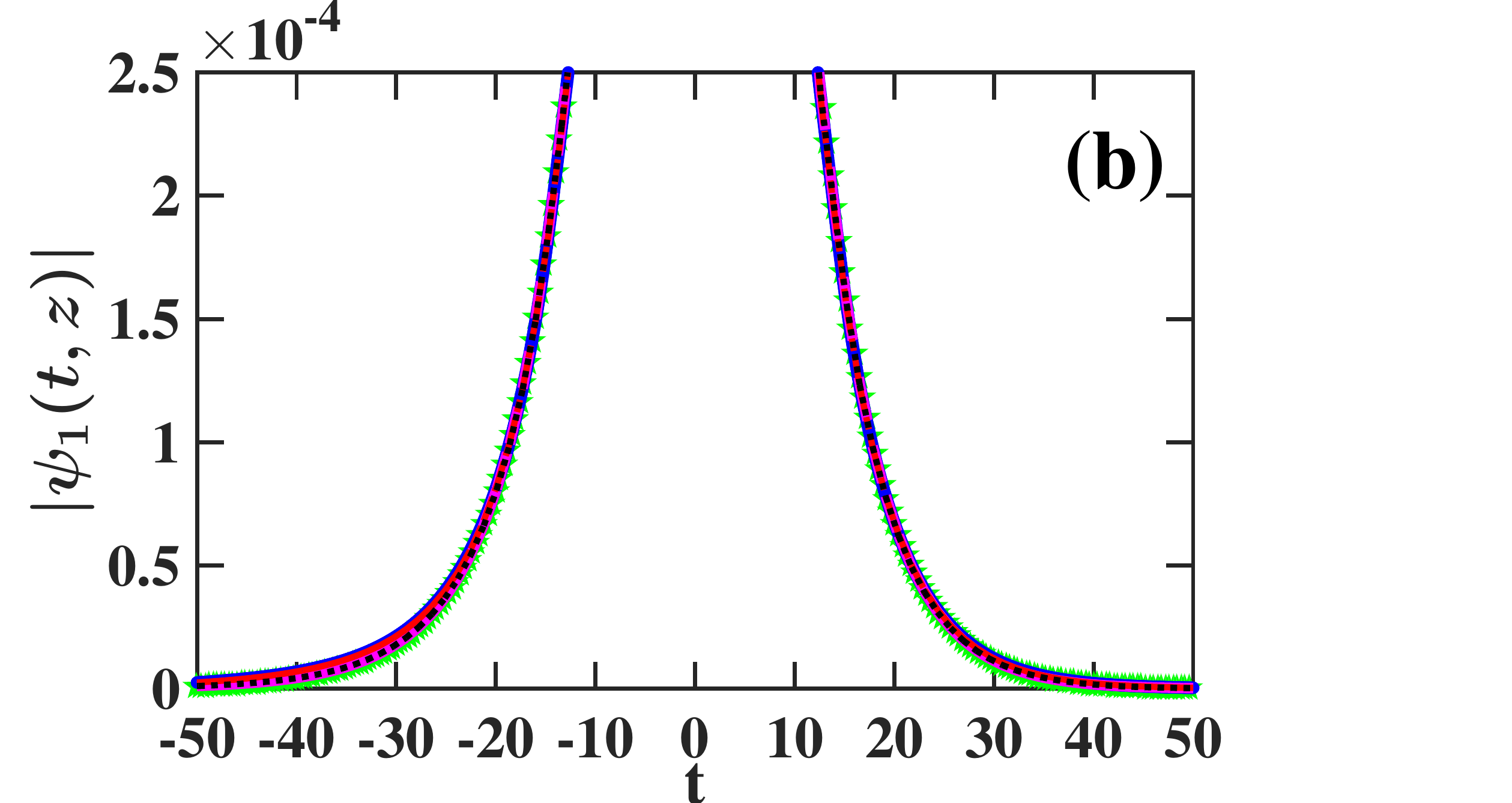} \\
\epsfxsize=9.5cm  \epsffile{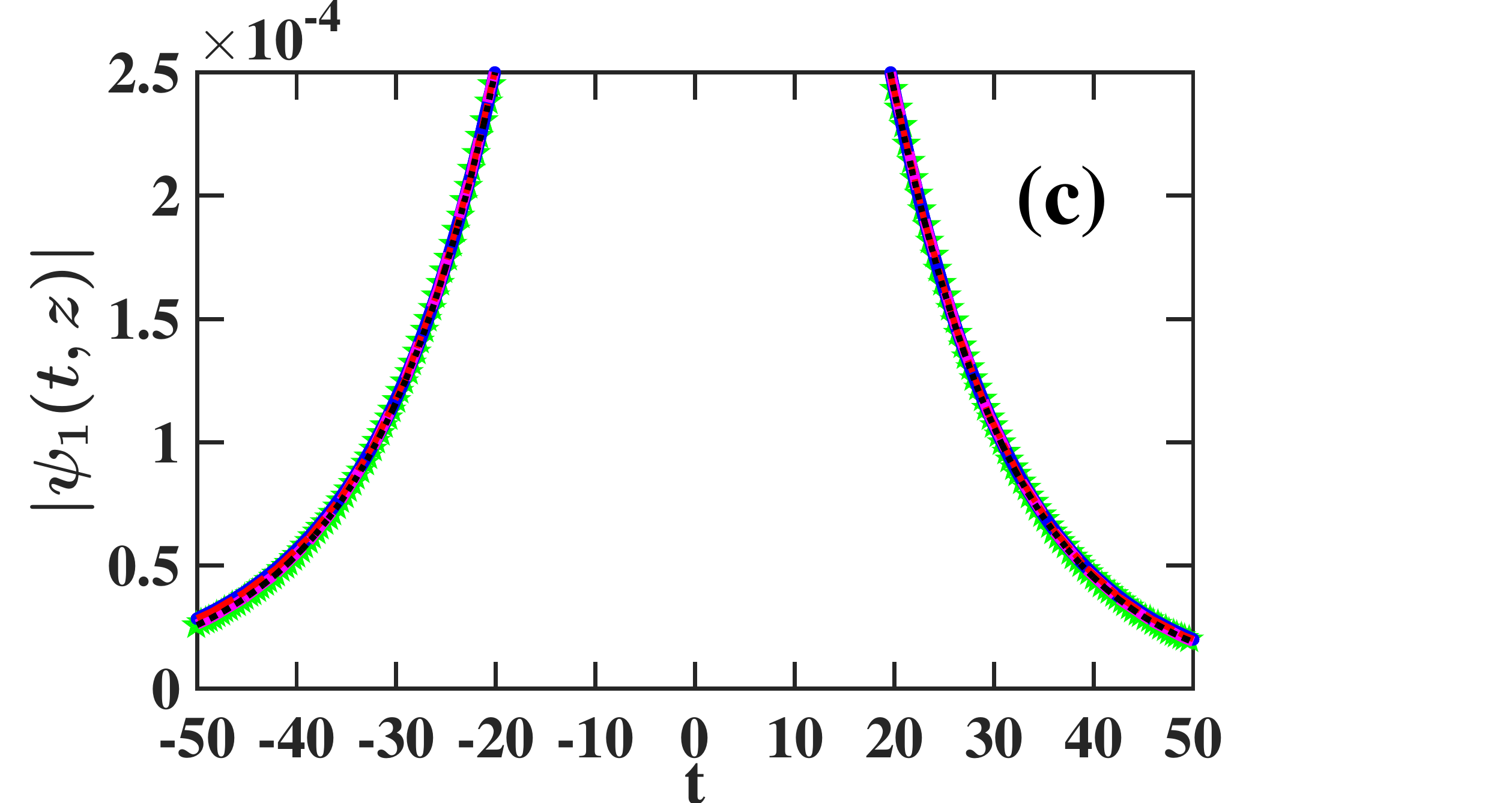} 
\end{tabular}
\end{center}
\caption{Magnified versions of the graphs in Fig. \ref{fig_add1} for small 
$|\psi_{1}(t,z)|$ values. The symbols are the same as in Fig. \ref{fig_add1}.}        
 \label{fig_add2}
\end{figure}

It is helpful to compare the predictions of the perturbation theory  
with results of numerical simulations with the four simplified NLS models (\ref{rad44}), 
(\ref{rad46}), (\ref{rad51}), and (\ref{rad53}).  Figure \ref{fig_add1} shows the 
pulse profile $|\psi_{1}(t,z)|$ obtained by the improved perturbation approach 
and by numerical solution of Eqs. (\ref{rad44}), (\ref{rad46}), (\ref{rad51}), 
and (\ref{rad53}) at $z=z_{c}+2$, $z=z_{c}+5$, and $z=z_{c}+10$.   
Figure \ref{fig_add2} shows magnified versions of the graphs 
in Fig. \ref{fig_add1} for small $|\psi_{1}(t,z)|$ values. 
We observe very good agreement between the improved perturbation theory's prediction 
and the results of the four simplified NLS models for all three distances. 
We recall that the perturbation theory and the models (\ref{rad46}) and (\ref{rad53}) 
neglect the effects of Kerr-induced interpulse interaction on radiation dynamics, 
but that these effects are taken into account in the models (\ref{rad44}) and (\ref{rad51}).      
Therefore, the comparison presented in Figs. \ref{fig_add1} and \ref{fig_add2} 
indicates that the effects of interpulse interaction due to Kerr nonlinearity on the collision-induced  
radiation dynamics are unimportant for $\epsilon_{3}=0.02$ and $\beta=20$.

\begin{figure}[ptb]
\begin{center}
\begin{tabular}{cc}
\epsfxsize=9.5cm  \epsffile{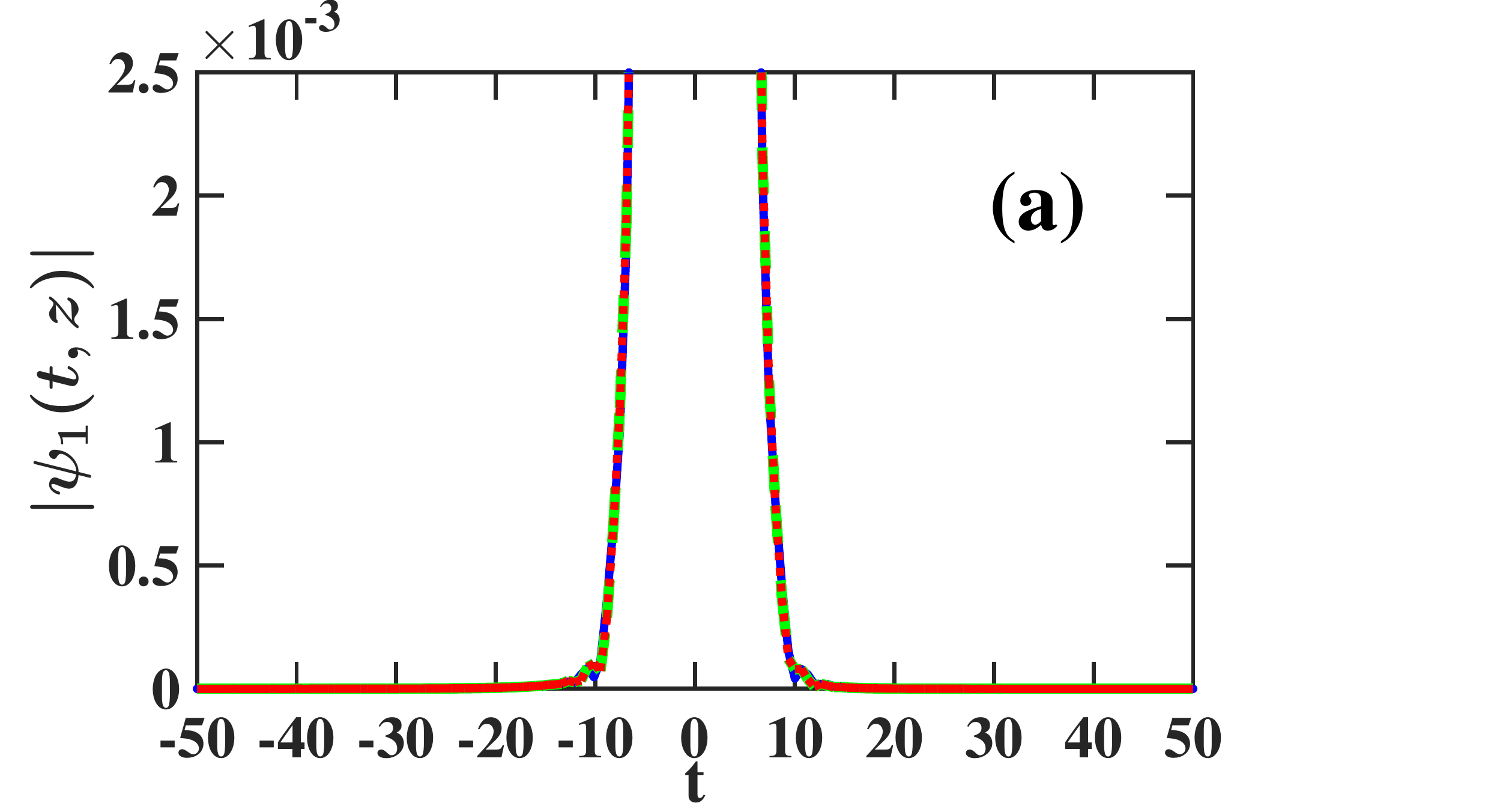} \\
\epsfxsize=9.5cm  \epsffile{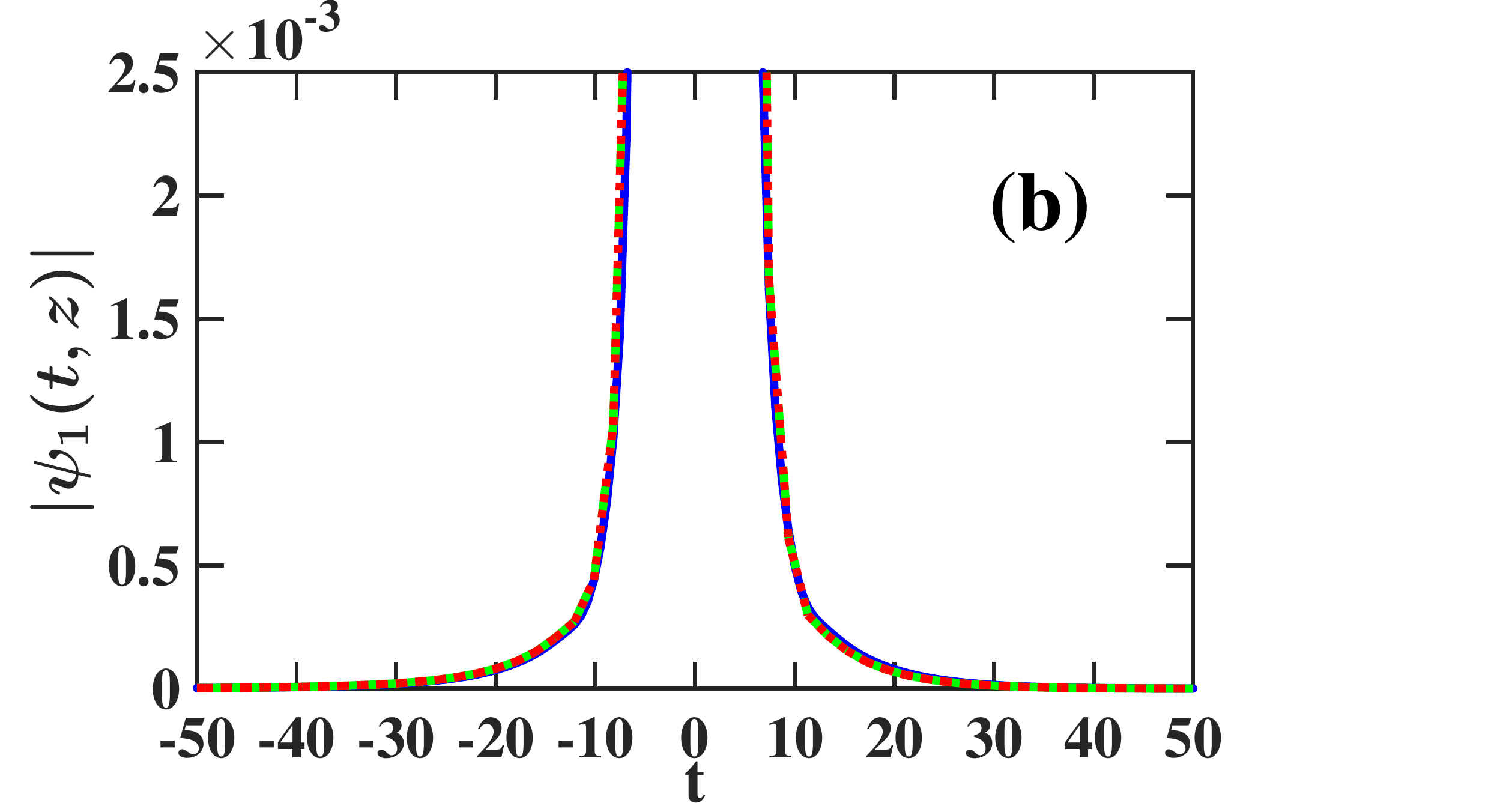} \\
\epsfxsize=9.5cm  \epsffile{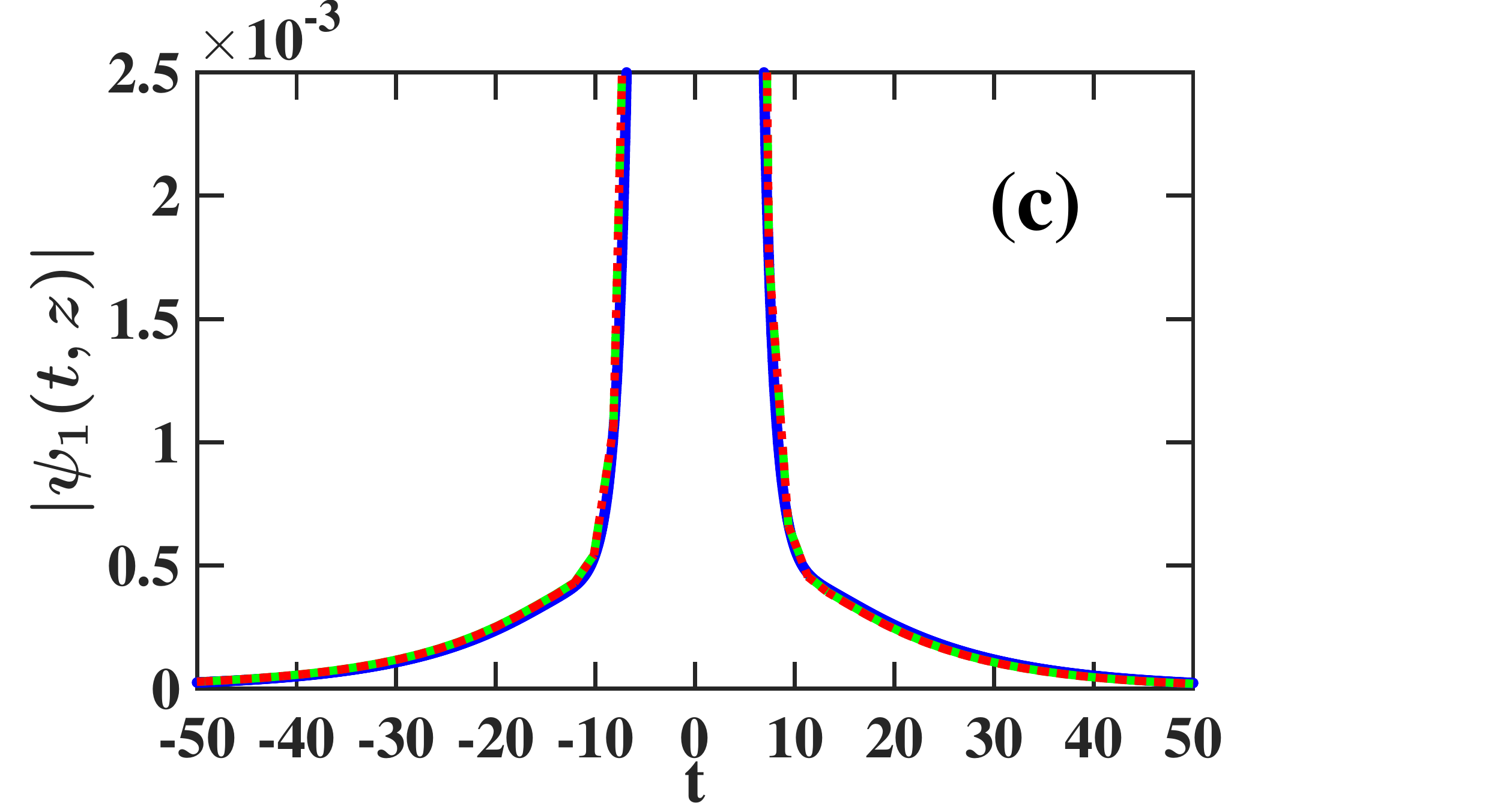} 
\end{tabular}
\end{center}
\caption{A comparison between the $t$ dependences of the pulse profile 
$|\psi_{1}(t,z)|$ obtained in numerical simulations with the three 
perturbed NLS  models (\ref{rad1}), (\ref{rad44}), and (\ref{rad51}).   
The physical parameter values are $\epsilon_{3}=0.02$ and $\beta=20$ 
and the distances are $z=z_{c}+2$ in (a), $z=z_{c}+5$ in (b), 
and $z=z_{c}+10$ in (c). The solid blue, dashed green
and dotted red curves represent $|\psi_{1}(t,z)|$ 
obtained by numerical solution of Eqs. (\ref{rad1}), (\ref{rad44}), 
and (\ref{rad51}), respectively.}                   
 \label{fig7}
\end{figure}
\begin{figure}[ptb]
\begin{center}
\begin{tabular}{cc}
\epsfxsize=9.5cm  \epsffile{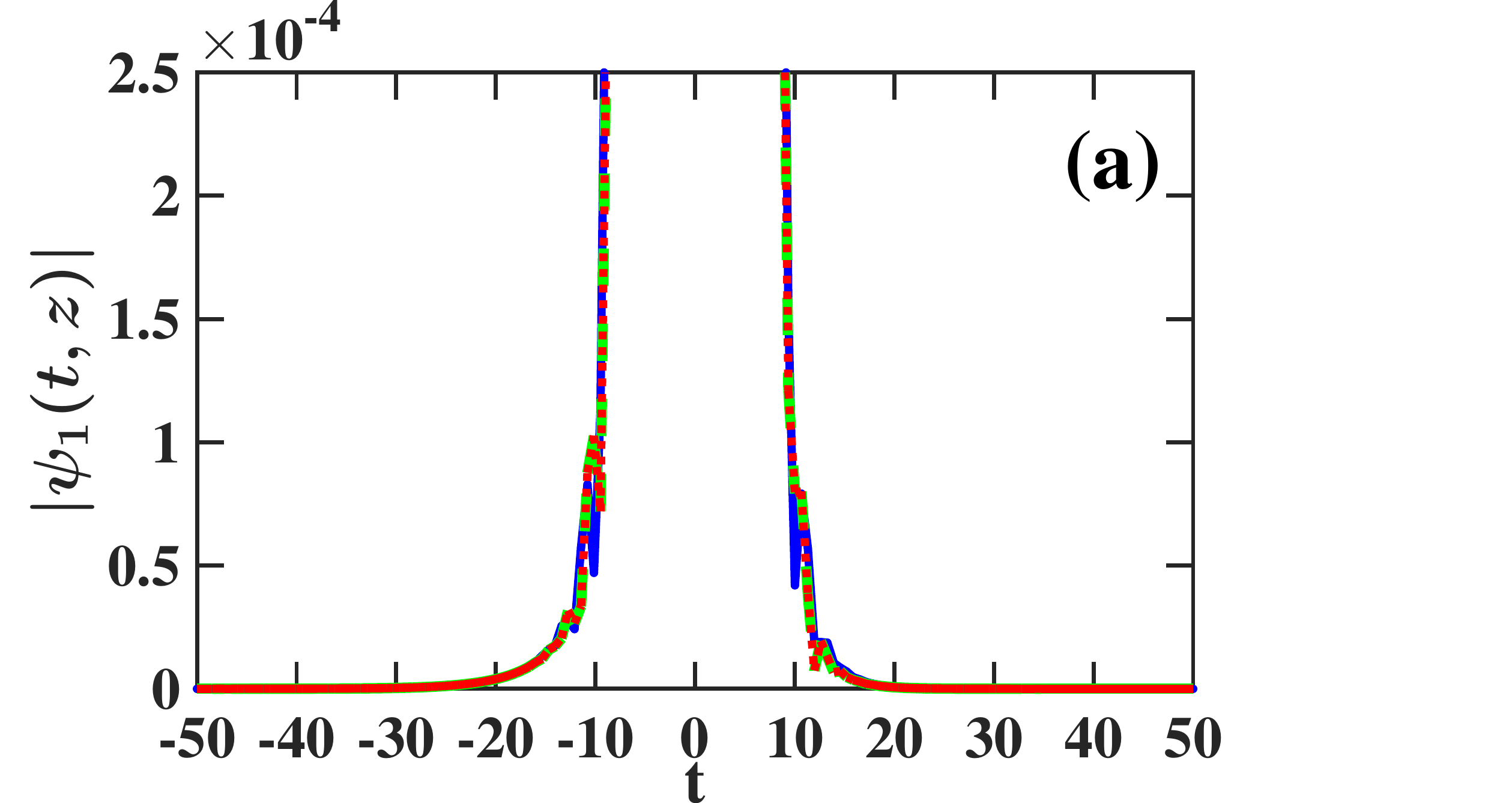} \\
\epsfxsize=9.5cm  \epsffile{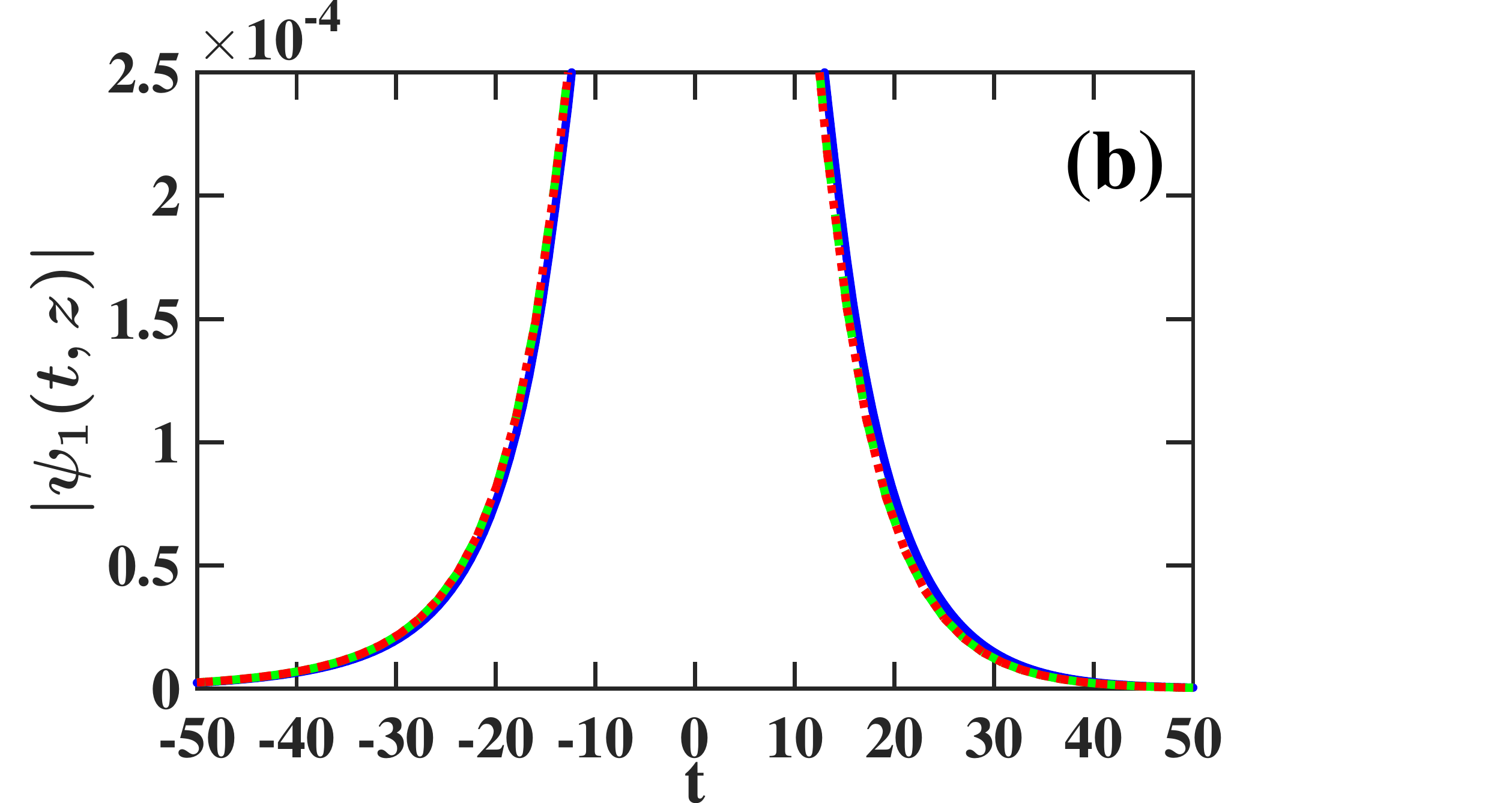} \\
\epsfxsize=9.5cm  \epsffile{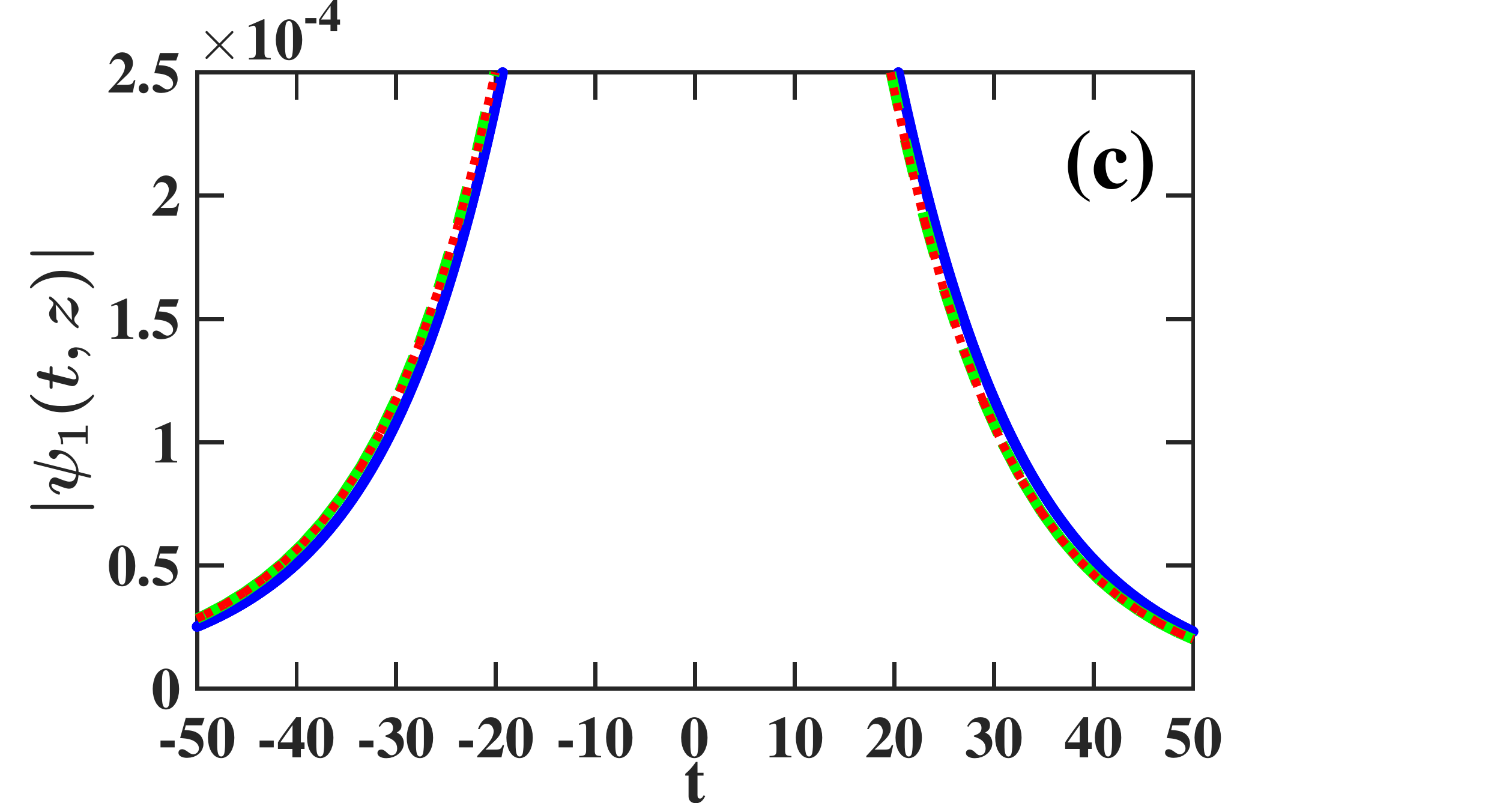} 
\end{tabular}
\end{center}
\caption{Magnified versions of the graphs in Fig. \ref{fig7} for small 
$|\psi_{1}(t,z)|$ values. The symbols are the same as in Fig. \ref{fig7}.}        
 \label{fig8}
\end{figure}

To gain further insight into the physical processes that govern 
radiation dynamics in the collision we compare the pulse profile 
$|\psi_{1}^{(num)}(t,z)|$ obtained in simulations with the 
full coupled-NLS model (\ref{rad1}) with the pulse profiles
$|\psi_{1}^{(num,s)}(t,z)|$ and $|\psi_{1}^{(num,p)}(t,z)|$ 
obtained in simulations with the two simplified NLS models 
(\ref{rad44}) and (\ref{rad51}), respectively. 
Figure \ref{fig7} shows the comparison of the pulse profiles obtained 
in numerical simulations with Eqs. (\ref{rad1}),  (\ref{rad44}), and (\ref{rad51}) 
at $z=z_{c}+2$, $z=z_{c}+5$, and $z=z_{c}+10$. 
Figure \ref{fig8} shows magnified versions of the graphs in Fig. \ref{fig7} 
for small $|\psi_{1}(t,z)|$ values. We observe that the pulse profiles 
obtained with the simplified models (\ref{rad44}) and (\ref{rad51}) are very close 
to each other at the three distances. 
Furthermore, as seen in Figs. \ref{fig7} and \ref{fig8}, the pulse profiles 
obtained in numerical simulations with the two simplified models 
are in very good agreement with the 
result obtained in simulations with the full coupled-NLS model (\ref{rad1}). 
More specifically, the deviations $||\psi_{1}^{(num)}(t,z)| - |\psi_{1}^{(num,s)}(t,z)||$ 
and $||\psi_{1}^{(num)}(t,z)| - |\psi_{1}^{(num,p)}(t,z)||$ are smaller 
than $1.87 \times 10^{-4}$ and $1.95 \times 10^{-4}$ for all $t$ values at $z=z_{c}+2$, 
and are smaller than $7.57 \times 10^{-4}$ and 
$7.48 \times 10^{-4}$ for all $t$ values at $z=z_{c}+10$.
Moreover, the deviations for $|t|>6$ are both smaller than $1.03 \times 10^{-4}$ 
at $z=z_{c}+2$, and are smaller than $5.45 \times 10^{-5}$ and 
$5.31 \times 10^{-5}$ at $z=z_{c}+10$.
Thus, the agreement between the results of the two simplified NLS models 
for the pulse tails, where radiation is dominant, and the result of the full 
coupled-NLS model improves with increasing propagation distance. 
Recall that distortion of soliton 2 is completely neglected in Eq. (\ref{rad44}). 
Therefore, the good agreement between the results obtained with Eqs. (\ref{rad44}) 
and (\ref{rad1}) means that for $\epsilon_{3}=0.02$ and $\beta=20$, distortion 
of soliton 2 does not play an important role in the collision-induced radiation dynamics 
of soliton 1. Additionally, as seen in Figs. \ref{fig_add1} and \ref{fig_add2}, 
for $\beta=20$ and $\epsilon_{3}=0.02$, 
the results of numerical simulations with Eqs. (\ref{rad44}) and (\ref{rad51}) 
are in very good agreement with the results obtained with the simplified models 
(\ref{rad46}) and (\ref{rad53}), which neglect radiation emission due 
to Kerr-induced interpulse interaction. Therefore, based on the comparison in Figs. 
\ref{fig_add1} - \ref{fig8} we conclude that the contribution of Kerr-induced interpulse interaction 
to radiation dynamics is very small for $\beta=20$ and $\epsilon_{3}=0.02$. 
It follows that in this case, we can use the two simplified NLS models (\ref{rad46}) 
and (\ref{rad53}) to analyze radiation dynamics with very good accuracy.

\begin{figure}[ptb]
\begin{center}
\begin{tabular}{cc}
\epsfxsize=9.5cm  \epsffile{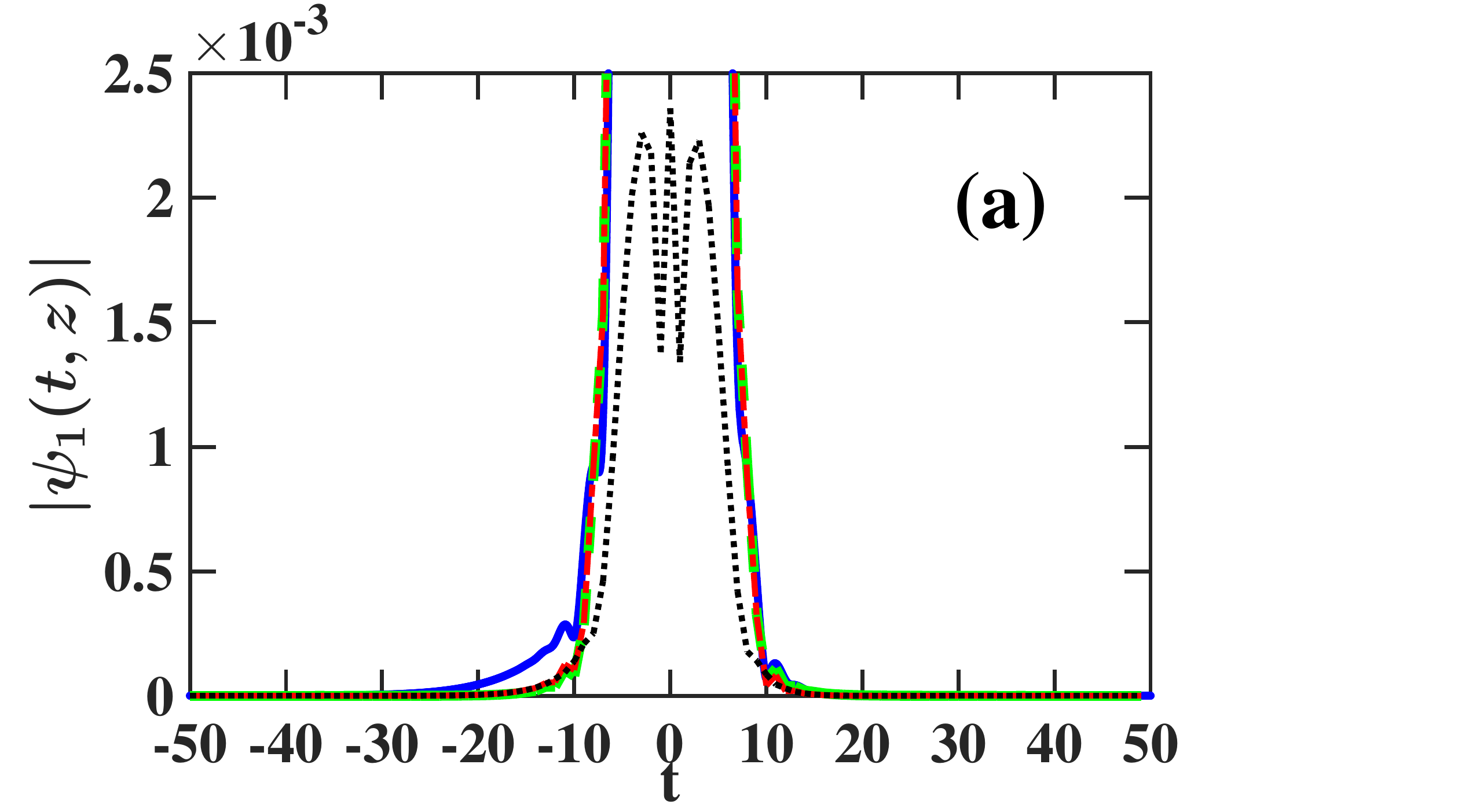} \\
\epsfxsize=9.5cm  \epsffile{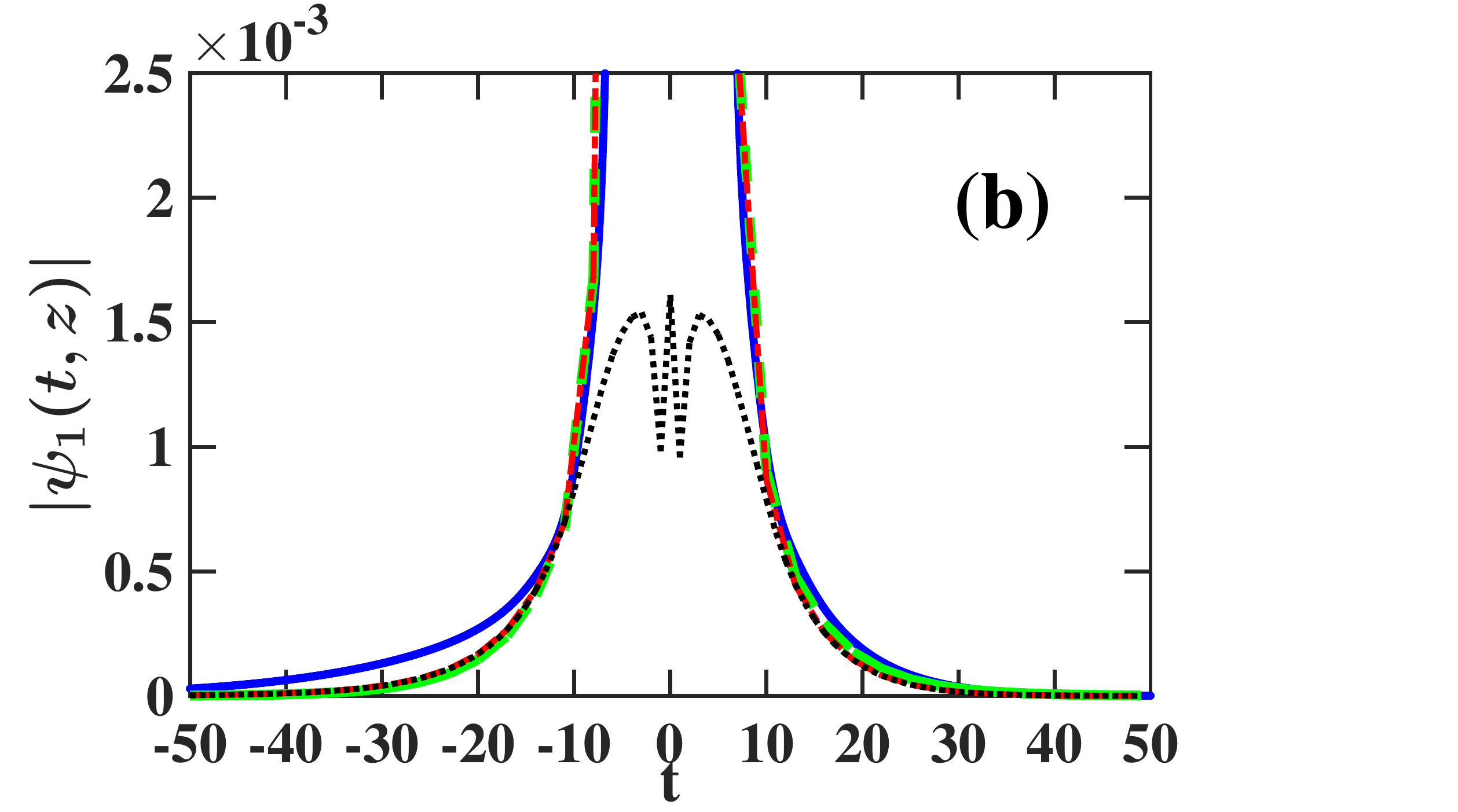} \\
\epsfxsize=9.5cm  \epsffile{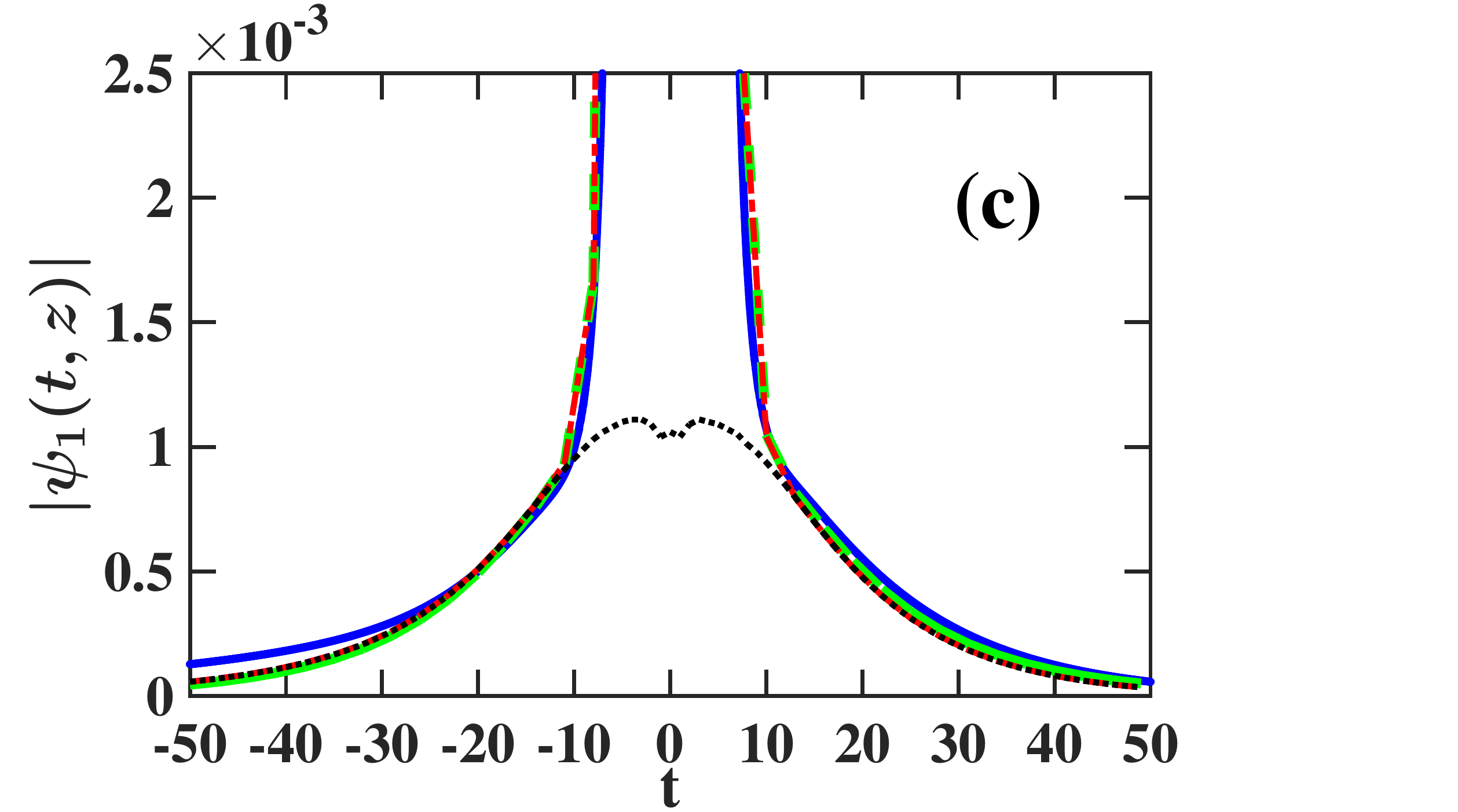} 
\end{tabular}
\end{center}
\caption{A comparison between the perturbation theory's predictions and 
the result of numerical simulation with Eq. (\ref{rad1}) for   
the $t$ dependence of the pulse profile $|\psi_{1}(t,z)|$.
The physical parameter values are $\epsilon_{3}=0.02$ and $\beta=10$ 
and the distances are $z=z_{c}+2$ in (a), $z=z_{c}+5$ in (b), 
and $z=z_{c}+10$ in (c). 
The solid blue curve represents the result obtained by numerical solution 
of Eq. (\ref{rad1}). The dashed green and dashed-dotted red 
curves correspond to the predictions of the basic and improved versions of the 
perturbation approach $|\psi_{1b}(t,z)|$ and $|\psi_{1c}(t,z)|$, respectively. 
The dotted black curve corresponds to the radiation profile $|v_{12c}(t,z)|$ 
obtained with the improved perturbation procedure.}              
 \label{fig9}
\end{figure}

\begin{figure}[ptb]
\begin{center}
\begin{tabular}{cc}
\epsfxsize=9.5cm  \epsffile{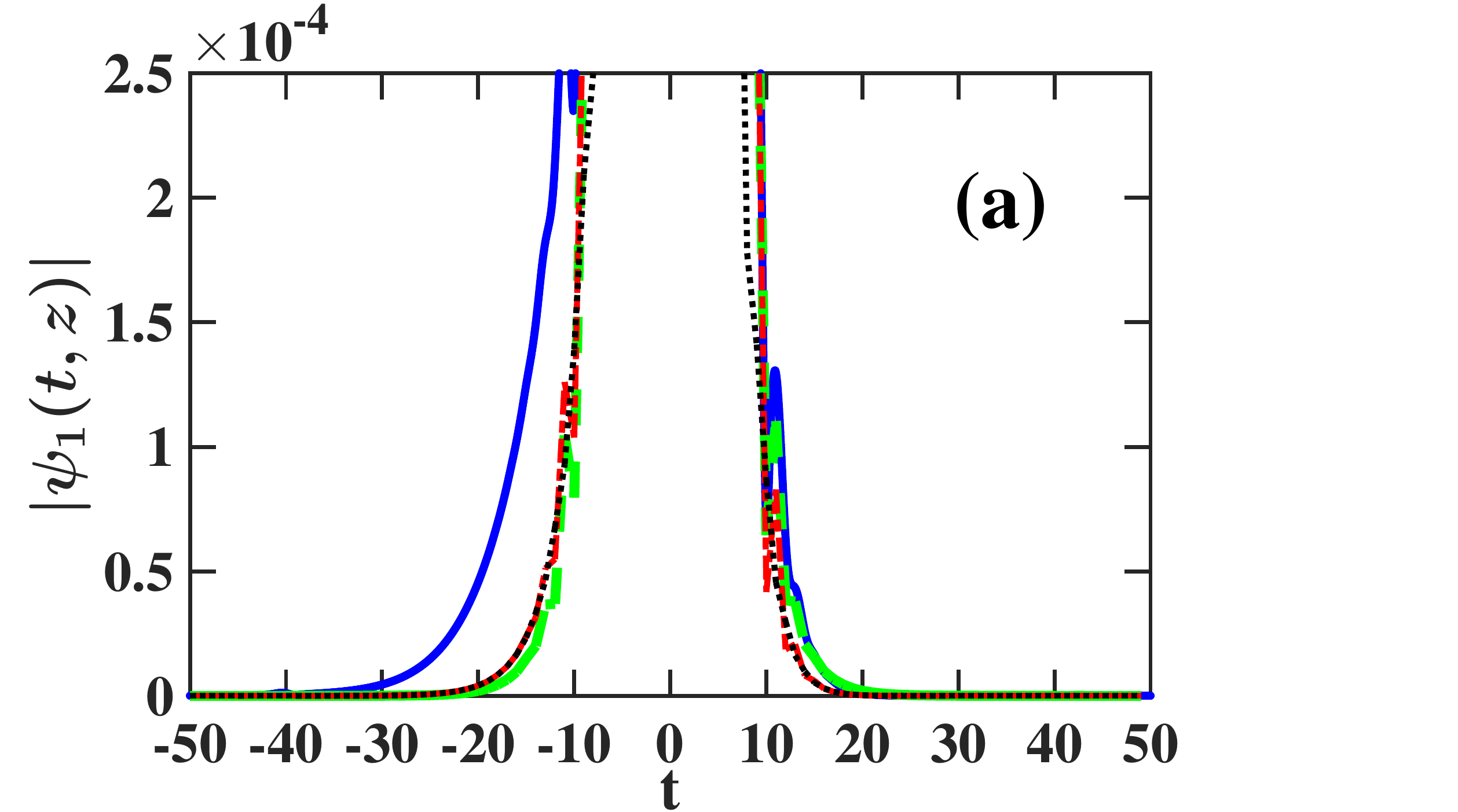} \\
\epsfxsize=9.5cm  \epsffile{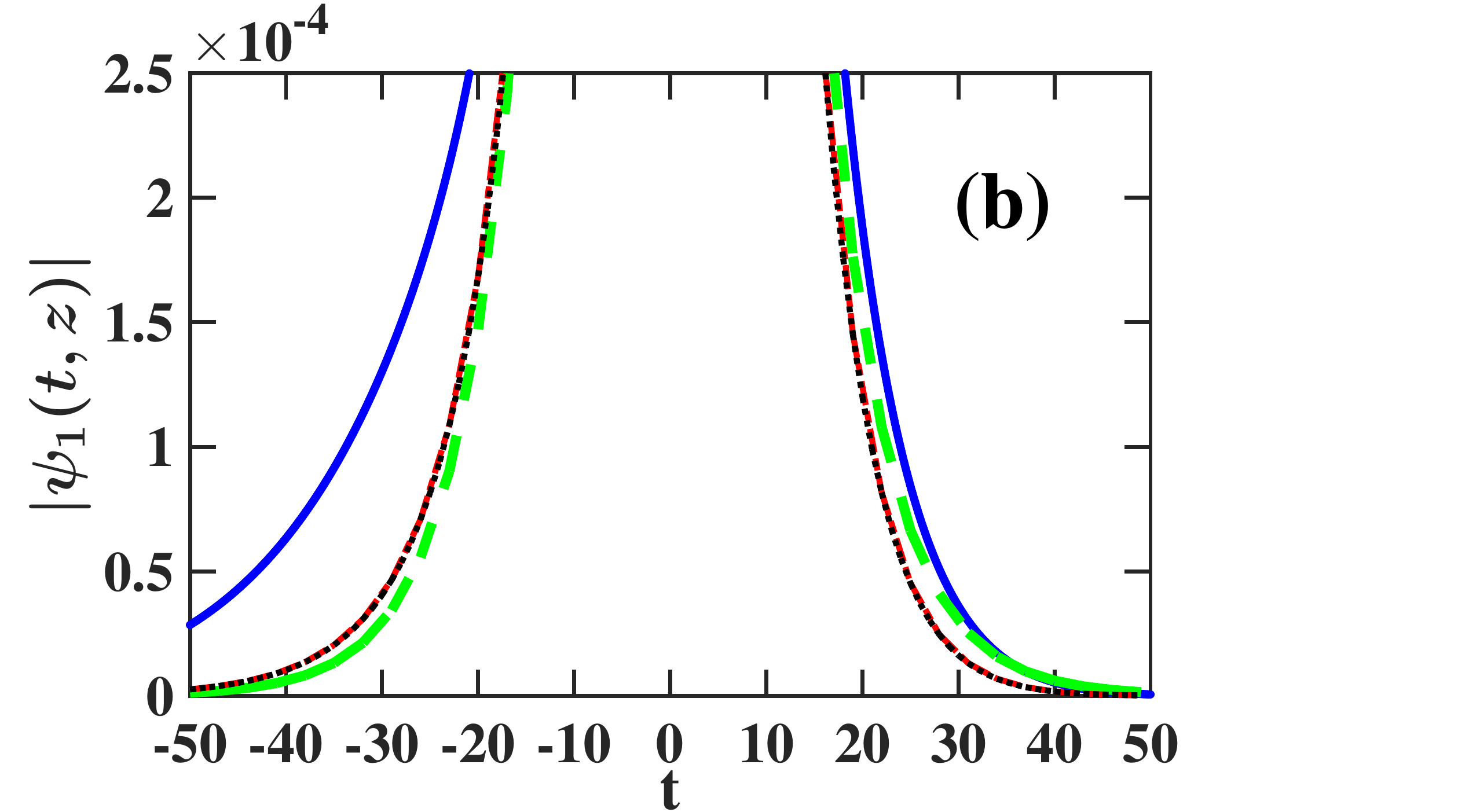} \\
\epsfxsize=9.5cm  \epsffile{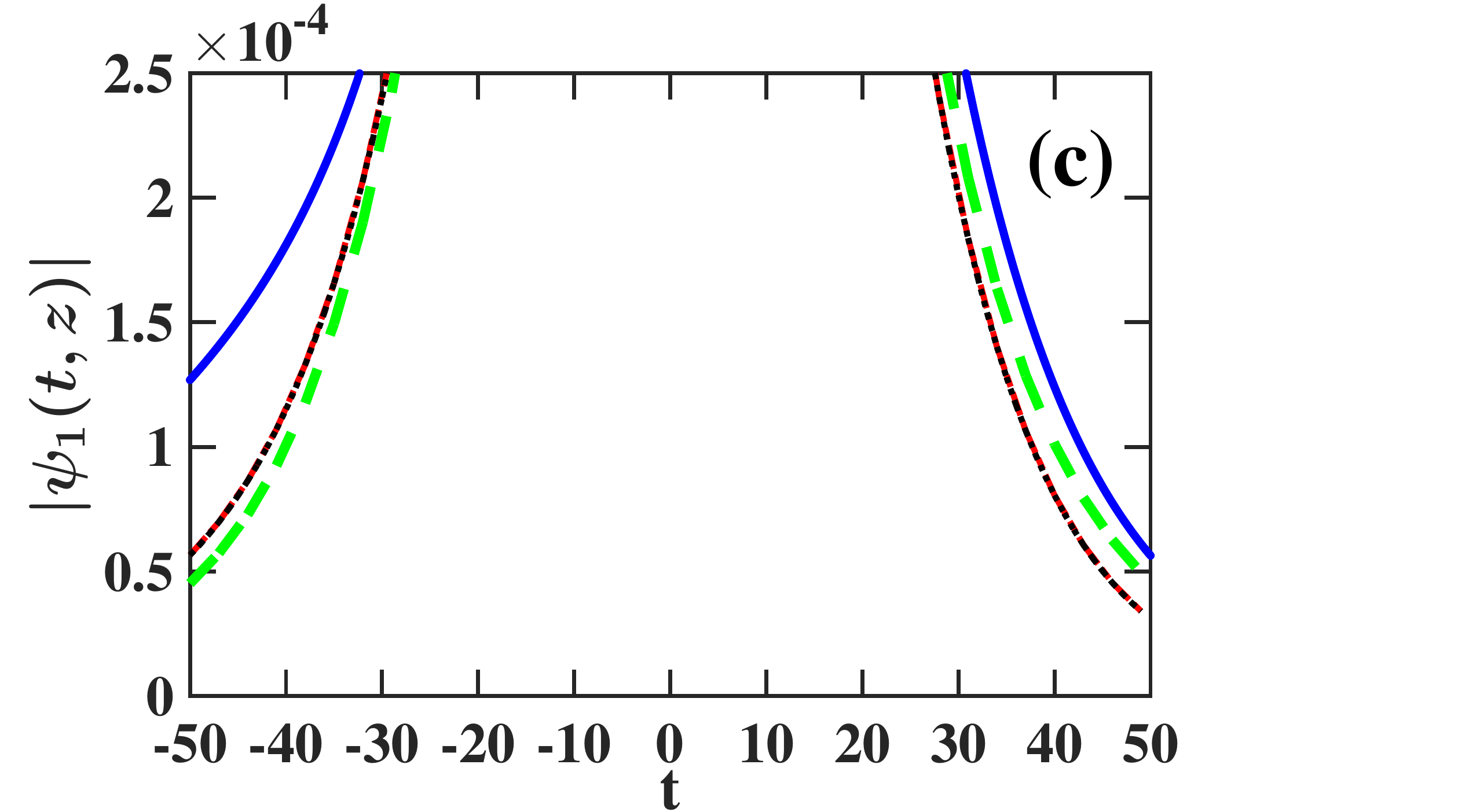} 
\end{tabular}
\end{center}
\caption{Magnified versions of the graphs in Fig. \ref{fig9} for small 
$|\psi_{1}(t,z)|$ values. The symbols are the same as in Fig. \ref{fig9}.}        
 \label{fig10}
\end{figure}

We now turn to analyze the results of numerical simulations with 
$\epsilon_{3}=0.02$ and $\beta=10$ in comparison with the predictions 
of the perturbation theory. Figure \ref{fig9} shows the $t$ dependence 
of the pulse profile  obtained in the simulation with Eq. (\ref{rad1}) 
$|\psi_{1}^{(num)}(t,z)|$ at $z=z_{c}+2$, $z=z_{c}+5$, 
and $z=z_{c}+10$. The predictions of the simple and improved perturbation approaches 
$|\psi_{1b}(t,z)|$ and $|\psi_{1c}(t,z)|$, and the radiation profile  
obtained with the improved perturbation approach $|v_{12c}(t,z)|$ are also shown. 
Figure \ref{fig10} shows magnified versions of the graphs in Fig. \ref{fig9} 
for small $|\psi_{1}(t,z)|$ values. 
We observe that the tails of the pulse profile obtained by the simulation 
with Eq. (\ref{rad1}) for $\epsilon_{3}=0.02$ and $\beta=10$ are larger 
than the tails obtained with Eq. (\ref{rad1}) for $\epsilon_{3}=0.02$ and $\beta=20$. 
This finding coincides with the expected increase in the strength of the collision-induced 
effects with decreasing value of $\beta$, see, e.g., Refs. \cite{PNC2010,CPN2016}. 
Furthermore, we find good agreement between the numerical simulation's result and the perturbation 
theory predictions. However, the agreement is not as good as the one obtained for 
$\epsilon_{3}=0.02$ and $\beta=20$. In particular, the deviations of 
the numerically obtained pulse shape from the perturbation theory predictions are more 
significant for negative $t$ values and are smaller for positive $t$ values (see Fig. \ref{fig10}).     
For example, the deviations $||\psi_{1}^{(num)}(t,z)| - |\psi_{1b}(t,z)||$ and 
$||\psi_{1}^{(num)}(t,z)| - |\psi_{1c}(t,z)||$ at $z=z_{c}+5$ are 
{\it larger} than $1.0 \times 10^{-4}$ in the intervals $-30.19 < t < -16.08$ 
and $-26.47 < t < -20.18$, respectively. In contrast, at the same distance 
these deviations are {\it smaller} than $1.0 \times 10^{-4}$ for all $t$ 
such that $t>9.76$ and $t>9.63$, respectively.      
We also observe that the prediction of the simple perturbation approach 
for the pulse tails is in better agreement with the simulation's result compared with the improved version 
for positive $t$ values. The improved perturbation 
approach is in slightly better agreement with the numerical result for negative 
$t$ values. Additionally, similar to the situation for $\epsilon_{3}=0.02$ and $\beta=20$, 
the agreement between the two perturbative predictions and the numerical simulation's 
result for the pulse tails improves with increasing distance. For example, 
for $|t| > 6$, the deviations $||\psi_{1}^{(num)}(t,z)| - |\psi_{1b}(t,z)||$ and 
$||\psi_{1}^{(num)}(t,z)| - |\psi_{1c}(t,z)||$ are smaller than $3.81 \times 10^{-4}$ 
and $3.90 \times 10^{-4}$ at $z=z_{c}+2$, and are smaller than $1.20 \times 10^{-4}$ 
and $1.14 \times 10^{-4}$ at $z=z_{c}+10$.

\begin{figure}[ptb]
\begin{center}
\begin{tabular}{cc}
\epsfxsize=9.5cm  \epsffile{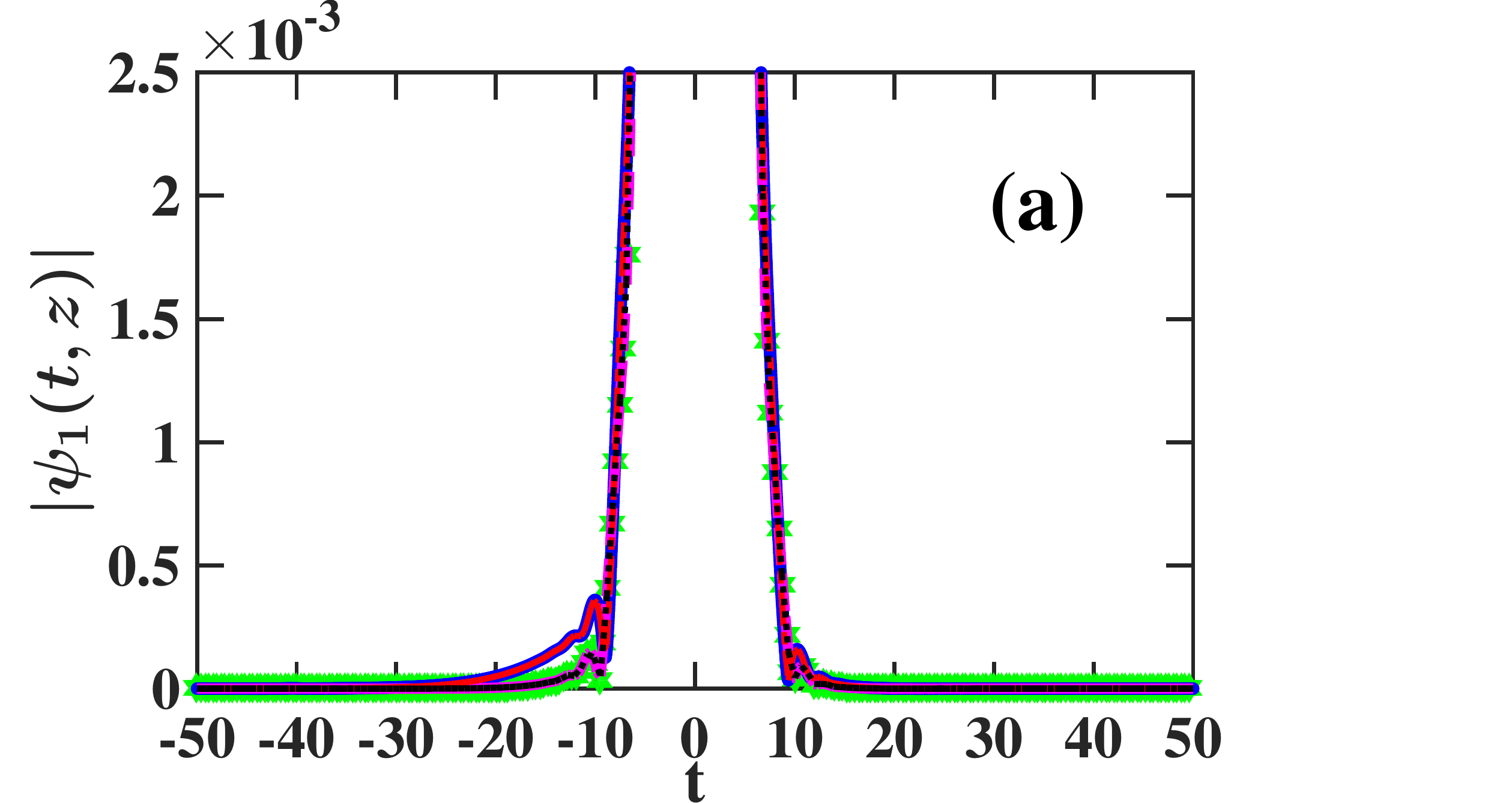} \\
\epsfxsize=9.5cm  \epsffile{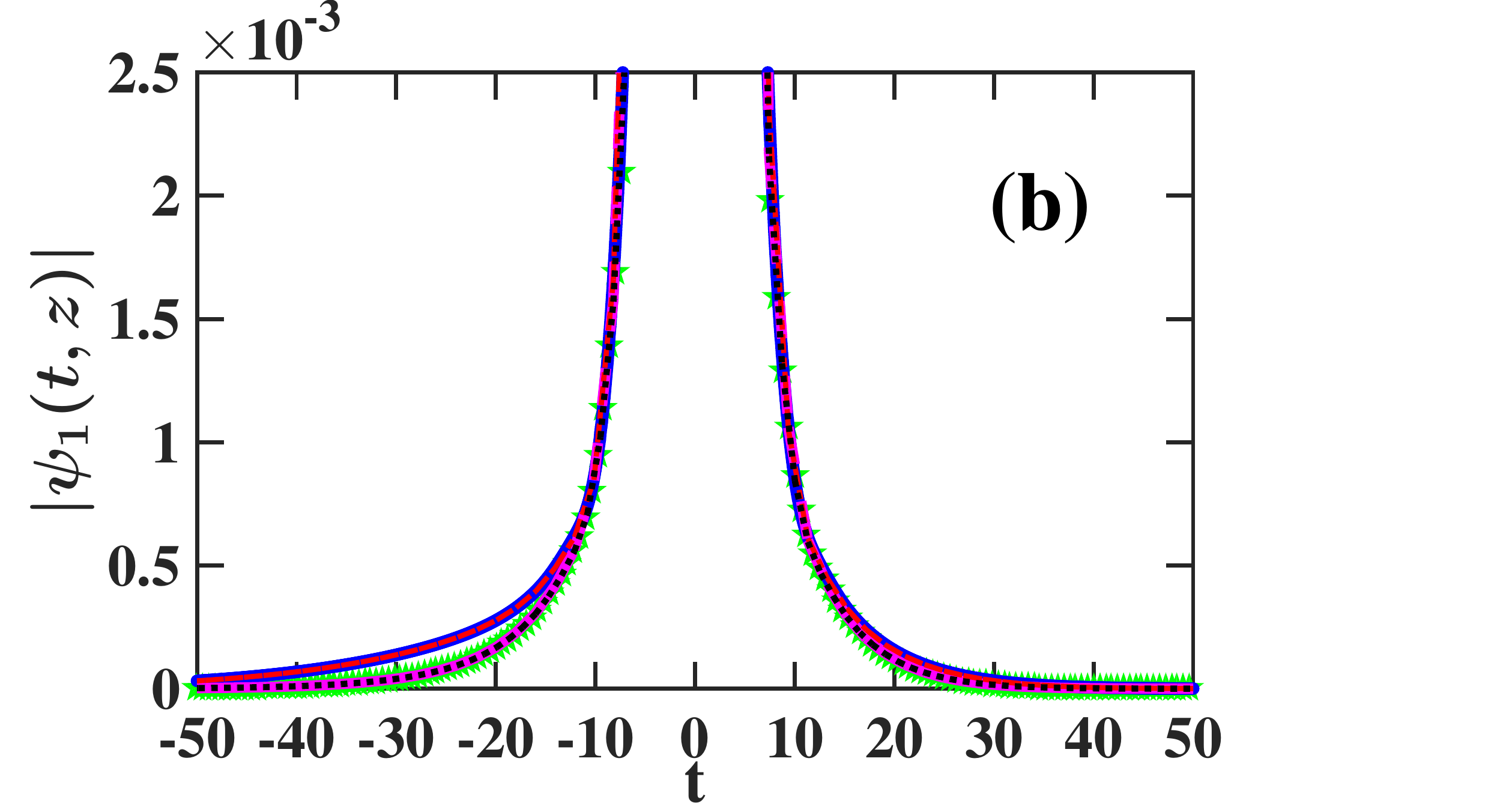} \\
\epsfxsize=9.5cm  \epsffile{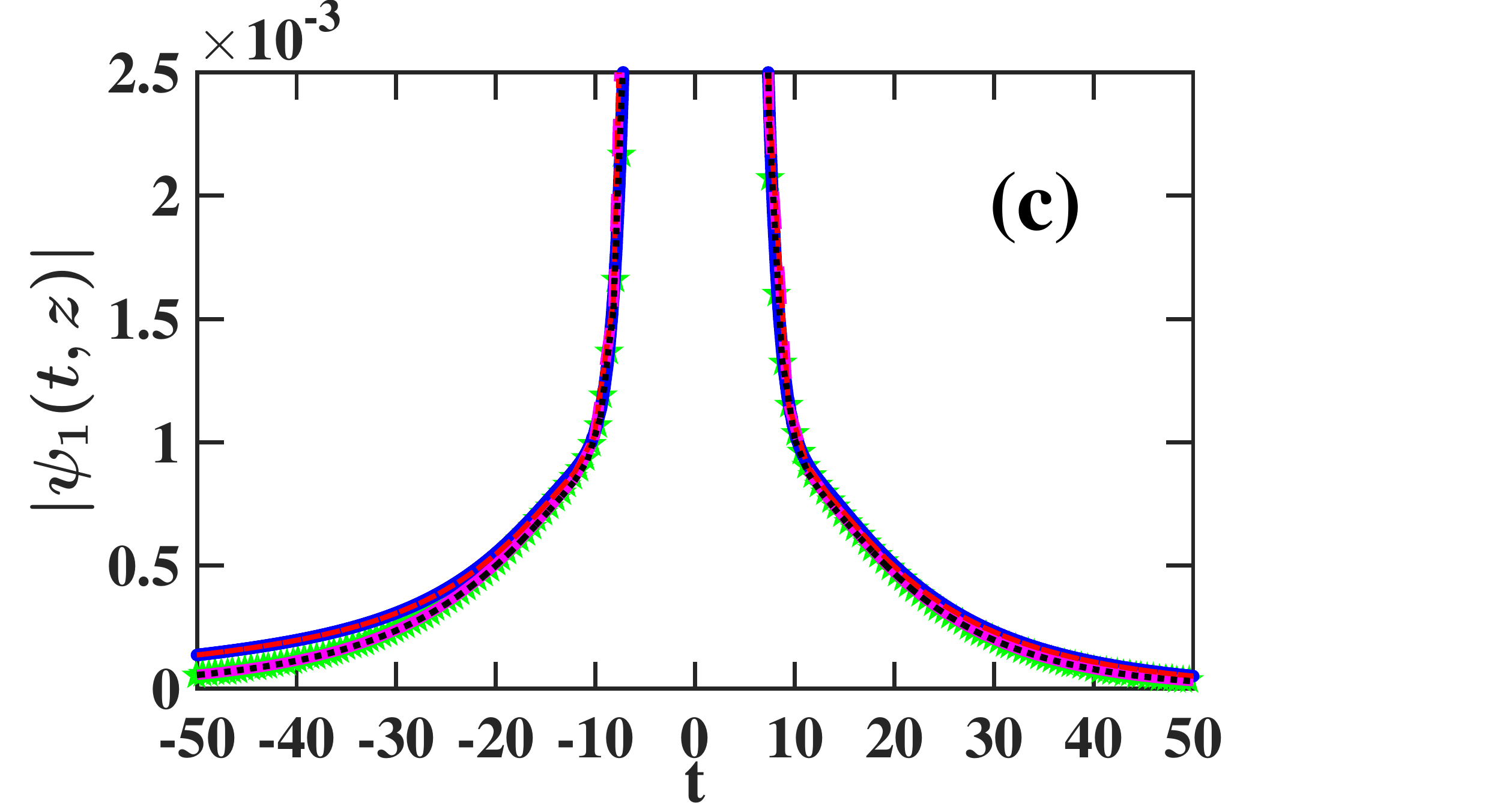} 
\end{tabular}
\end{center}
\caption{A comparison between the $t$ dependences of the pulse profile 
$|\psi_{1}(t,z)|$ obtained by the improved perturbation theory and the 
pulse profiles obtained by numerical solution of the four simplified NLS 
models (\ref{rad44}), (\ref{rad46}), (\ref{rad51}), and (\ref{rad53}). 
The parameter values are $\epsilon_{3}=0.02$ and $\beta=10$ 
and the distances are $z=z_{c}+2$ in (a), $z=z_{c}+5$ in (b), 
and $z=z_{c}+10$ in (c). The green stars represent the prediction of 
the improved perturbation approach. The solid-blue, dashed magenta, 
dashed-dotted red, and dotted black curves correspond to the pulse 
profiles obtained by numerical solution of Eqs. (\ref{rad44}), 
(\ref{rad46}), (\ref{rad51}), and (\ref{rad53}).} 
 \label{fig_add3}
\end{figure}

\begin{figure}[ptb]
\begin{center}
\begin{tabular}{cc}
\epsfxsize=9.5cm  \epsffile{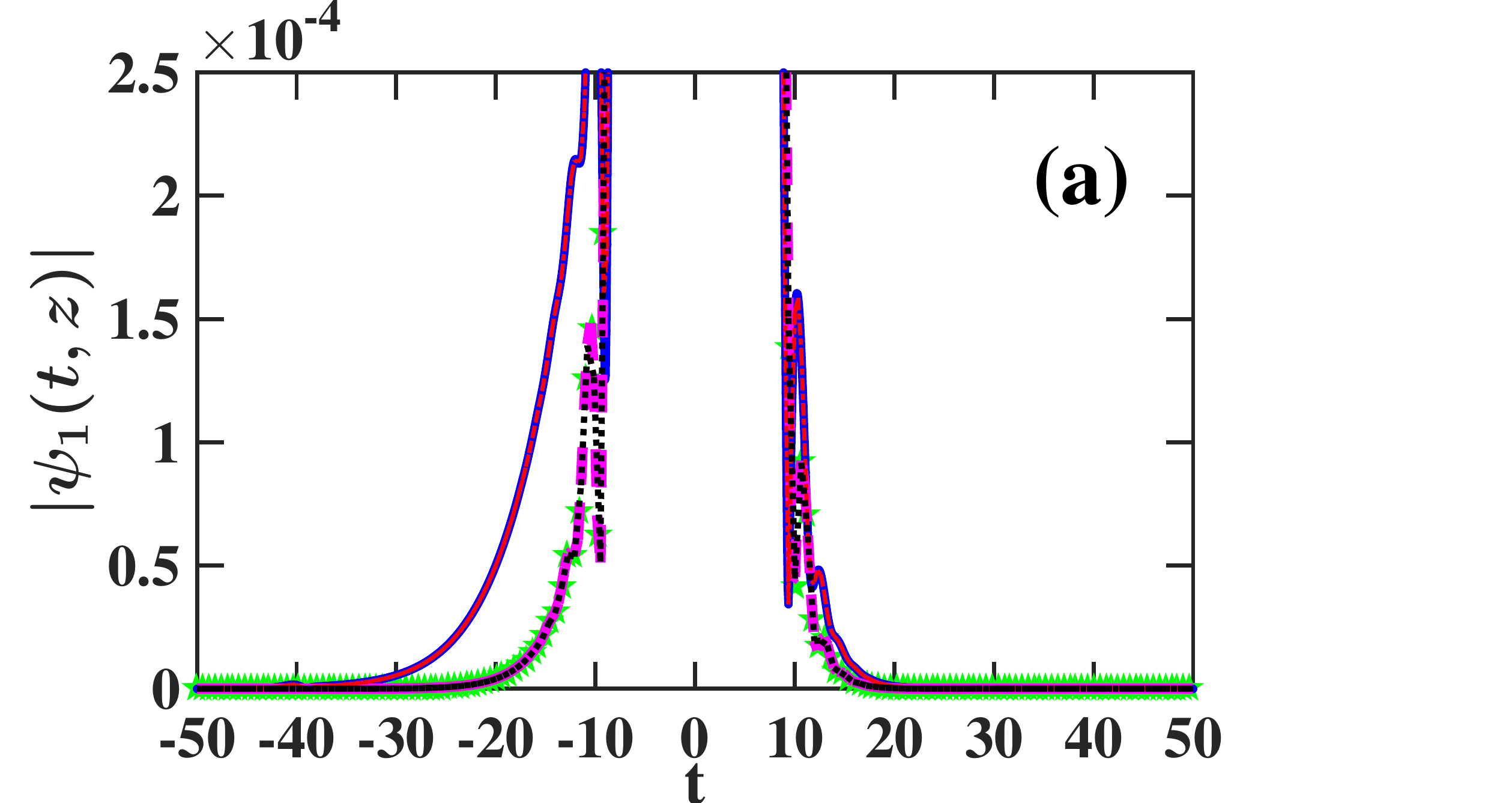} \\
\epsfxsize=9.5cm  \epsffile{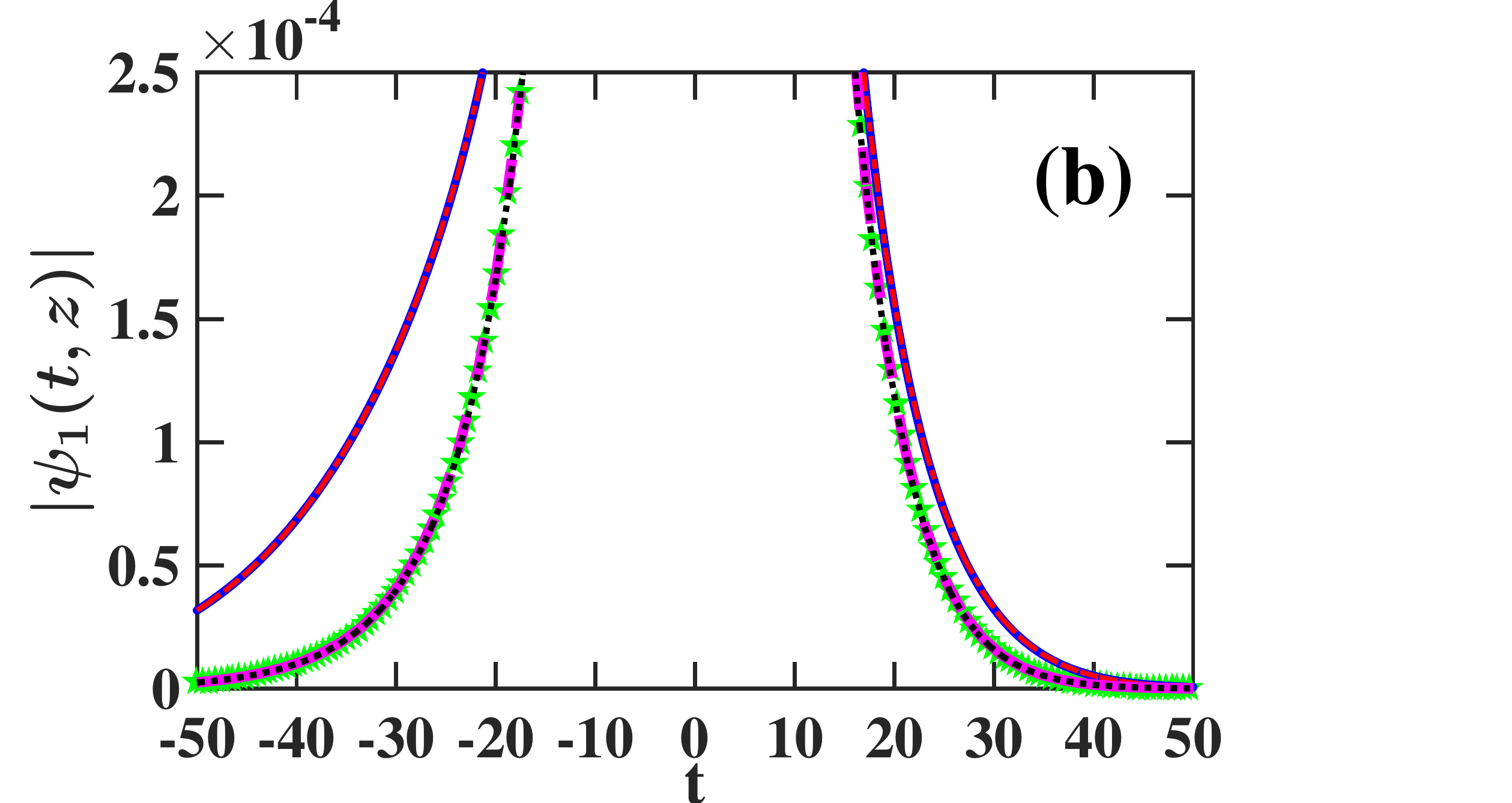} \\
\epsfxsize=9.5cm  \epsffile{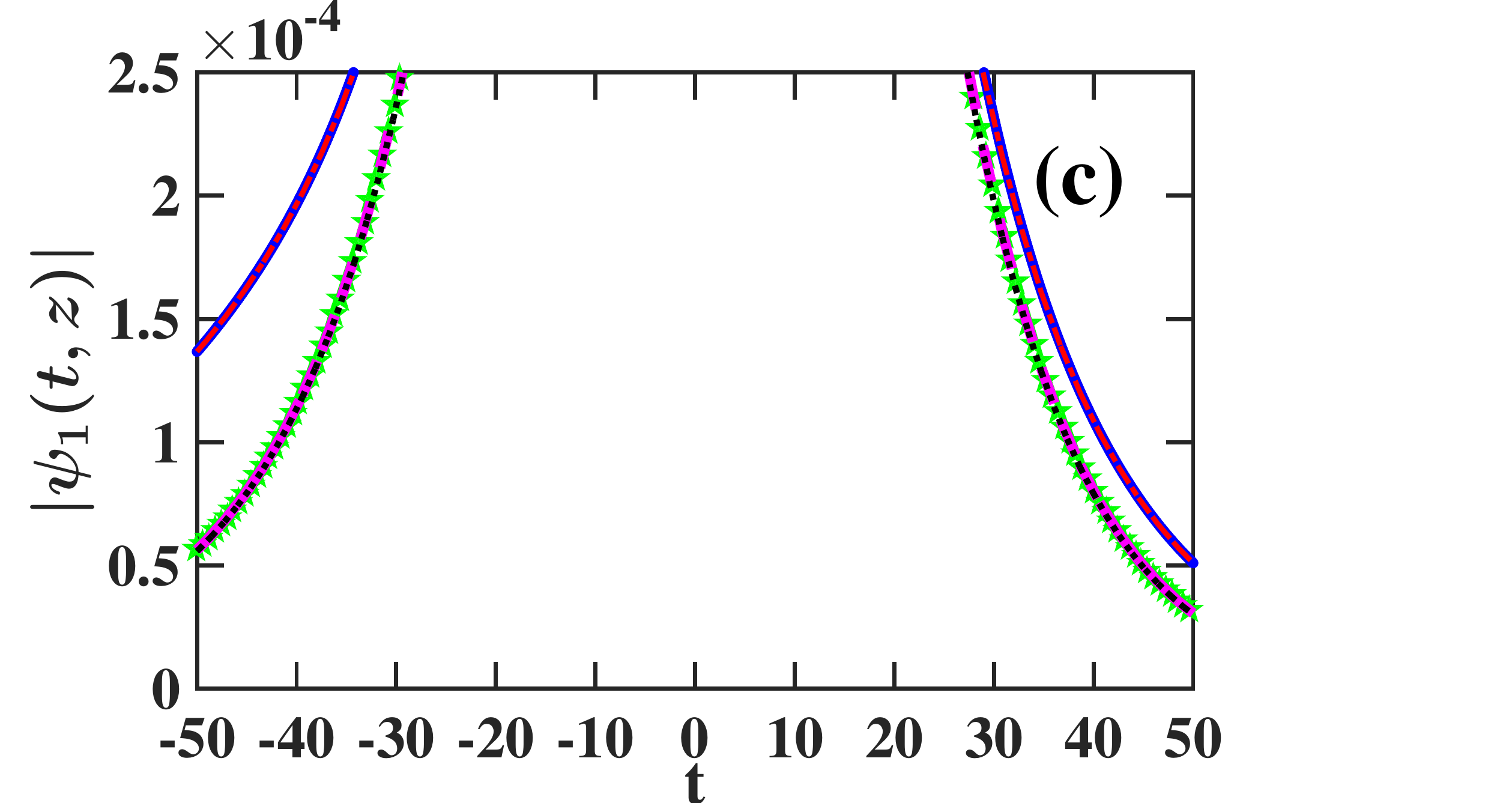} 
\end{tabular}
\end{center}
\caption{Magnified versions of the graphs in Fig. \ref{fig_add3} for small 
$|\psi_{1}(t,z)|$ values. The symbols are the same as in Fig. \ref{fig_add3}.}        
 \label{fig_add4}
\end{figure}

Further insight into the collision-induced radiation dynamics can be gained 
by comparing the predictions of the perturbation theory with results of 
numerical simulations with the four simplified NLS models (\ref{rad44}), 
(\ref{rad46}), (\ref{rad51}), and (\ref{rad53}). Figure \ref{fig_add3} shows the 
pulse profile $|\psi_{1}(t,z)|$ obtained by the improved perturbation approach 
and by numerical solution of Eqs. (\ref{rad44}), (\ref{rad46}), (\ref{rad51}), 
and (\ref{rad53}) at $z=z_{c}+2$, $z=z_{c}+5$, and $z=z_{c}+10$.   
Figure \ref{fig_add4} shows magnified versions of the graphs 
in Fig. \ref{fig_add3} for small $|\psi_{1}(t,z)|$ values. 
We observe very good agreement between the perturbation theory's prediction 
and the results of numerical simulations with the NLS models (\ref{rad46}) and 
(\ref{rad53}). There is also very good agreement between the results of the 
NLS models  (\ref{rad44}) and (\ref{rad51}). 
In contrast, the values of $|\psi_{1}(t,z)|$ at the pulse tails obtained with Eqs. (\ref{rad44}) and (\ref{rad51}) 
are noticeably larger than the $|\psi_{1}(t,z)|$ values at the pulse tails obtained with the perturbation theory 
and with the NLS models (\ref{rad46}) and (\ref{rad53}) (see Fig. \ref{fig_add4}). 
These deviations are more significant for  negative $t$ values, and they decrease 
with increasing propagation distance. 
As explained in subsection \ref{simu_1}, the NLS models 
(\ref{rad44}) and (\ref{rad51}) [and Eq. (\ref{rad1})] take into account the effects of Kerr-induced 
interpulse interaction on radiation dynamics, while the perturbation theory and 
Eqs. (\ref{rad46}) and (\ref{rad53}) neglect these effects. 
Therefore, based on the results shown in Figs. \ref{fig_add3}-\ref{fig_add4},   
we conclude that the effects of Kerr-induced interpulse interaction 
on radiation dynamics are significant for $\epsilon_{3}=0.02$ and $\beta=10$.

\begin{figure}[ptb]
\begin{center}
\begin{tabular}{cc}
\epsfxsize=9.5cm  \epsffile{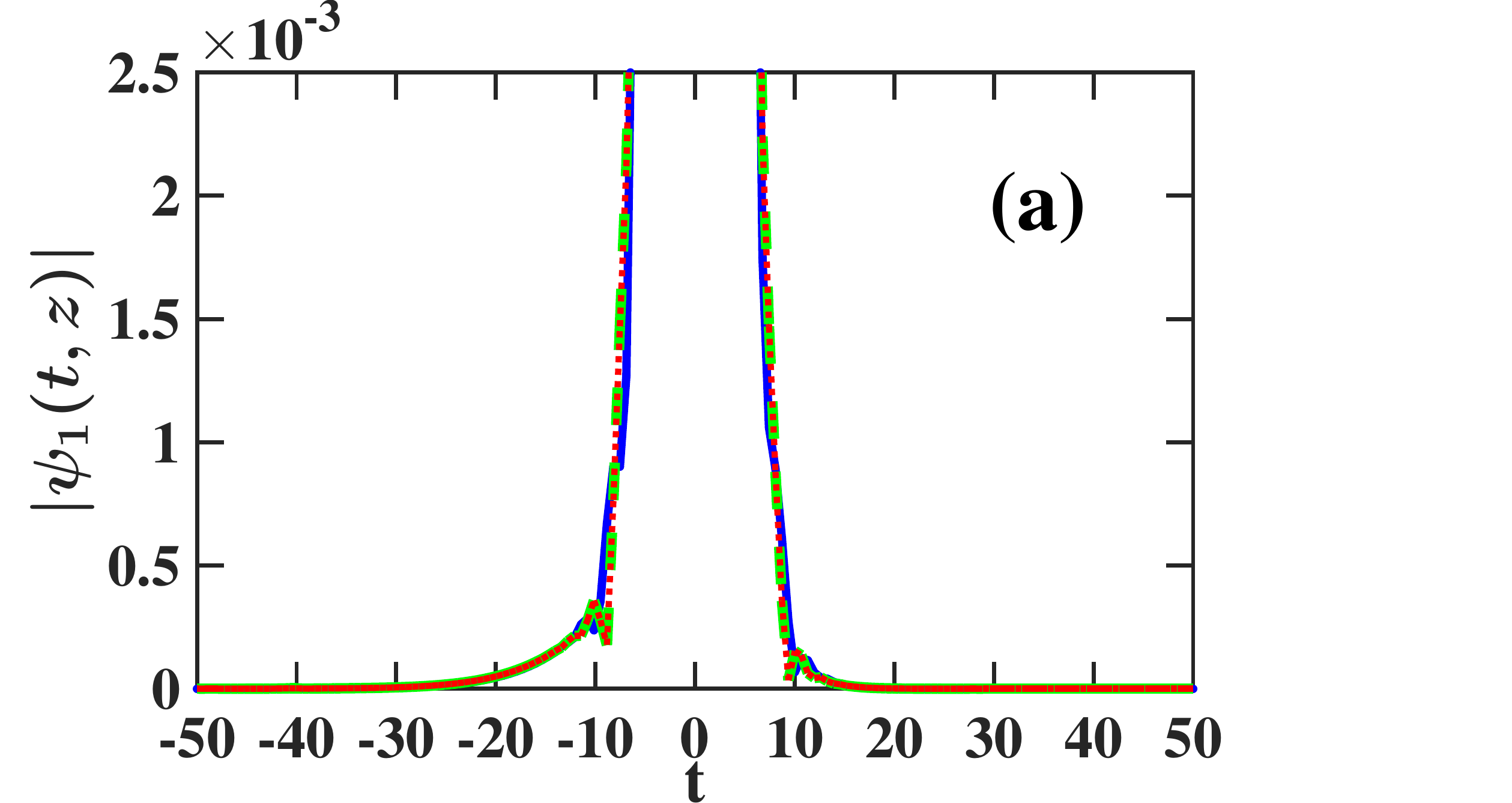} \\
\epsfxsize=9.5cm  \epsffile{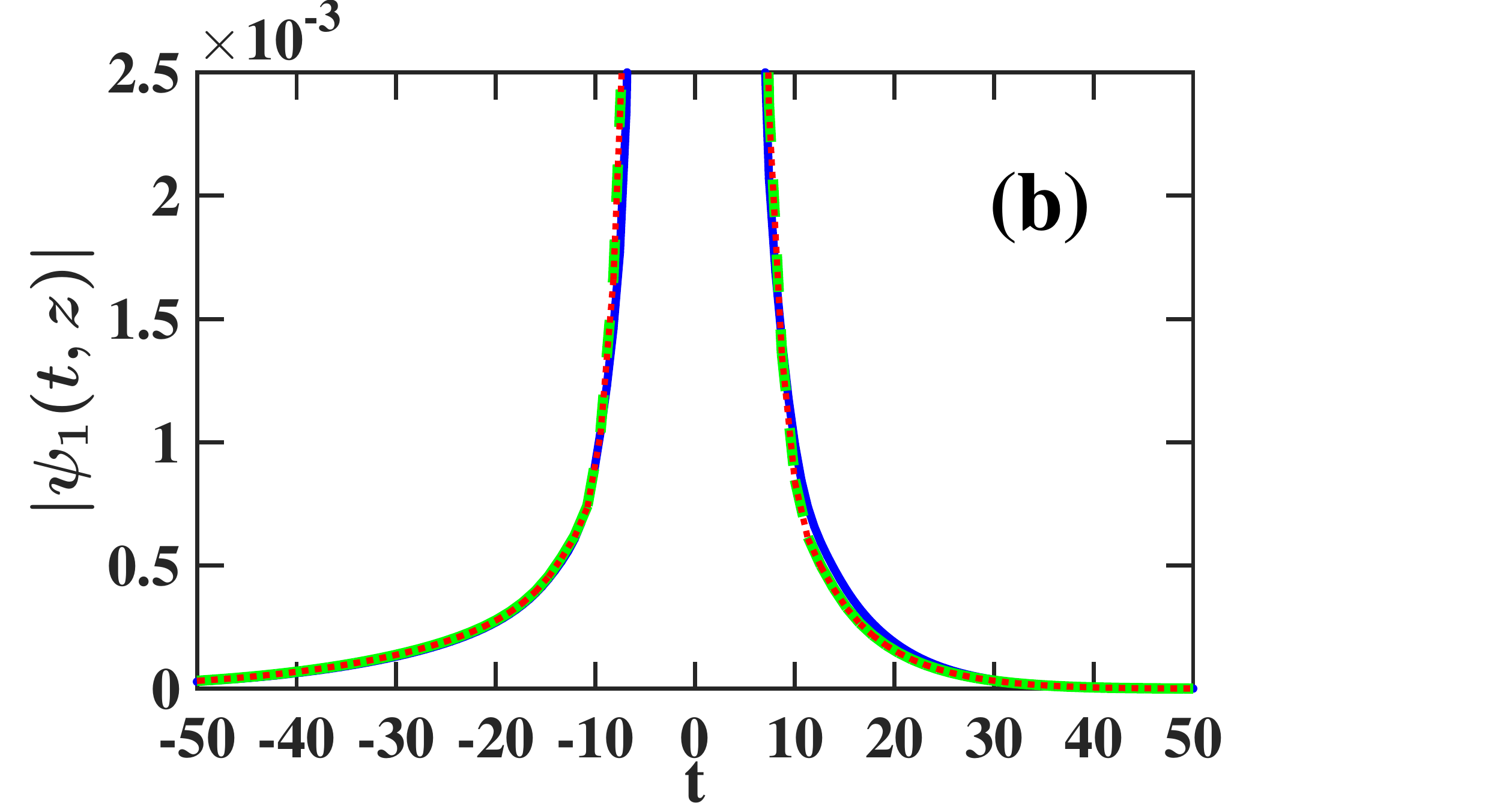} \\
\epsfxsize=9.5cm  \epsffile{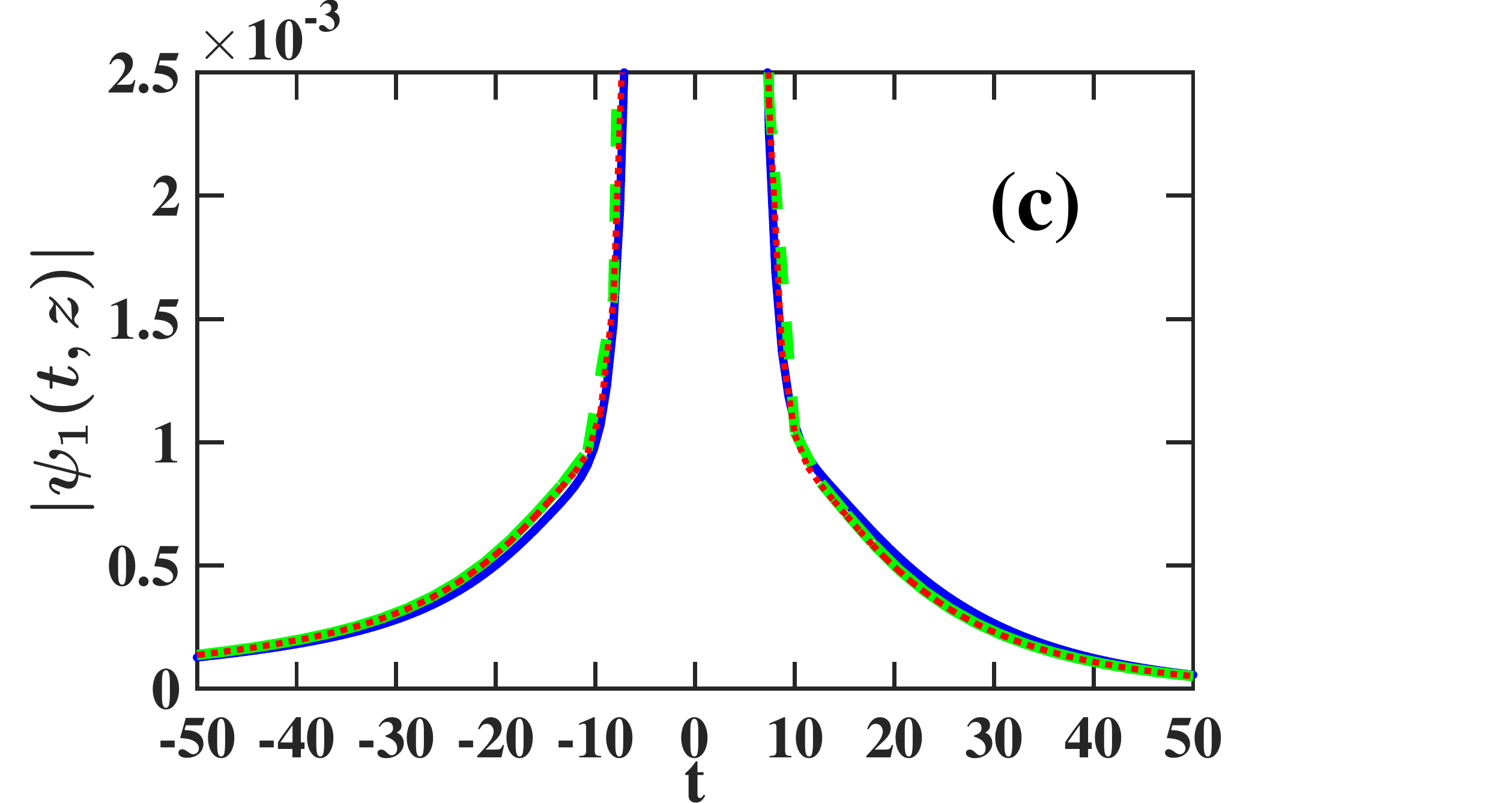} 
\end{tabular}
\end{center}
\caption{A comparison between the $t$ dependences of the pulse profile 
$|\psi_{1}(t,z)|$ obtained in numerical simulations with the three 
perturbed NLS  models (\ref{rad1}), (\ref{rad44}), and (\ref{rad51}).   
The physical parameter values are $\epsilon_{3}=0.02$ and $\beta=10$ 
and the distances are $z=z_{c}+2$ in (a), $z=z_{c}+5$ in (b), 
and $z=z_{c}+10$ in (c). The solid blue, dashed green,
and dotted red curves represent $|\psi_{1}(t,z)|$ 
obtained by numerical solution of Eqs. (\ref{rad1}), (\ref{rad44}), 
and (\ref{rad51}), respectively.}                   
 \label{fig11}
\end{figure}

\begin{figure}[ptb]
\begin{center}
\begin{tabular}{cc}
\epsfxsize=9.5cm  \epsffile{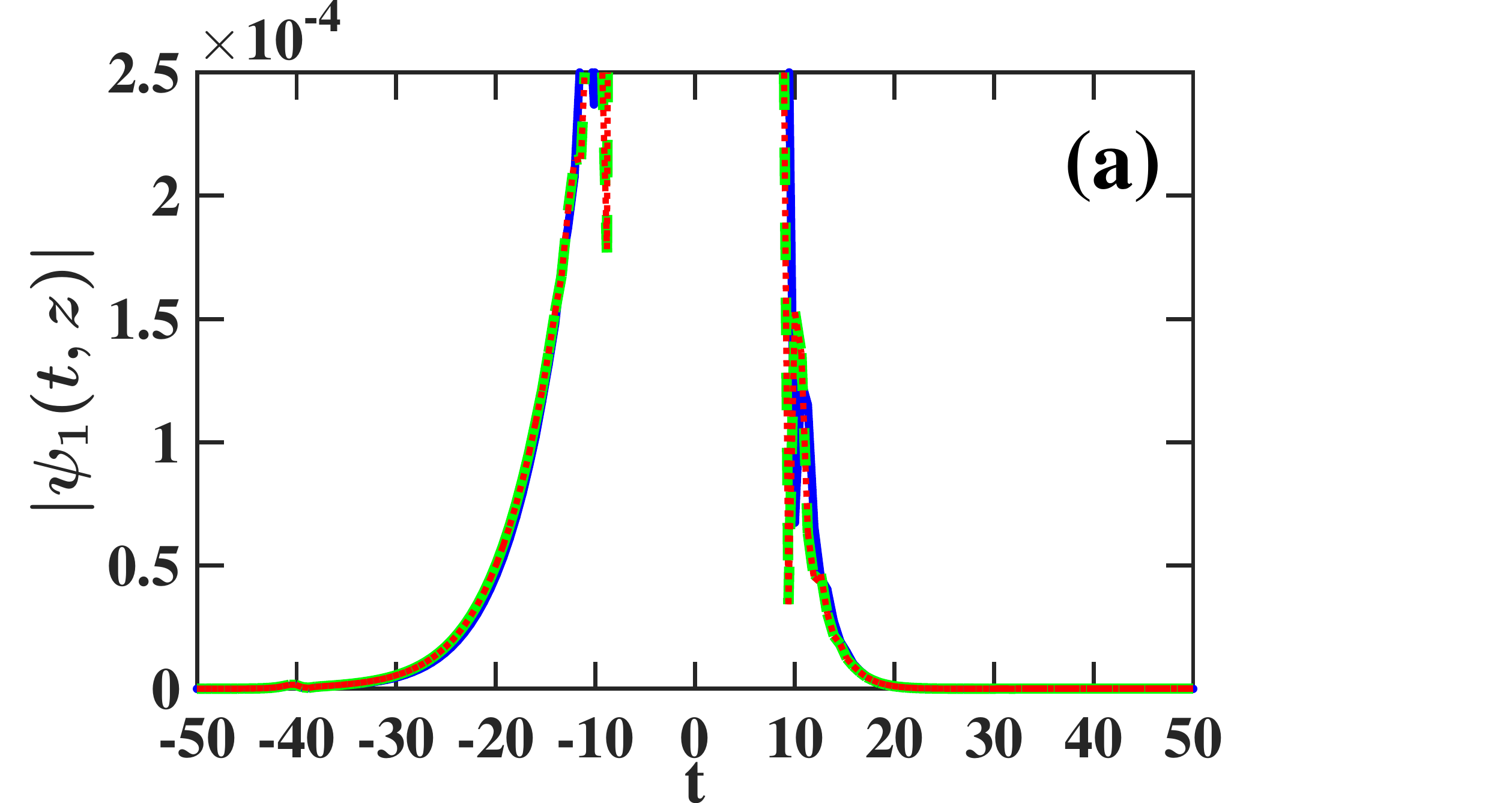} \\
\epsfxsize=9.5cm  \epsffile{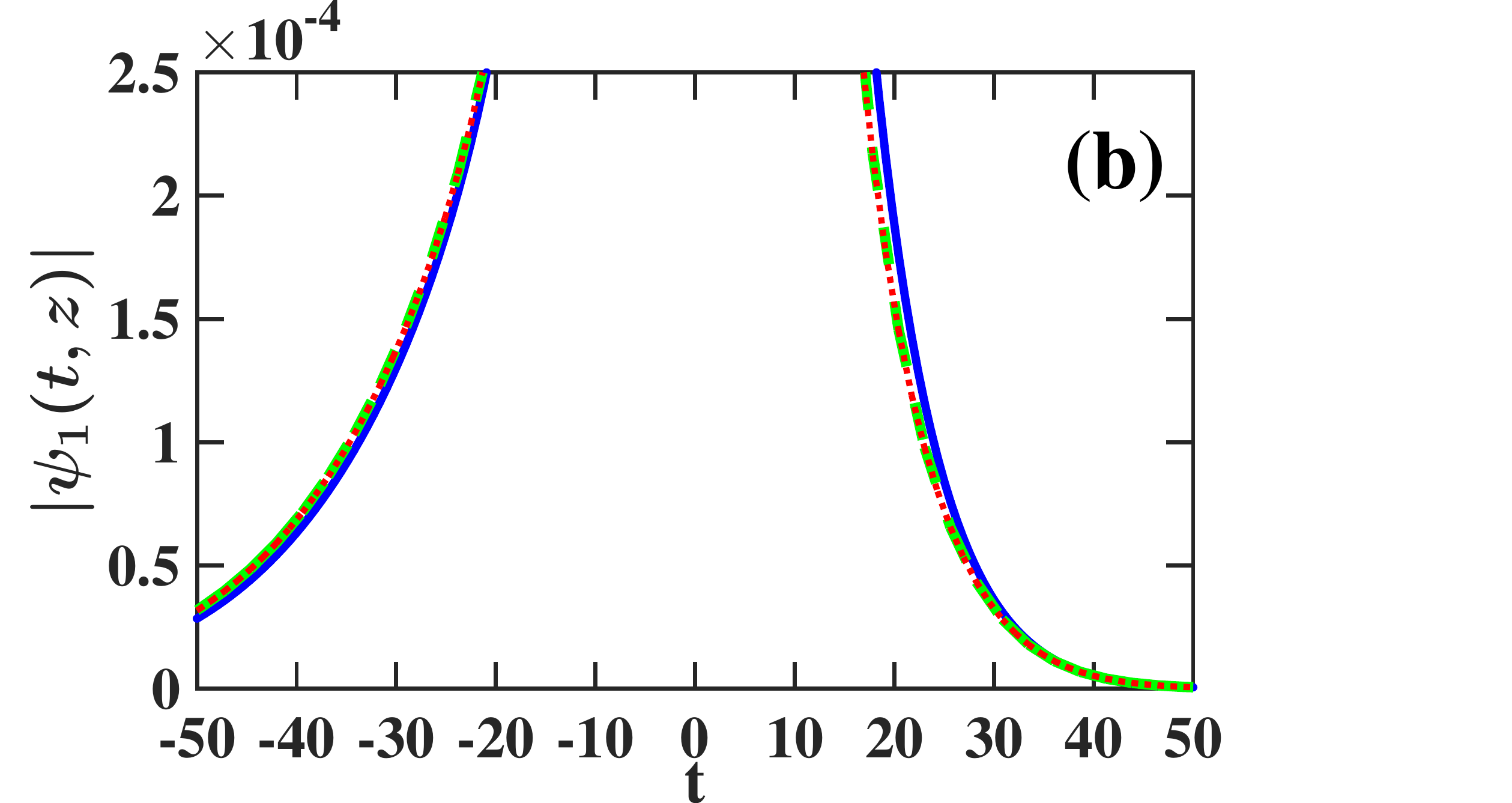} \\
\epsfxsize=9.5cm  \epsffile{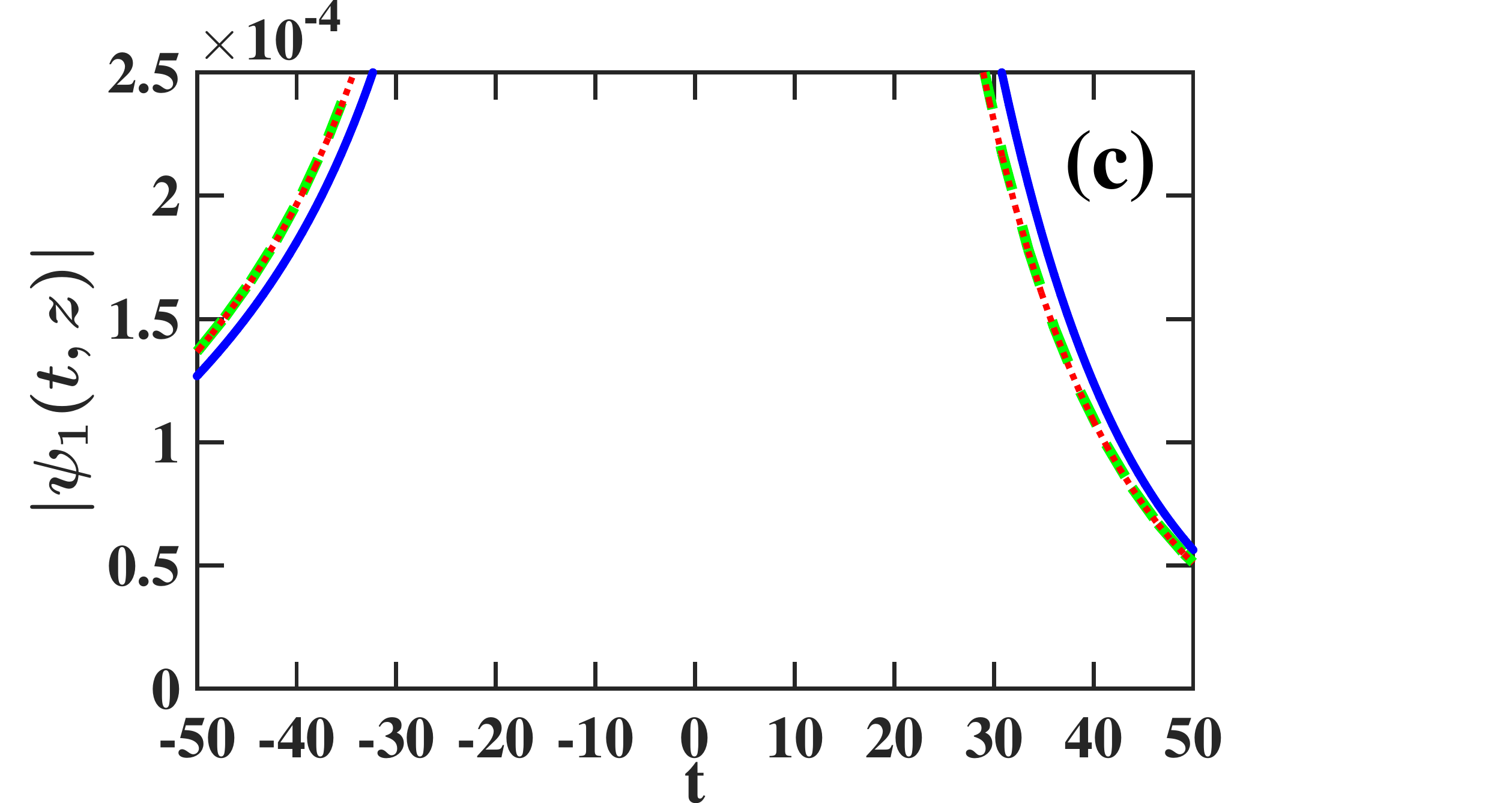} 
\end{tabular}
\end{center}
\caption{Magnified versions of the graphs in Fig. \ref{fig11} for small 
$|\psi_{1}(t,z)|$ values. The symbols are the same as in Fig. \ref{fig11}.}        
 \label{fig12}
\end{figure}

We complete the analysis of the collision-induced radiation dynamics 
for $\epsilon_{3}=0.02$ and $\beta=10$ by comparing 
the pulse profiles $|\psi_{1}^{(num)}(t,z)|$, 
$|\psi_{1}^{(num,s)}(t,z)|$, and $|\psi_{1}^{(num,p)}(t,z)|$ obtained in 
numerical simulations with the perturbed NLS models (\ref{rad1}), (\ref{rad44}), and (\ref{rad51}), respectively. 
Figure \ref{fig11} shows the comparison of these pulse profiles at 
$z=z_{c}+2$, $z=z_{c}+5$, and $z=z_{c}+10$.  
Figure \ref{fig12} shows magnified versions of the graphs in Fig. \ref{fig11} 
for small $|\psi_{1}(t,z)|$ values. We observe that the pulse profiles 
obtained with the simplified NLS models (\ref{rad44}) and (\ref{rad51}) are 
in good agreement with the result obtained in simulations with the 
full coupled-NLS model (\ref{rad1}), although the agreement is not as good as 
the one obtained for $\epsilon_{3}=0.02$ and $\beta=20$   
(compare Figs. \ref{fig11} and \ref{fig12} with  Figs. \ref{fig7} and \ref{fig8}).    
In particular, the deviations $||\psi_{1}^{(num)}(t,z)| - |\psi_{1}^{(num,s)}(t,z)||$ 
and $||\psi_{1}^{(num)}(t,z)| - |\psi_{1}^{(num,p)}(t,z)||$ are smaller 
than $1.36 \times 10^{-3}$ and $1.37 \times 10^{-3}$ for all $t$ values at $z=z_{c}+2$, 
and are smaller than $3.79 \times 10^{-3}$ and 
$3.74 \times 10^{-3}$ for all $t$ values at $z=z_{c}+10$.
Furthermore, the deviations for $|t|>6$ are smaller than 
$8.95 \times 10^{-4}$ and $8.96 \times 10^{-4}$ at $z=z_{c}+2$, 
and are smaller than $3.67 \times 10^{-4}$ and $3.50 \times 10^{-4}$ at $z=z_{c}+10$.
Similar to the situation for $\epsilon_{3}=0.02$ and $\beta=20$, 
the good agreement between the pulse profiles obtained with 
Eqs. (\ref{rad1}) and (\ref{rad44}) shows that distortion of soliton 2 
does not play an important role in the collision-induced radiation dynamics of soliton 1.    
Furthermore, based on the comparison in Figs. \ref{fig11} and \ref{fig12}, 
we conclude that for $\epsilon_{3}=0.02$ and $\beta=10$, 
we can use the two simplified models (\ref{rad44}) and (\ref{rad51}) to describe 
the collision-induced radiation dynamics with good accuracy.


The differences between the perturbation theory predictions for the pulse 
tails and the pulse tails obtained in the simulations with Eq. (\ref{rad1}) 
can be explained with the help of the results shown in Figs. \ref{fig_add3} - \ref{fig12}. 
In particular, in Figs. \ref{fig_add3} - \ref{fig_add4}, we found similar differences 
between the pulse tails obtained with Eqs. (\ref{rad44}) and (\ref{rad51}) and 
the perturbation theory predictions. In contrast, the latter predictions were in 
good agreement with results of simulations with Eqs. (\ref{rad46}) and (\ref{rad53}). 
Furthermore, in Figs. \ref{fig11} - \ref{fig12}, we found good agreement between 
the results of the simulations with Eqs. (\ref{rad1}), (\ref{rad44}), and (\ref{rad51}). 
Recall that the perturbed NLS models (\ref{rad1}), (\ref{rad44}), and (\ref{rad51}) 
take into account the effects of interpulse interaction due to Kerr nonlinearity 
on radiation dynamics, while the perturbation theory and the perturbed NLS models           
(\ref{rad46}) and (\ref{rad53}) neglect these effects. 
Therefore, based on the comparisons in Figs. \ref{fig_add3} - \ref{fig12}, 
we conclude that the effects of Kerr-induced interpulse interaction 
on radiation dynamics are the main cause for the observed differences 
between the pulse tails obtained with the full coupled-NLS model (\ref{rad1}) 
and the pulse tails predicted by the perturbation theory for $\epsilon_{3}=0.02$ and $\beta=10$.     
                                                                                                                                                
\section{Conclusions}
\label{conclusions}

We studied the dynamics of emission of radiation in fast collisions 
between solitons of the NLS equation in the presence of weak cubic loss. 
We calculated the radiation dynamics and the pulse profile by a 
perturbation technique with two small parameters: the cubic loss coefficient 
$\epsilon_{3}$ and the reciprocal of the group velocity difference $1/\beta$.
We then compared the predictions of the perturbation theory with results of 
numerical simulations with the full propagation equation [the full coupled-NLS 
model (\ref{rad1})]. The comparison showed very good agreement between 
the perturbation theory predictions and the simulations with Eq. (\ref{rad1}) 
for large values of $\beta$ ($\beta=20$). 
For intermediate values of $\beta$ ($\beta=10$), we obtained good agreement 
between the predictions of the perturbation theory and the results of Eq. (\ref{rad1}), 
but the agreement was not as good as the one obtained for $\beta=20$. 
Therefore, our study provides the first demonstration that the perturbation technique 
developed in Refs. \cite{PCG2003,PCG2004,SP2004} for studying radiation dynamics 
in fast soliton collisions in the presence of conservative perturbations can in fact be 
employed for soliton collisions in the presence of dissipative perturbations.

To gain further insight into the reasons for the differences between the perturbation 
theory predictions and the results of simulations with Eq. (\ref{rad1}), we carried out 
numerical simulations with four simplified propagation models: two single-NLS models 
and two simpler coupled-NLS models. The first single-NLS model [Eq. (\ref{rad44})] 
and the first simplified coupled-NLS model [Eq. (\ref{rad51})] take into account both 
collision-induced emission of radiation and collision-induced position shift due to Kerr nonlinearity. 
The second single-NLS model [Eq. (\ref{rad46})] and the second simplified coupled-NLS model 
[Eq. (\ref{rad53})] take into account only the latter effect. 
For large $\beta$ values ($\beta=20$), the pulse profiles obtained with the four simplified models 
were in very good agreement with the predictions of the perturbation theory 
and with the pulse profile obtained with the full coupled-NLS model (\ref{rad1}).                               
In contrast, for intermediate values of  $\beta$ ($\beta=10$), 
only the two simplified models that take into account the effects of Kerr nonlinearity 
on the collision-induced radiation dynamics [Eqs. (\ref{rad44}) and (\ref{rad51})] 
were in very good agreement with the result of the full coupled-NLS model (\ref{rad1}). 
It follows that the effects of Kerr nonlinearity on the collision-induced radiation dynamics 
are unimportant for $\beta=20$, but are significant for $\beta=10$, 
i.e., the strength of these effects increases with decreasing value of $\beta$. 
Furthermore, it follows that the main reason for the observed differences between  
the perturbation theory calculations and the simulations with Eq. (\ref{rad1}) for $\beta=10$ 
was the additional emission of radiation due to the effects of Kerr nonlinearity on the collision. 
We also point out that the simulations with the four simplified NLS models also 
provided a clear demonstration of the significance of the interplay between 
Kerr nonlinearity and dissipative processes in dynamics of radiation emitted in
collisions between NLS solitons.

As explained in sections \ref{theory_4} and \ref{simu_1}, the simplified NLS models (\ref{rad34}), 
(\ref{rad44}), and (\ref{rad46}) provide a description of the collision-induced 
dynamics in terms of a single perturbed NLS equation. 
In previous studies, we used a generalized form of such perturbed NLS 
models to describe soliton propagation in multisequence nonlinear optical 
waveguide systems, where each soliton undergoes many collisions with 
solitons from all other pulse sequences \cite{CP2005,CP2008,PC2012a}. 
The generalized model used in Refs. \cite{CP2005,CP2008,PC2012a} had 
the form of a perturbed stochastic NLS equation with a linear gain-loss 
term and a stochastic (distance-dependent) linear gain-loss coefficient.        
The linear gain-loss term described the energy exchange of a given soliton 
in many collisions with solitons from all other pulse sequences. 
The distance-dependent stochastic coefficient of this term described the 
bit-pattern randomness of the pulse sequences \cite{CP2005,CP2008,PC2012a}. 
It was assumed in these studies that the stochastic linear gain-loss term accurately 
describes collision-induced radiation dynamics. The results of the current paper 
provide the first strong evidence in favor of the validity of this assumption. 
More specifically, the good agreement between the pulse profiles obtained 
in numerical simulations with the single-NLS model (\ref{rad46}) and with 
the full coupled-NLS model (\ref{rad1}) shows that the stochastic 
linear gain-loss term of the NLS models used in Refs. \cite{CP2005,CP2008,PC2012a} 
correctly captures the dynamics of the radiation emitted in soliton collisions.

\section*{Acknowledgments}
D.C. is grateful to the Mathematics Department of NJCU 
for providing technological support for the computations.

\appendix
\section{The adiabatic perturbation theory for the fundamental NLS soliton} 
\label{appendA}

We provide a summary of the adiabatic perturbation theory 
for the fundamental NLS soliton, which was developed by Kaup \cite{Kaup90,Kaup91,Kaup76}. 
The theory was used for studying soliton propagation in the presence of perturbations 
in a variety of nonlinear optical waveguide systems (see, e.g., 
Refs. \cite{Hasegawa95,CP2005,CCDG2003} and references therein).

We illustrate the approach by considering the following perturbed NLS equation 
\begin{eqnarray}
i\partial_z\psi+\partial_t^2\psi+2|\psi|^2\psi = \epsilon h(t,z),
\label{perturbation1}
\end{eqnarray}
where $0 < |\epsilon| \ll 1$. 
We look for a solution of Eq. (\ref{perturbation1}) in the form: 
\begin{eqnarray}
\!\!\!\!\!\!\!\!\! 
\psi(t,z)=\psi_{s}(t,z)+\psi_{rad}(t,z)=\eta(z)\frac{\exp[i\chi(t,z)]}{\cosh(x)}
\nonumber \\
+v(t,z)\exp[i\chi(t,z)],
\label{perturbation2}
\end{eqnarray}
where $x=\eta(z)\left[t-y(z)\right]$, $\chi(t,z)=\alpha(z)-\beta(z)\left[t-y(z)\right]$, 
$y(z)=y(0)-2\int_{0}^{z} dz' \beta(z')$, 
and $\alpha(z)=\alpha(0)+\int_{0}^{z} dz' \left[\eta^{2}(z')+\beta^{2}(z')\right]$. 
The first term on the right hand side of Eq. (\ref{perturbation2}) 
has the form of a fundamental soliton solution of the NLS equation 
with slow varying parameters, while the second term, 
which is of $O(\epsilon)$, is the radiation part.   
We now substitute Eq. (\ref{perturbation2}) into Eq. (\ref{perturbation1}) 
and keep terms up to $O(\epsilon)$. 
The equation obtained by this substitution and the complex conjugate 
of this equation can be written in the following vector form:     
\begin{eqnarray} &&
\frac{i}{\cosh(x)} {1 \choose -1}\eta \left(\frac{d\alpha}{dz}+\beta\frac{dy}{dz}
-\eta^{2}+\beta^{2}\right)
\nonumber \\&&
+\frac{\tanh(x)}{\cosh(x)} {1 \choose 1}\eta^{2} \left(\frac{dy}{dz}+2\beta\right)
-\frac{ix}{\cosh(x)} {1 \choose -1} \frac{d\beta}{dz}
\nonumber \\&&
-\frac{\left[x\tanh(x)-1\right]}{\cosh(x)} {1 \choose 1}\frac{d\eta}{dz} 
+ \partial _{z} {v  \choose v^{\ast }} -i\eta^{2}{\cal L} {v  \choose v^{\ast }} 
\nonumber \\&&
-2\beta \partial _{t} {v  \choose v^{\ast }}= 
-i \epsilon{h(t,z) e^{-i\chi} \choose - h^{\ast}(t,z) e^{i\chi}} .
\label{perturbation3} 
\end{eqnarray}
The linear operator ${\cal L}$ in Eq. (\ref{perturbation3}) is defined by: 
\begin{eqnarray}
{\cal L}=\left(\partial _{x}^{2}-1\right) \pmb{\sigma_{3}}
+\frac{2}{\cosh^{2}(x)} \left(2\pmb{\sigma_{3}}+i\pmb{\sigma_{2}}\right) ,
\label{perturbation4} 
\end{eqnarray}
where $\pmb{\sigma_{j}}$ with $1 \le j \le 3$ are the Pauli spin matrices.

The complete set of orthogonal eigenfunctions of ${\cal L}$ was found in Refs. \cite{Kaup90,Kaup91,Kaup76}. 
It includes four localized eigenfunctions, which appear in the first four terms on the 
left hand side of Eq. (\ref{perturbation3}):  
\begin{eqnarray} && 
\!\!\!\!\!\!\!\!\!\!\!\!\!\!
f_{0}(x)=\frac{1}{\cosh(x)} {1 \choose -1}, \;
f_{1}(x)=\frac{\tanh(x)}{\cosh(x)} {1 \choose 1},  
\nonumber \\&&
\!\!\!\!\!\!\!\!\!\!\!\!\!\!
f_{2}(x)=\frac{x}{\cosh(x)} {1 \choose -1}, \;
f_{3}(x)=\frac{x\tanh(x)-1}{\cosh(x)}{1 \choose 1} .
\label{perturbation5} 
\end{eqnarray}
The eigenfunctions $f_{0}(x)$ and $f_{1}(x)$ have a zero eigenvalue, 
while $f_{2}(x)$ and $f_{3}(x)$ satisfy ${\cal L}f_{2}=-2f_{1}$ and ${\cal L}f_{3}=-2f_{0}$ \cite{Kaup90,Kaup91,Kaup76}. The left localized eigenfunctions of ${\cal L}$, 
which are given by $f_{m}^{T}\pmb{\sigma_{3}}$ for $0 \le m \le 3$, 
satisfy the following relations \cite{Kaup90,Kaup91,Kaup76}:  
\begin{eqnarray} && 
\int\limits_{-\infty }^{+\infty } dx f_{2}^{T}(x) \pmb{\sigma_{3}}f_{1}(x)=2, 
\label{perturbation6a} \\ &&
\int\limits_{-\infty }^{+\infty } dx f_{0}^{T}(x) \pmb{\sigma_{3}}f_{3}(x)=-2.
\label{perturbation6b} 
\end{eqnarray}
We obtain the dynamic equations for the four soliton parameters by projecting 
both sides of Eq. (\ref{perturbation3}) on the four left localized eigenfunctions of ${\cal L}$.

The set of eigenfunctions of ${\cal L}$ also contains an infinite set of nonlocalized 
eigenfunctions $f_{s}(x)$ and $\bar{f}_{s}(x)$ with eigenvalues $s^{2}+1$ and 
$-(s^{2}+1)$, respectively, where $-\infty < s < \infty$, and 
$\bar{f}_{s}(x) \equiv \pmb{\sigma}_{1}\,f_{s}^{\ast}(x)$.  
The eigenfunctions $f_{s}(x)$ and $\bar{f}_{s}(x)$ 
are given by \cite{Kaup90,Kaup91,Kaup76}:    
\begin{eqnarray} &&
f_{s}(x)=\exp (isx)\left [ 1-\frac{2is\exp (-x)}
{(s+i)^{2}\cosh (x)}\right ] {0\choose 1}
\nonumber \\ &&
+\frac{\exp (isx)}{(s+i)^{2}\cosh ^{2}(x)}{1  \choose 1} ,
\label{perturbation7} 
\end{eqnarray}
and 
\begin{eqnarray} &&
\bar{f}_{s}(x)=\exp (-isx)\left [ 1+\frac{2is\exp (-x)}
{(s-i)^{2}\cosh (x)}\right ] {1\choose 0}
\nonumber \\ &&
+\frac{\exp (-isx)}{(s-i)^{2}\cosh ^{2}(x)}{1  \choose 1} .
\label{perturbation8} 
\end{eqnarray}
The corresponding left eigenfunctions of ${\cal L}$ are given by 
$f_{s}^{+}(x)\pmb{\sigma}_{3}$ and $\bar{f}_{s}^{+}(x)\pmb{\sigma}_{3}$.  
These eigenfunctions satisfy the relations \cite{Kaup90,Kaup91,Kaup76} 
\begin{eqnarray} && 
\int\limits_{-\infty }^{+\infty } dx 
f_{s}^{+}(x)\pmb{\sigma}_{3}f_{s'}(x)=-2\pi \delta (s-s') ,
\label{perturbation9a} \\ &&
\int\limits_{-\infty }^{+\infty } dx 
\bar f_{s}^{+}(x)\pmb{\sigma}_{3}\bar{f}_{s'}(x)=2\pi \delta (s-s') . 
\label{perturbation9b} 
\end{eqnarray}
To obtain the dynamics of the emitted radiation, we expand $v(t,z)$ 
and $\epsilon h(t,z)$ in the nonlocalized eigenfunctions of ${\cal L}$: 
\begin{eqnarray} &&
{v(t,z)\choose v^{\ast }(t,z)}=
\int_{-\infty}^{+\infty} \frac{ds}{2\pi }
\left[a_{s}(z) f_{s}(x)+a_{s}^{\ast}(z) \bar{f}_{s}(x)\right] ,
\label{perturbation10}
\end{eqnarray}  
and
\begin{eqnarray} &&
-i \epsilon{h(t,z) e^{-i\chi} \choose - h^{\ast}(t,z) e^{i\chi}} = 
\int_{-\infty}^{+\infty} \frac{ds}{2\pi }
\left[b_{s}(z) f_{s}(x)+b_{s}^{\ast}(z) \bar{f}_{s}(x)\right] .  
\label{perturbation11}  
\end{eqnarray}  
We then substitute the expansions (\ref{perturbation10}) and (\ref{perturbation11}) 
into Eq. (\ref{perturbation3}) and project both sides of the resultant equation 
on the left nonlocalized eigenfunctions of ${\cal L}$, $f_{s}^{+}(x)\pmb{\sigma}_{3}$. 
This calculation yields the evolution equation for the expansion coefficients $a_{s}(z)$:  
\begin{equation}
\frac{d a_{s}(z)}{dz} - i\eta^{2}(s^{2}+1)a_{s}(z)=b_{s}(z).
\label{perturbation12}
\end{equation}

\section{The form of the four simplified NLS models of section \ref{simu}} 
\label{appendB}
In this appendix, we explain the form of the four simplified NLS models 
(\ref{rad44}), (\ref{rad46}), (\ref{rad51}), and (\ref{rad53}) that are 
used in section \ref{simu} to gain further insight into 
the collision-induced radiation dynamics. 
More specifically, the simplified models are used for the following two reasons. 
(a) To enable accurate identification of the most important corrections to the 
perturbative clacluations of radiation dynamics in section \ref{theory}. 
(b) To help determine the significance of radiation-induced distortion of 
soliton 2 for radiation dynamics of soliton 1. It turns out (from numerical 
simulations with the simplified models) that the most important correction 
to the perturbative calculations of section \ref{theory} is due to the effects 
of interpulse interaction induced by Kerr nonlinearity. Therefore, the form 
of the simplified NLS models that we use must help identify the role 
of interpulse interaction due to Kerr nonlinearity in radiation dynamics. 
In addition, in order to enable accurate comparison with the results of 
numerical simulations with the full coupled-NLS model (\ref{rad1}) and with the perturbation 
theory predictions, the simplified NLS models must take into account 
the main collision-induced position shifts.

We achieve the two goals specified in the beginning of the preceding paragraph 
by using two perturbed single-NLS models and two perturbed coupled-NLS models 
as the simplified propagation models. The two perturbed single-NLS models 
[Eqs. (\ref{rad44}) and (\ref{rad46})] neglect the effects of radiation-induced 
distortion of soliton 2 on radiation dynamics of soliton 1. 
Therefore, in these models, $\psi_{2}(t,z)$ is replaced by the corresponding 
fundamental soliton solution of the unperturbed NLS equation $\psi_{s2}(t,z)$. 
As a result, the term $-2i\epsilon_{3}|\psi_{2}|^2\psi_{1}$ in Eq. (\ref{rad1}) 
turns into the term $-(2i\epsilon_{3}\eta_{2}^{2}\psi_{1})/[\cosh^{2}(x_{2})]$   
in Eqs. (\ref{rad44}) and (\ref{rad46}). The single-NLS model (\ref{rad44}) takes 
into account the effects of Kerr nonlinearity on both collision-induced radiation dynamics 
and collision-induced position shifts. 
This is achieved by turning the term $4|\psi_{2}|^2\psi_{1}$ on the 
left hand side of Eq. (\ref{rad1}), which is associated with the latter effects, 
into the term $-(4\eta_{2}^{2}\psi_{1})/[\cosh^{2}(x_{2})]$ on the 
right hand side of Eq. (\ref{rad44}).     
Next, we incorporate the position shift arising from the collision-induced 
frequency shift experienced by soliton 1 into the model. For this purpose, we first note that 
the latter frequency shift is given by \cite{PNC2010}: 
$\Delta\beta_{13}^{(1)}=-(20\epsilon_{3}\eta_{1}^{2}\eta_{2})/(3|\beta|\beta)$.   
Therefore, by the adiabatic perturbation theory for the fundamental NLS soliton, 
the associated position shift of soliton 1, $y_{1}^{(C)}(z)$, satisfies the equations 
$dy_{1}^{(C)}/dz = 0$ for $z<z_{c}$, and $dy_{1}^{(C)}/dz = -2\Delta\beta_{13}^{(1)}=(40\epsilon_{3}\eta_{1}^{2}\eta_{2})/(3|\beta|\beta)$
for $z \ge z_{c}$. 
Consequently, $y_{1}^{(C)}(z)$ is given by: 
\begin{eqnarray} &&
y_{1}^{(C)}(z)=
\left\{ \begin{array}{l l}
0 &  \mbox{for} \;\;   z<z_{c} \, ,\\
\frac{40\epsilon_{3}\eta_{1}^{2}\eta_{2}}{3|\beta|\beta}(z-z_{c})
& \mbox{for} \;\;   z \ge z_{c} \, ,\\
\end{array} \right. 
\label{simple_NLS_1A}
\end{eqnarray}        
a result that we use in  Eqs. (\ref{rad17}) and (\ref{rad30_add1}) 
for $\psi_{1b}(t,z)$ and $\psi_{1c}(t,z)$.  
On the other hand, it is known that a position shift, 
which is not accompanied by a change in the soliton's shape, 
amplitude, and frequency, is described by a term 
of the form $-iC_{1}(z)\partial_{t}\psi_{1}$ (see, e.g., Ref. \cite{PC2012a}). 
The position shift induced by the latter perturbation term satisfies 
$dy_{1}^{(C)}/dz = C_{1}(z)$. Equating the right hand sides of the 
two equations for $dy_{1}^{(C)}/dz$, we arrive at: 
\begin{eqnarray} &&
C_{1}(z)=
\left\{ \begin{array}{l l}
0 &  \mbox{for} \;\;   z<z_{c} \, ,\\
\frac{40\epsilon_{3}\eta_{1}^{2}\eta_{2}}{3|\beta|\beta}
& \mbox{for} \;\;   z \ge z_{c} \, ,\\
\end{array} \right. 
\label{simple_NLS_1}
\end{eqnarray}      
which is Eq. (\ref{rad45}). Combining all the aforementioned approximations 
for the full coupled-NLS propagation model, we find that the first simplified 
single-NLS model is given by Eq. (\ref{rad44}), that is:  
\begin{eqnarray} &&
i\partial_z\psi_{1}+\partial_{t}^2\psi_{1}+2|\psi_{1}|^2\psi_{1}=
-\frac{2i\epsilon_{3}\eta_{2}^{2}}
{\cosh^{2}(x_{2})}\psi_{1}
-\frac{4\eta_{2}^{2}}{\cosh^{2}(x_{2})}\psi_{1}
-iC_{1}(z)\partial_{t}\psi_{1}.
\label{simple_NLS_2}
\end{eqnarray}

The second simplified single-NLS model [Eq. (\ref{rad46})] takes into 
account the effects of cubic loss on collision-induced radiation dynamics 
and the position shift arising from the collision-induced frequency shift in 
exactly the same manner as in the first single-NLS model [Eq. (\ref{rad44})]. 
Thus, these effects are described by the terms 
$-(2i\epsilon_{3}\eta_{2}^{2}\psi_{1})/[\cosh^{2}(x_{2})]$       
and $-iC_{1}(z)\partial_{t}\psi_{1}$ on the right hand side of Eq. (\ref{rad46}), 
where $C_{1}(z)$ is given by Eq. (\ref{rad45}). 
In addition, the second single-NLS model takes into account the effects of 
Kerr nonlinearity on the collision-induced position shift, but neglects the effects 
of Kerr nonlinearity on the collision-induced radiation dynamics. 
Therefore, in this case, the term $4|\psi_{2}|^2\psi_{1}$ on the 
left hand side of Eq. (\ref{rad1}) is replaced by a term of the form 
$-iC_{2}(z)\partial_{t}\psi_{1}$ on the right hand side of Eq. (\ref{rad46}). 
Combining all the approximations for the full coupled-NLS propagation model, 
we find that the second simplified single-NLS model is given by:        
\begin{eqnarray} &&
i\partial_z\psi_{1}+\partial_{t}^2\psi_{1}+2|\psi_{1}|^2\psi_{1}=
-\frac{2i\epsilon_{3}\eta_{2}^{2}}
{\cosh^{2}(x_{2})}\psi_{1}
-iC_{2}(z)\partial_{t}\psi_{1}
-iC_{1}(z)\partial_{t}\psi_{1},
\label{simple_NLS_3}
\end{eqnarray}                                      
which is Eq. (\ref{rad46}). The function $C_{2}(z)$ is calculated in a similar 
manner to $C_{1}(z)$. For this purpose, we note that the collision-induced 
position shift experienced by soliton 1 due to the effects of Kerr nonlinearity 
is given by Eq. (\ref{rad5_add2}). For simplicity, we assume that 
the contribution of the collision-induced position shift 
to the total position shift of soliton 1, $y_{1}^{(K)}(z)$,    
changes linearly from 0 to $4\eta_{2}/(\beta|\beta|)$ in the small 
interval $[z_{c}-1/(2|\beta|),z_{c}+1/(2|\beta|)]$ about the collision point $z_{c}$. 
It follows that $y_{1}^{(K)}(z)$ is given by: $y_{1}^{(K)}(z)=0$ for $z < z_{c}-1/(2|\beta|)$,   
$y_{1}^{(K)}(z)=4\eta_{2}z/|\beta| - \{ 2\eta_{2}[y_{2}(0)-1] \}/(|\beta|\beta)$ 
for $z_{c}-1/(2|\beta|) \le z \le z_{c}+1/(2|\beta|)$, and 
$y_{1}^{(K)}(z)=4\eta_{2}/(\beta|\beta|)$ for $z > z_{c} + 1/(2|\beta|)$.
As a result, $dy_{1}^{(K)}/dz=0$ for $z < z_{c}-1/(2|\beta|)$ and $z > z_{c} + 1/(2|\beta|)$, 
and $dy_{1}^{(K)}/dz=4\eta_{2}/|\beta|$ for $z_{c}-1/(2|\beta|) \le z \le z_{c}+1/(2|\beta|)$. 
On the other hand, $dy_{1}^{(K)}/dz = C_{2}(z)$.         
Equating the right hand sides of the eqations for $dy_{1}^{(K)}/dz$, we obtain 
\begin{eqnarray} &&
C_{2}(z)=
\left\{ \begin{array}{l l}
0 &  \mbox{for} \;\;   z<z_{c} - \frac{1}{2|\beta|} \, ,\\
\frac{4\eta_{2}}{|\beta|}
& \mbox{for} \;\; z_{c} - \frac{1}{2|\beta|} \le  z \le z_{c} + \frac{1}{2|\beta|} \, ,\\ 
0 & \mbox{for} \;\;   z > z_{c}+ \frac{1}{2|\beta|} \, ,  \\
\end{array} \right. 
\label{simple_NLS_4}
\end{eqnarray}        
which is Eq. (\ref{rad47}).

The two simplified coupled-NLS models [Eqs. (\ref{rad51}) and (\ref{rad53})] 
take into account the effects of radiation-induced distortion of soliton 2 on 
radiation dynamics of soliton 1 in the leading order. That is, these effects are 
taken into account for radiation dynamics induced by cubic loss, but are neglected 
for radiation dynamics induced by Kerr nonlinearity. Consequently, the cubic loss 
terms of the full coupled-NLS model (\ref{rad1}), $-2i\epsilon_{3}|\psi_{2}|^2\psi_{1}$ 
and $-2i\epsilon_{3}|\psi_{1}|^2\psi_{2}$, appear unchanged 
in Eqs. (\ref{rad51}) and (\ref{rad53}). The first simplified 
coupled-NLS model [Eq. (\ref{rad51})] takes into account the effects of 
Kerr nonlinearity on both collision-induced radiation dynamics and collision-induced 
position shifts of each soliton. Therefore, the terms $4|\psi_{2}|^2\psi_{1}$ 
and $4|\psi_{1}|^2\psi_{2}$ on the left hand side of Eq. (\ref{rad1}),  
turn into the terms $-(4\eta_{2}^{2}\psi_{1})/[\cosh^{2}(x_{2})]$ 
and $-(4\eta_{1}^{2}\psi_{2})/[\cosh^{2}(x_{1})]$ on the 
right hand side of Eq. (\ref{rad51}). The position shifts arising from the 
collision-induced frequency shifts are taken into account in the same manner 
as in the simplified single-NLS models (\ref{rad44}) and (\ref{rad46}). 
That is, we include terms of the form $-iC_{1}(z)\partial_{t}\psi_{1}$ 
and $-iC_{3}(z)\partial_{t}\psi_{2}$ on the right hand side of Eq. (\ref{rad51}), 
where $C_{1}(z)$ and $C_{3}(z)$ are given by 
Eqs. (\ref{rad45}) and (\ref{rad52}), respectively \cite{C_3_z}. 
Taking into account all the approximations for the full coupled-NLS model, 
we find that the first simplified coupled-NLS model is given by: 
\begin{eqnarray} &&
i\partial_z\psi_{1}+\partial_{t}^2\psi_{1}+2|\psi_{1}|^2\psi_{1}
=-2i\epsilon_{3}|\psi_{2}|^2\psi_{1}
-\frac{4\eta_{2}^{2}}{\cosh^{2}(x_{2})}\psi_{1}
-iC_{1}(z)\partial_{t}\psi_{1},
\nonumber \\&&
i\partial_z\psi_{2}+\partial_{t}^2\psi_{2}
+2|\psi_{2}|^2\psi_{2}=
-2i\epsilon_{3}|\psi_{1}|^2\psi_{2}
-\frac{4\eta_{1}^{2}}{\cosh^{2}(x_{1})}\psi_{2}
-iC_{3}(z)\partial_{t}\psi_{2}, 
\label{simple_NLS_5}
\end{eqnarray}    
which is Eq. (\ref{rad51}).

The only difference between the second and first simplified coupled-NLS 
models is in the description of the effects of Kerr nonlinearity on 
the collision-induced dynamics. More specifically, the second simplified 
coupled-NLS model [Eq. (\ref{rad53})] takes into account only the 
effects of Kerr nonlinearity on the collision-induced position shifts, while 
the effects of Kerr nonlinearity on the collision-induced radiation dynamics 
are neglected. Therefore, the terms $4|\psi_{2}|^2\psi_{1}$ 
and $4|\psi_{1}|^2\psi_{2}$ on the left hand side of Eq. (\ref{rad1}) 
are replaced by the terms $-iC_{2}(z)\partial_{t}\psi_{1}$ and  
$-iC_{4}(z)\partial_{t}\psi_{2}$ on the right hand side of Eq. (\ref{rad53}), 
where $C_{2}(z)$ and $C_{4}(z)$ are given by Eqs. (\ref{rad47}) 
and (\ref{rad54}), respectively \cite{C_4_z}. 
Combining all the approximations for Eq. (\ref{rad1}), we find that the 
second simplified coupled-NLS model is given by:    
\begin{eqnarray} &&
i\partial_z\psi_{1}+\partial_{t}^2\psi_{1}+2|\psi_{1}|^2\psi_{1}
=-2i\epsilon_{3}|\psi_{2}|^2\psi_{1}
-iC_{2}(z)\partial_{t}\psi_{1}
-iC_{1}(z)\partial_{t}\psi_{1},
\nonumber \\&&
i\partial_z\psi_{2}+\partial_{t}^2\psi_{2}
+2|\psi_{2}|^2\psi_{2}=
-2i\epsilon_{3}|\psi_{1}|^2\psi_{2}
-iC_{4}(z)\partial_{t}\psi_{2}
-iC_{3}(z)\partial_{t}\psi_{2}, 
\label{simple_NLS_6}
\end{eqnarray}    
which is Eq. (\ref{rad53}).


\begin{thebibliography}{}
\bibitem{Zakharov84} S. Novikov, S.V. Manakov, L.P. Pitaevskii, and V.E. Zakharov, 
 {\it Theory of Solitons: The Inverse Scattering Method}  (Plenum, New York, 1984). 
 
\bibitem{Newell85} A.C. Newell, {\it Solitons in Mathematics and Physics} 
(SIAM, Philadelphia, 1985).

\bibitem{Asano69} N. Asano, T. Taniuti, and N. Yajima, 
J. Math. Phys. {\bf 10}, 2020 (1969)

\bibitem{Horton96} W. Horton and Y.H. Ichikawa, 
{\it Chaos and Structure in Nonlinear Plasmas}  
(World Scientific, Singapore, 1996).    

\bibitem{Dalfovo99} F. Dalfovo, S. Giorgini, L.P. Pitaevskii, 
and S. Stringari, Rev. Mod. Phys. {\bf 71}, 463 (1999).  

\bibitem{BEC2008} R. Carretero-Gonz\'alez, D.J. Frantzeskakis, 
and P.G. Kevrekidis, Nonlinearity {\bf 21}, R139 (2008).  

\bibitem{Agrawal2001} G.P. Agrawal, {\it Nonlinear Fiber Optics} (Academic, San Diego, CA, 2001).

\bibitem{Hasegawa95} A. Hasegawa and Y. Kodama, 
{\it Solitons in Optical Communications} (Clarendon, Oxford, 1995). 

\bibitem{Iannone98} E. Iannone, F. Matera, A. Mecozzi, and M. Settembre, 
{\it Nonlinear Optical Communication Networks} (Wiley, New York, 1998).

\bibitem{Mollenauer2006} L.F. Mollenauer and J.P. Gordon,  
{\it Solitons in Optical Fibers: Fundamentals and Applications} (Academic, San Diego, CA, 2006).  

\bibitem{Agrawal2001B} G.P. Agrawal, {\it Applications of Nonlinear Fiber Optics} 
(Academic, San Diego, CA, 2001).

\bibitem{Tkach97} F. Forghieri, R.W. Tkach, and A.R. Chraplyvy,
 in {\it Optical Fiber Telecommunications III}, I.P. 
Kaminow and T.L. Koch, eds., (Academic, San Diego, CA, 1997), Chapter 8.

\bibitem{Essiambre2010} R.-J. Essiambre, G. Kramer, P.J. Winzer, 
G.J. Foschini, and B. Goebel, J. Lightwave Technol. {\bf 28}, 
662 (2010).  

\bibitem{MM98} L.F. Mollenauer and P.V. Mamyshev,
IEEE J. Quantum Electron. {\bf 34}, 2089  (1998).

\bibitem{PNH2017B} A. Peleg, Q.M. Nguyen, and T.T. Huynh, 
Eur. Phys. J. D {\bf 71}, 315 (2017). 

\bibitem{CP2005} Y. Chung and A. Peleg, Nonlinearity {\bf 18}, 1555 (2005).

\bibitem{PCG2003} A. Peleg, M. Chertkov, and I. Gabitov, 
Phys. Rev. E {\bf 68}, 026605 (2003).

\bibitem{Malomed89} Y.S. Kivshar and B.A. Malomed, 
Rev. Mod. Phys. {\bf 61}, 763 (1989). 

\bibitem{Malomed86} Y.S. Kivshar and B.A. Malomed, 
J. Phys. A {\bf 19}, L967 (1986).  

\bibitem{Malomed91a} B.A. Malomed, Phys. Rev. E {\bf 43}, 3114 (1991).  

\bibitem{Malomed91b} B.A. Malomed, Phys. Rev. A {\bf 44}, 1412 (1991).

\bibitem{Kumar98} S. Kumar, Opt. Lett. {\bf 23}, 1450 (1998).  

\bibitem{SP2004} J. Soneson and A. Peleg, Physica D {\bf 195}, 123 (2004).  

\bibitem{CPN2016} D. Chakraborty, A. Peleg, and Q.M. Nguyen, 
Opt. Commun. {\bf 371}, 252 (2016).  

\bibitem{PNT2016} A. Peleg, Q.M. Nguyen, and T.P. Tran, 
Opt. Commun. {\bf 380}, 41  (2016).   

\bibitem{CCDG2003} M. Chertkov, Y. Chung, A. Dyachenko, I. Gabitov, 
I. Kolokolov, and V. Lebedev, Phys. Rev. E {\bf 67}, 036615 (2003).

\bibitem{PCG2004} A. Peleg, M. Chertkov, and I. Gabitov, 
J. Opt. Soc. Am. B {\bf 21}, 18 (2004).

\bibitem{PC2003} A. Peleg and Y. Chung, J. Phys. A {\bf 36}, 
10039 (2003).

\bibitem{PNC2010} A. Peleg, Q.M. Nguyen, and Y. Chung, 
Phys. Rev. A {\bf 82}, 053830 (2010).

\bibitem{PC2012b} A. Peleg and Y. Chung, Phys. Rev. A {\bf 85}, 063828 (2012).

\bibitem{P2004} A. Peleg, Opt. Lett. {\bf 29}, 1980 (2004).

\bibitem{NP2010b} Q.M. Nguyen and A. Peleg,  
J. Opt. Soc. Am. B {\bf 27}, 1985 (2010).   

\bibitem{cubic_loss_1} This assumption is justified in the penultimate 
paragraph of the introduction. 

\bibitem{CP2008} Y. Chung and A. Peleg, Phys. Rev. A  {\bf 77}, 063835 (2008).

\bibitem{PC2012a} A. Peleg and Y. Chung, Opt. Commun. {\bf 285}, 
1429 (2012).  

\bibitem{Boyd2008} R.W. Boyd, {\it Nonlinear Optics} (Academic, San Diego, CA, 2008). 

\bibitem{Agrawal2007a} Q. Lin, O.J. Painter, and G.P. Agrawal, 
Opt. Express {\bf 15}, 16604 (2007).

\bibitem{Dekker2007} R. Dekker, N. Usechak, M. F\"orst, and A. Driessen, 
J. Phys. D {\bf 40}, R249 (2007). 

\bibitem{Borghi2017} M. Borghi, C. Castellan, S. Signorini, A. Trenti, and L. Pavesi, 
J. Opt. {\bf 19}, 093002 (2017).  

\bibitem{Stegeman89} V. Mizrahi, K.W. DeLong, G.I. Stegeman, 
M.A. Saifi, and M.J. Andrejco, Opt. Lett. {\bf 14}, 1140 (1989). 

\bibitem{Silberberg90} Y. Silberberg, Opt. Lett. {\bf 15}, 1005 (1990).

\bibitem{Aceves92} A.B. Aceves and J.V. Moloney, Opt. Lett. {\bf 17}, 
1488 (1992).

\bibitem{Kivshar95} V.V. Afanasjev, J.S. Aitchison. and Y.S. Kivshar, 
Opt. Commun. {\bf 116}, 331 (1995).   
 
\bibitem{Tsoy2001} E.N. Tsoy, C.M. de Sterke, and F.Kh. Abdullaev, 
J. Opt. Soc. Am. B {\bf 18}, 1144 (2001).  

\bibitem{Silberberg2008} O. Katz, Y. Lahini, and Y. Silberberg, 
Opt. Lett. {\bf 33}, 2830 (2008). 

\bibitem{PCDN2009} A. Peleg, Y. Chung, T. Dohnal, and Q.M. Nguyen, 
Phys. Rev. E {\bf 80}, 026602 (2009).

\bibitem{Gaeta2012} Y. Okawachi, O. Kuzucu, M.A. Foster, R. Salem, 
A.C. Turner-Foster, A. Biberman, N. Ophir, K. Bergman, M. Lipson, 
and A.L. Gaeta, IEEE Photon. Technol. 
Lett. {\bf 24}, 185 (2012).

\bibitem{Gaeta2008} M.A. Foster, A.C. Turner, M. Lipson, and A.L. Gaeta, 
Opt. Express {\bf 16}, 1300 (2008).

\bibitem{Soref2006} R. Soref, IEEE J. Sel. Top. Quantum Electron. {\bf 12}, 1678 (2006). 

\bibitem{Hagan2002} R.A. Negres, J.M. Hales, A. Kobyakov, D.J. Hagan, 
and E.W. Van Stryland, IEEE J. Quantum Electron. {\bf 38}, 1205 (2002).   

\bibitem{Hagan2011} C.M. Cirloganu, L.A. Padilha, D.A. Fishman, S.Webster, 
D.J. Hagan, and E.W. Van Stryland, Opt. Express {\bf 19}, 22951 (2011). 

\bibitem{Rauscher97} C. Rauscher and R. Laenen, J. Appl. Phys. {\bf 81}, 
2818 (1997).  


\bibitem{Islam2004} M.N. Islam (Ed.), {\it Raman Amplifiers for Telecommunications 1: Physical Principles} 
(Springer, New York, 2004).

\bibitem{Agrawal2005} C. Headley and G.P. Agrawal (Eds.), 
{\it Raman Amplification in Fiber Optical Communication Systems} (Elsevier, San Diego, CA, 2005).

\bibitem{Jalali2003} R. Claps, D. Dimitropoulos, V. Raghunathan, Y. Han, 
and B. Jalali, Opt. Express {\bf 11}, 1731 (2003). 

\bibitem{Lipson2004} Q. Xu, V.R. Almeida, and M. Lipson, 
Opt. Express {\bf 12}, 4437 (2004).

\bibitem{Cohen2005b} R. Jones, H. Rong, A. Liu, A. Fang, M. Paniccia, 
D. Hak, and O. Cohen, Opt. Express {\bf 13}, 519 (2005)     

\bibitem{dimensions1} The dimensionless distance $z$ in Eq. (\ref{rad1})  is 
$z=X/(2L_{D})$, where $X$ is the dimensional distance, 
$L_{D}=\tau_{0}^{2}/|\tilde\beta_{2}|$ is the dispersion length,
$\tau_{0}$ is the soliton width, and $\tilde\beta_{2}$ is the second-order
dispersion coefficient. The dimensionless time is
$t=\tau/\tau_{0}$, where $\tau$ is time. 
$\psi_{j}=(\gamma \tau_{0}^{2}/|\tilde\beta_{2}|)^{1/2}E_{j}$, 
where $E_{j}$ is the electric field of the $j$th pulse, $j=1,2$, 
and $\gamma$ is the Kerr nonlinearity coefficient.
The coefficient $\epsilon_{3}$ is related to the dimensional cubic loss 
coefficient $\rho_{3}$ by $\epsilon_{3}=2\rho_{3}/\gamma$. 

\bibitem{Nakazawa2000} M. Nakazawa, 
IEEE J.  Sel. Top. Quantum Electron. {\bf 6}, 1332 (2000).     

\bibitem{Kaup90} D.J. Kaup, Phys. Rev. A {\bf 42}, 5689 (1990).

\bibitem{Kaup91} D.J. Kaup, Phys. Rev. A {\bf 44}, 4582 (1991). 

\bibitem{Yang2010} J. Yang, {\it Nonlinear Waves in Integrable and Nonintegrable Systems} 
(SIAM, Philadelphia, 2010).    

\bibitem{numerics1} The interval $-10.77 \le t \le 21.44$ consists of several subintervals 
on which one of the two approaches is more accurate than the other. 

\bibitem{numerics2} The interval $-6.28 \le t \le 6.54$ consists of several subintervals on which 
one of the two approaches is more accurate than the other.                                                                                    

\bibitem{Kaup76} D.J. Kaup, J. Math. Anal. Appl. {\bf 54}, 849 (1976).

\bibitem{C_3_z} Note that the collision-induced frequency shift experienced by 
soliton 2 is \cite{PNC2010}: 
$\Delta\beta_{23}^{(1)}=(20\epsilon_{3}\eta_{1}\eta_{2}^{2})/(3|\beta|\beta)$. 
Therefore, Eq. (\ref{rad52}) for $C_{3}(z)$ is similar to Eq. (\ref{rad45}) for $C_{1}(z)$, 
except that $C_{3}(z)=-(40\epsilon_{3}\eta_{1}\eta_{2}^{2})/(3|\beta|\beta)$ 
for $z \ge z_{c}$. 

\bibitem{C_4_z} Note that the collision-induced position shift experienced by 
soliton 2 due to the effects of Kerr nonlinearity is \cite{MM98,CP2005,PCG2003}:      
$\Delta y_{22}^{(0)}=-4\eta_{1}/(\beta|\beta|)$.
Therefore, Eq. (\ref{rad54}) for $C_{4}(z)$ is similar to Eq. (\ref{rad47}) for $C_{2}(z)$, 
except that $C_{4}(z)=-4\eta_{1}/|\beta|$ 
for $z_{c} - 1/(2|\beta|) \le  z \le z_{c} + 1/(2|\beta|)$.


\end{thebibliography}
\end{document}